\documentclass[aps,preprint,floats,epsf,epsfig,nofootinbib,letter]{revtex4}
\usepackage{graphicx}
\usepackage{dcolumn}
\usepackage{bm}
\usepackage{subfigure}
\usepackage{mathrsfs}
\usepackage{amssymb}
\usepackage{color}
\usepackage{caption}
\usepackage{ragged2e}
\usepackage{multirow}
\usepackage{hyperref} 
\usepackage{color}

\def\be{\begin{eqnarray}}
\def\en{\end{eqnarray}}
\def\non{\nonumber}
\def\la{\langle}
\def\ra{\rangle}
\def\pslash{\!\not{\!\partial}}

\def\pslash{\!\not{\! p}}
\def\kslash{\!\not{\! k}}
\def\Dslash{\!\not{\! \! D}}

\begin{document}

\renewcommand{\baselinestretch}{1.10}
\font\el=cmbx10 scaled \magstep2{\obeylines\hfill Oct., 2017}

\vskip 1.5 cm

\centerline{\Large\bf  Confronting Dirac Fermionic Dark Matter with Recent Data}

\bigskip
\centerline{\bf Chun-Khiang Chua, Gwo-Guang Wong}
\medskip
\centerline{Department of Physics and Chung Yuan Center for High Energy Physics,}
\centerline{Chung Yuan Christian University,}
\centerline{Taoyuan, Taiwan 32023, Republic of China}
\medskip

\centerline{\bf Abstract}
\bigskip

We study Dirac fermionic dark matter (DM, $\chi^0$) and confront it with recent data. To evade the stringent direct search limits from PandaX-II, XENON1T and LUX experiments, the quantum numbers of the Dirac DM are taken to be $I_3=Y=0$ to remove the tree-level $Z$-exchange diagram. Loop amplitudes can contribute to the elastic scattering cross section. We find that there are cancellations in the one-loop diagrams, which largely reduce the cross section and make the Dirac DM viable in the direct search. For a generic isospin $I$, we survey the Dirac DM mass constrained by the latest results of PandaX-II, XENON1T and LUX experiments, the observed DM relic density, and the H.E.S.S. and the Fermi-LAT astrophysical observations. Sommerfeld enhancement effects on DM annihilation processes are investigated. We find that the cross section of $\chi^0\bar\chi^0$ annihilating to the standard model (SM) gauge bosons are in general significantly enhanced, and the Fermi-LAT, the H.E.S.S. upper limits on $\langle\sigma v\rangle({W^+W^-,\gamma\gamma})$ and the observed relic density become serious constraints on the Dirac DM mass. The $I<4$ cases are ruled out and for $I\geq 4$, the lower bound on Dirac DM mass are forced to be $\gtrsim$ 60 TeV. The elastic scattering cross section for $m_\chi$ of few tens TeV with a generic $I$ is found to be $\sigma^{SI}\simeq I^2(I+1)^2\times7\times10^{-49}$~cm$^2$. The predicted $\la\sigma (\chi^0\bar\chi^0\to Z^0Z^0, Z^0\gamma, \gamma\gamma)v\ra$ and $\sigma^{SI}$ are sizable and they will be useful to search for DM in astrophysical observation and in direct search in near future.

\bigskip
\small


\maketitle

\section{Introduction}

The existence of dark matter (DM) in different astrophysical scales of the Universe~\cite{RF,BBSnBxxx,PDG1,Carroll,Cxxx,WMAPa,SDSS} provides strong evidence for physics beyond the Standard Model (SM). 
Weakly interacting massive particles (WIMPs) are promising DM candidates.
The WIMPs are assumed to be created thermally during the big bang, and froze out of thermal equilibrium escaping the Boltzmann suppression in the early Universe.
The DM relic density is approximately related to the velocity averaged DM annihilation cross section by a simple relation~\cite{JKG},
\be
\Omega_\chi h^2\approx {\frac{0.1 {\rm pb} \times c}{<\sigma v>}}.
\en
Comparing to the measured value of CDM relic density is~\cite{PDG}
\be
\Omega_{\rm obs} h^2=0.1186\pm 0.0020,
\en
it suggests the case of DM with mass in the range of 100 GeV to few TeV and governed by an electroweak size interaction.
That is well known as the WIMP miracle~\cite{DG}. 

There are three complementary searching strategies to detect the DM particles in experiments (see~\cite{DG,Arcadi:2017kky} for brief reviews).
They are the direct detection of DM-nucleus scattering in underground laboratories, 
the indirect detection of DM annihilation processes in astrophysical observation 
and the DM direct production at colliders~\cite{DG,Arcadi:2017kky,LHC-0,LHC-1,LHC-2,LHC-2.5,LHC-3,LHC-4,Mitsou,GIST,FHKT}. 
So far these searches only provide upper limits.
Recently upper limits on spin-independent (SI),  spin-dependent (SD)
DM-nucleus scattering cross sections are reported by
PandaX-II~\cite{PandaX-II}, XENON1T~\cite{Xenon1T2017-SI}, LUX~\cite{LUX2016-SI, LUX2016-SD} experiment groups, while that on velocity averaged DM annihilation cross sections are from H.E.S.S~\cite{HESS2016,HESS2017rr} and Fermi-LAT~\cite{Fermi-LAT2015} groups.
In spite of the fact that DM contains about $85\%$ for the total mass in the universe~\cite{WMAPb,Plank}, we still do not know much about its nature. 
In the literature, the possibilities of DM particle as a scalar~\cite{BPV,PW,He1,BBEG,He2,DS,GWZ}, fermion~\cite{JKG,Haber,RSW,Fowlie,CS,Cirelli2006,BB,Dutra,ChuaWong,chua,APQ,Yaguna,Arcadi,Dedes,Agrawal} or a vector boson~\cite{BBP,DFT,KMY,BKMNT,DGM,CT,KMNO,LLM,KT1,KT2,Agrawal} are considered.


In this paper, we study the case of a Dirac fermionic DM. 
We investigate a renormalizable DM model by introducing a
pure weak eigenstate Dirac fermion as a DM candidate.
The DM only interacts through gravity and weak interactions~\cite{CS,Cirelli2006,chua}. 
We will confront the model to the recent experimental data of relic density, direct and indirect detection
experiments.

This paper is organized as follows. In Sec. II we introduce the model. 
In Sec. III we calculate the cross sections of DM-nucleus elastic scattering and compare them with the results of recent LUX, XENON1T and PandaX-II experiments in the direct search. 
In Sec. IV we calculate the velocity averaged cross section of DM annihilating to the SM particles and the DM relic density incorporating the Sommerfeld enhancement effect. 
We compare our results with the Fermi-LAT and the H.E.S.S data and the observed relic density, respectively.  
Conclusions are given in Sec. V.
We collect the loop integral functions in Appendix~A, followed by a comprehensive derivation of Sommerfeld factors in Appendix~B.

\section{A Dirac fermionic dark matter model}

We introduce a Dirac fermionic dark matter model by adding on SM a Dirac fermionic multiplet $\chi$ with arbitrary $I$ and $Y$,
and the Lagrangian can be written as~\cite{CS,Cirelli2006,chua}
\be
{\cal L}&=&{\cal L}_{SM}+{\bar \chi}(i\gamma^\mu D_\mu -m_\chi)\chi
\non\\
&=&{\cal L}_{SM}+{\bar \chi^j}\gamma^\mu(i\partial_\mu-gW^a_\mu T^a_{jk}+g'B_\mu Y_j \delta_{jk})\chi^k-m_\chi\bar\chi^i \chi^i,
\label{eq: L1}
\en
where $W^a_\mu$ and $B_\mu$ are the familiar electroweak $\rm {SU(2)_L}$ and $\rm{U(1)}$  gauge fields, respectively. 
In this model, the SM is minimally extended so that the DM can only interact with the SM gauge bosons. 
Besides, there is only one free parameter, the DM mass $m_\chi$ that makes the model predictive. 
As in most DM models, an addition discrete symmetry is needed to assure the stability of DM. 
For example, we may assign the Dirac DM with $Z_2$-odd quantum number and the SM particles with a $Z_2$-even quantum number to maintain the stability of DM. 

The Lagrangian for the WIMPs interacting with the SM gauge bosons can be extracted from Eq.(\ref{eq: L1}), 
giving
\be
{\cal L'}&=&-\frac{g}{\sqrt 2}{\bar \chi^j}\gamma^\mu(W^+_\mu T^+_{jk}+W^-_\mu T^-_{jk})\chi^k
-{\bar \chi^j}\gamma^\mu [ ( g\cos\theta_W T^3_{jk}-g'\sin\theta_W Y_j \delta_{jk})Z_\mu+eA_\mu Q) ] \chi^k.
\label{eq: L2}
\en
Here we have used the weak mixing angle $\theta_W$ and the relation $e=g\sin\theta_W=g'\cos\theta_W$ with electric charge $Q=T^3+Y$. 
As the DM is electrically neutral, the Lagrangian for DM interacting with $Z$-boson can be further extracted as
\be
{\cal L'}_{\chi^0}=-\frac{g}{\cos\theta_W}T^3{\bar \chi^0}\gamma^\mu \chi^0 Z_\mu.
\label{eq: L3}
\en
The tree-level vector interaction of DM with $Z$-boson leads to the SI elastic cross section with a nucleus $N$:
\be
\sigma^{\rm SI}_A(\chi N \rightarrow \chi N)=\frac{\mu^2_N}{4\pi}\bigg(\frac{g}{\cos\theta_W M_Z}\bigg)^4 I_3^2 \bigg[-\frac{1}{4}(A-Z)+(\frac{1}{4}-\sin\theta_w^2)Z \bigg]^2,
\en
where $Z$ and $A$ are the numbers of protons and nucleons in the target nucleus respectively, $I_3$ is the weak isospin quantum number and $\mu_N$ is the of 
DM-nucleus reduced mass. It corresponds to the normalized cross section with a nucleon:
\be
\sigma^{SI}_{p,n}\simeq I^2_3\times 10^{-39} {\rm cm^2},
\en
for $m_\chi$ ranges from a few GeV to a few 100 TeV.

The above cross section has already been ruled out in the direct search of SI-experiments of DM-nucleus scattering for many years, unless
the quantum numbers of the Dirac DM are either (i)~$I\neq 0, I_3=Y=0$ or (ii) $I=Y=0$~\cite{chua}. 
In this model, we shall consider the first case, i.e. $I\neq 0$, but $I_3=Y=0$.
Nevertheless, we still need to check the contribution to the SI elastic cross section from loop diagram. 

As we know that the DM is highly nonrelativistic in the DM-nucleus elastic scattering. 
%
In literature, the loop contributions to DM-nucleus elastic cross section are explored in the Majorana fermionic DM case~\cite{Cirelli2009,HINT,Essig}. 
In this paper we will study the loop contribution to the elastic DM-nucleus scattering in the Dirac fermionic DM case.

\section{Dirac fermionic dark matter in the direct search}

\subsection{Effective Lagrangian}

In this paper, we only consider the case: $I\neq 0, I_3=Y=0$. Hence the elastic cross section of Dirac DM-nucleus scattering is vanishing at tree level. 
Note that the Majorana fermionic DM case were studied in \cite{Cirelli2006,Cirelli2009}. 
The effective Lagrangian of Dirac DM with quarks and gluon can be written as~\cite{JKG,HINT}
\be
{\cal L}^{\rm eff}&=&\sum_{q=u,d,s}{\cal L}^{\rm eff}_q+{\cal L}^{\rm eff}_g,
\\
{\cal L}^{\rm eff}_q&=&
a_q[{\bar \chi^0}\chi^0][{\bar q}q]+
b_q [{\bar \chi^0}\gamma^\mu \chi^0][{\bar q}\gamma_\mu q]+
d_q [{\bar \chi^0}\gamma^\mu\gamma^5 \chi^0][{\bar q}\gamma_\mu\gamma^5 q]
\non\\
&+&\frac{g_q^{(1)}}{m_\chi}\bar\chi^0i\partial^\mu\gamma^\nu\chi^0{\cal O}_{\mu\nu}^q+\frac{g_q^{(2)}}{m_\chi}\bar\chi^0(i\partial^\mu)(i\partial^\nu)\chi^0{\cal O}_{\mu\nu}^q,
\\
{\cal L}^{\rm eff}_g&=&
f_G{\bar \chi^0}\chi^0G_{\mu\nu}^aG^{a\mu\nu}.
\label{eq: Leff}
\en
where the quark twist-2 operator, ${\cal O}_{\mu\nu}^q$, is defined as
\be
{\cal O}_{\mu\nu}^q\equiv\frac{1}{2}\bar qi\bigg(D_\mu\gamma_\nu+D_\nu\gamma_\mu-\frac{1}{2}g_{\mu\nu}\Dslash\bigg)q.
\en
\begin{figure}[b!]
  \centering
  \includegraphics[width=13cm]{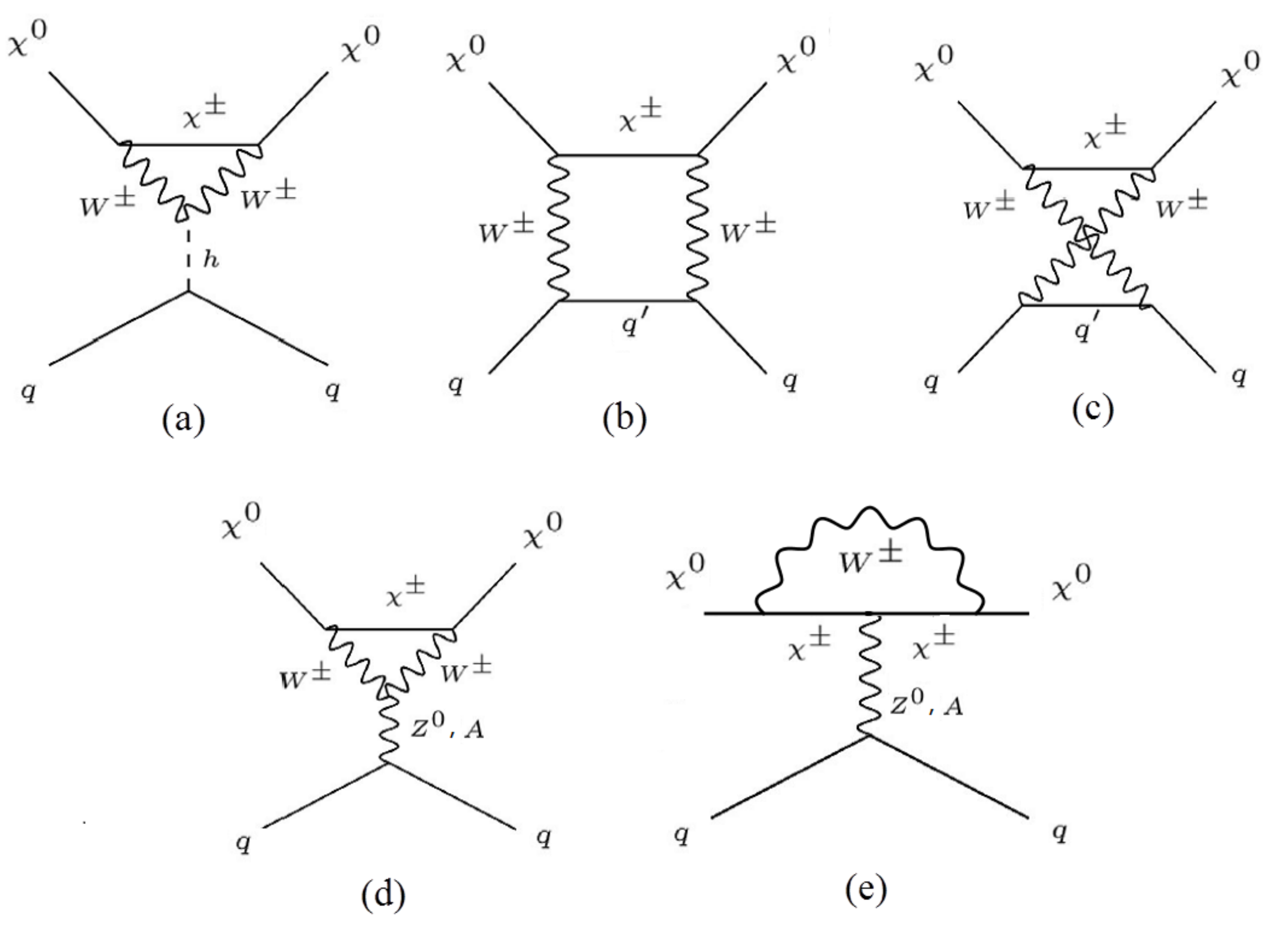}\\
  \caption{One-loop diagrams for the effective interaction of Dirac DM with quarks}
  \label{loop}
\end{figure}

Fig.~\ref{loop} shows the one-loop Feynman diagrams which induce the effective interactions of DM $\chi^0$ with quarks. 
Note that in the Dirac fermionic DM case, $\chi^0$ and $\bar\chi^0$ are different particles. 
There is an additional set of diagrams with $\chi^0$ in Fig.~\ref{loop} replaced by $\bar\chi^0$. These diagrams are related. 
In fact, one can show that the former amplitude $M_{\chi^0}$ is equal to latter one $M_{\bar\chi^0}$.
Now we let  $n_{\chi^0}$ and $n_{\bar\chi^0}$ be the number density of the particle $\chi^0$ and the antipaticle $\bar\chi^0$, respectively. 
They add up to the total number density of DM, $n_{DM}=n_{\chi^0}+n_{\bar\chi^0}$.
Using $M_{\chi^0}=M_{\bar\chi^0}$, we have 
$n_\chi^0|M_{\chi^0}|^2+n_{\bar\chi^0}|M_{\bar\chi^0}|^2=(n_\chi^0+n_{\bar\chi^0})|M_{\chi^0}|^2= n_{DM}|M_{\chi^0}|^2$. 
Hence we only need to calculate $M_{\chi^0}$.

Note that the amplitudes correspond to the Feynman diagrams in Figs.~\ref{loop} (d) and (e) turns out to be vanishing individually.  
Both diagrams have two sub-diagrams.
The relative sign differences between the trilinear couplings of the vector bosons $W^+W^+V^0$ and $W^-W^-V^0$ ($V^0=Z$ or $A$) are responsible for the cancelation of diagrams in Fig.~\ref{loop} (d), while relative sign differences between the couplings $\chi^-\chi^-V^0$ and $\chi^+\chi^+V^0$ are responsible for vanishing of the sum of diagrams in Fig.~\ref{loop} (e).
\footnote{
For example, the photon exchange diagram in Fig.~\ref{loop} (d) contains two sub-diagrams: with $\chi^+$, $W^-$ or $\chi^-$, $W^+$ running in the loop. The corresponding matrix elements are 
\be
M_{\pm}&=&\int\frac{d^4k}{(2\pi)^4}[{\bar \mu}(p_3)(\frac{-i}{\sqrt{2}}gT^{\mp}_{0\pm}\gamma^{\mu'})(\frac{i}{\pslash_1+\kslash-m_\chi})(\frac{-i}{\sqrt{2}}gT^{\pm}_{\pm0}\gamma^{\nu'})u(p_1)](\frac{-ig_{\mu\mu'}}{(k+p_1-p_3)^2-M_W^2})\times
\non\\
& &\{\mp i \sin\theta_W[(p_1-p_3)_\lambda g_{\mu\nu}+(k-p_3+p_1)_\mu g_{\nu\lambda}+(2p_3-2p_1-k)_\nu g_{\lambda\mu]}\}(\frac{-ig_{\nu\nu'}}{k^2-M_W^2})\times 
\non\\
& &\frac{-ig_{\lambda\kappa}}{(p_3-p_1)^2}[{\bar u}_q(p_4)(-iQ_qe\gamma^\kappa)u_q(p_2)],
\en 
where $p_{1(3)}$ and $p_{2(4)}$ are the momenta of incident (outgoing) $\chi$ and $q$, respectively, and 
$T^-_{0+}=T^+_{+0}=T^+_{0-}=T^-_{-0}=\sqrt{I(I+1)}$.
Consequently, we have $M_++M_-=0$ from the relative sign difference between the trilinear couplings of the vector bosons $W^+W^+A$ and $W^-W^-A$.}
Hence, only the Feynman diagrams in Figs.~\ref{loop}(a)-(c) contribute to the effective interactions.

We calculate the effective couplings $a_q$, $b_q$ and $d_q$ from Fig.~1 and use $g_q^{(1)}$, $g_q^{(2)}$ obtained in Ref.~\cite{HINT}. 
In the nonrelativistic limit with $m_\chi\gg m_t, m_W$, and ignoring the mass splitting among the neutral DM ($\chi^0$) and the single charged WIMP partners ($\chi^\pm$),
we have~\footnote{Our $a_q$ and $d_q$ agree with those in \cite{HINT} up to an additional factor 2 for $m_\chi\gg m_t, m_W$. 
Note that $b_q=0$ for Majorana DM in Ref.~\cite{HINT}.  
The additional factor is from the differences of the Lagrangians and amplitudes in the Dirac and Majorana cases. 
We apply the factor 2 to $g_q^{(1)}$ and $g_q^{(2)}$ when adapting the results from Ref.~\cite{HINT}. \label{fn: majorana->dorac}} 
\be
 a_q&=&I(I+1)\frac{g^4}{16\pi^2}\bigg[\frac{m_q}{M_W m^2_h}F^S_h(q,x,y)+\frac{m_q}{M^3_W}F^S_{W}(q,x,y)\bigg ], \non\\
 & & \non\\
 b_q&=&I(I+1)\frac{g^4}{16\pi^2} \bigg[\frac{m_q}{M_W^3}F^V_{W_1}(q,x,y)+\frac{1}{M^2_W} F^V_{W_2}(q,x,y)  \bigg], \non\\
 & & \non\\
 d_q&=&I(I+1)\frac{g^4}{16\pi^2} \bigg[\frac{1}{M_W^2}F^A_{W_1}(q,x,y)+\frac{1}{m^2_\chi}F^A_{W_2}(q,x,y) \bigg], \non\\
 g_q^{(1)}&=& I(I+1)\frac{g^4}{16\pi^2}\frac{1}{M_W^3}g_{T1}(x), \non\\
 g_q^{(2)}&=& I(I+1)\frac{g^4}{16\pi^2}\frac{1}{M_W^3}g_{T2}(x),
\label{eq:qcouplings}
\en 
where we define $x\equiv {M_W^2}/{m_\chi^2}$, $y\equiv {m_t^2}/{m_\chi^2}$ with $m_q$, $m_t$ and $m_W$ the quark $q$, $t$ and the $W$-boson masses, respectively. 
Note that $a_q$, $b_q$ and $d_q$ contribute to the so-called SS, VV and AA interactions, respectively, while
$g_q^{(1)}$ and $g_q^{(2)}$ also contribute to the SS interaction.
All loop integral functions $F^{S}_{h}(q,x,y)$, $F^{V,A}_{W_{1,2}}(q,x,y)$, $g_{T1}(x)$ and $g_{T2}(x)$ are collected in Appendix~A. 

The two-loop diagrams shown in Fig.~\ref{fig:gluon} produce the effective scalar coupling of Dirac DM with gluon in a nucleon, $f_G$ defined in Eq.~(\ref{eq: Leff}). 
These diagrams contribute to the effective scalar coupling, $f_G$ giving
\be
f_G=f_G^{(a)}+f_G^{(b)}+f_G^{(c)},
\en
where the superscripts correspond to the labels of the diagrams in Fig.~\ref{fig:gluon}.
We use the results obtained in Ref.~\cite{HINT} for these $f_G^{(i)}$. 

\begin{figure}[t]
  \centering
  \includegraphics[width=14cm]{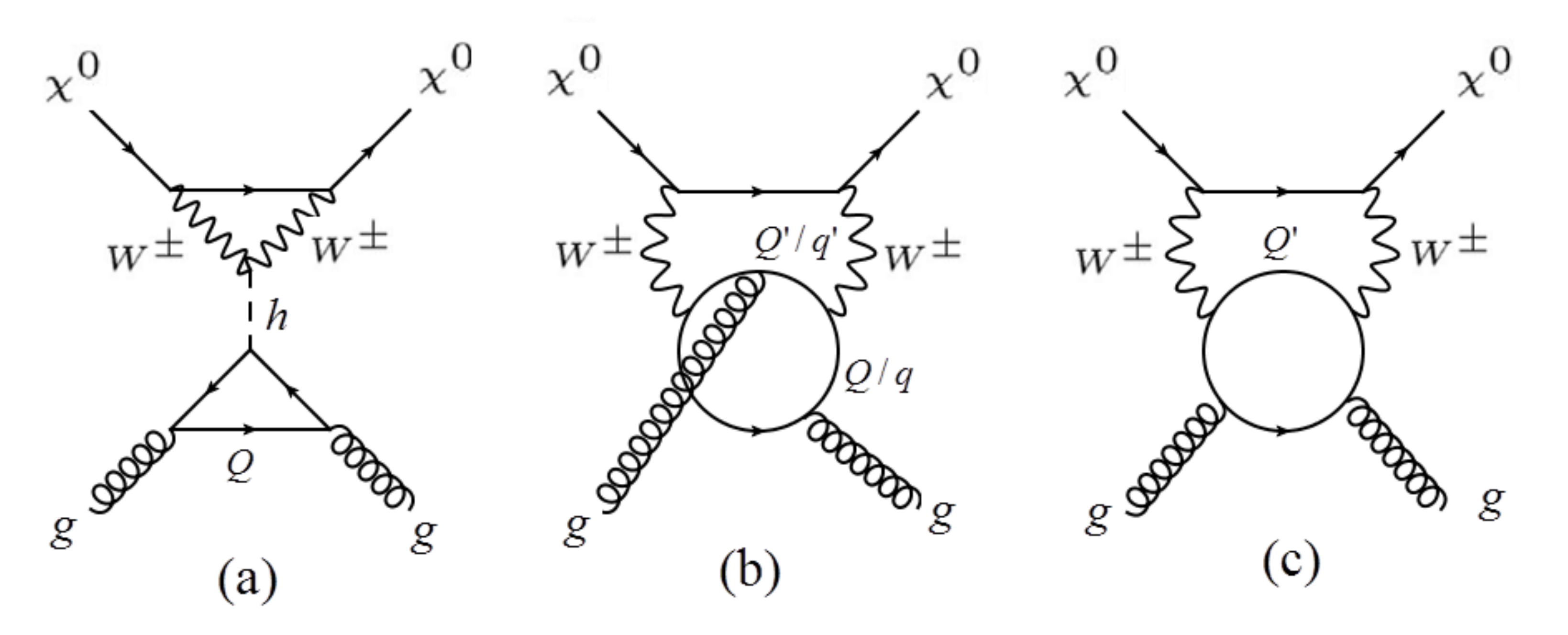}\\
  \caption{Two-loop diagrams contributing to the effective scalar coupling of Dirac DM with gluons.}   \label{fig:gluon}
\end{figure}

The heavy quark contribution to the mass of the nucleon through the triangle diagram~\cite{SVZ} shown in Fig.~\ref{fig:gluon}(a) gives the effective scalar coupling:
\be
f_G^{(a)}=-\frac{\alpha_s}{12\pi}\frac{g^4}{16\pi^2}I(I+1)c_Q\frac{1}{4 m_W}g_H(x)
\simeq -\frac{\alpha_s}{12\pi}\frac{a_Q}{m_Q},
\en 
where the loop integral function $g_H(x)$ with $x\equiv {M_W^2}/{m_\chi^2}$ is basically from~\cite{HINT} and is collected in Appendix A,~\footnote{Similarly, $f_G^{(i)}\ (i=a,b,c)$ used here have an additional factor of two compared to those in Ref.~\cite{HINT}. 
See footnote \ref{fn: majorana->dorac}.} 
$c_Q=1+11\alpha_s(m_Q)/4\pi$ with $Q=c,b,t$ and $a_Q$ is defined in Eq.~(\ref{eq:qcouplings}).
Note that in the above equation we have taken $c_Q=1$ for $Q=c,b,t$ for simplicity, while in Ref.~\cite{HINT},
$c_c=1.32, c_b=1.19$ and $c_t=1$ for  $\alpha_s(m_z)=0.118$ are used. 
From Figs.~\ref{fig:gluon}(b) and (c), 
we have~\cite{HINT}
\be
f_G^{(b)} + f_G^{(c)}=\frac{\alpha_s}{4\pi}\frac{g^4}{16\pi^2}I(I+1)\frac{1}{M_W^3}g_W(x,y),
\en
the loop integral function $g_W(x,y)$ with $x\equiv {M_W^2}/{m_\chi^2}$, $y\equiv {m_t^2}/{m_\chi^2}$ is collected in Appendix A.
Note that the contribution of the twist-2 operators of gluon which are suppressed by the strong coupling constant $\alpha_s$ have been ignored 
(see~Ref.\cite{Hisano2012}).

In Appendix~A, the loop integral functions $F^{S,V,A}_{h,W_{1,2}}(q,x,y)$ can be classified into three classes as whether the external line in Fig.~\ref{loop} involves light, $b$ or $t$ quarks. 
For the light quark case, we only keep the effective coupling up to the leading order in $m_q$. 
For the $b$-quark case, the top quark is in the loop of the box diagrams 
and the loop integral functions can be analytically solved under the assumption: $m_\chi\gg m_t$. 
For the $t$-quark case, the bottom quark is in the loop of the box diagrams 
and the loop integral functions can only be numerically solved (with the assumption: $m_\chi\gg m_t$). 
Note that there is a cancellation in the VV interaction so that the large component $(\sim 1/M^2_W)$ of the effective coupling $b_q$ is vanishing for all quark flavors; in other words, $F^V_{W2}(q,x,y)=0$. We will see later that it makes the Dirac DM model viable in the direct search. 
Note that a similar cancellation in the Majorana case was reported in Ref.~\cite{HINT}.

\subsection{Spin-independent and Spin-dependent Cross Sections}

To compare with the results of recent PandaX-II~\cite{PandaX-II}, XENON1T~\cite{Xenon1T2017-SI} and 
LUX~\cite{LUX2016-SI,LUX2016-SD} experiments, we calculate the normalized SI and SD cross sections of Dirac DM elastic scattering off $^{129,131}{\rm Xe}$ nuclei. 
We shall obtain the averaged unpolarized amplitude squared $\overline{\sum} |M_{fi}|^2$ at $q^2=0$ first. 
Note that there is no interference between various interaction terms in $i M_{fi}$. 
From the effective Lagrangian of DM-nucleus elastic scattering, the unpolarized matrix element square can be written as~\cite{ChuaWong} 
\be
\overline{\sum}|M_{fi}|^2(q^2=0)
&=&\overline{\sum}|M^{SI}_{fi}|^2(q^2=0)+\overline{\sum}|M^{SD}_{fi}|^2(q^2=0)
\non\\
&=&
16 m_{\cal N}^2 m^2_\chi
\bigg[\bigg(f^2_{s\cal N}
+Q_{V\cal N}^2\bigg)
+4 Q_{A\cal N}^2 J_{\cal N}(J_{\cal N}+1)\bigg].
\label{eq: iM^2}
\en
The first two terms are from the effective SS and VV interactions, respectively, 
contributing to $|M^{SI}|^2$, 
while the last term from the AA interaction contributing to $|M^{SD}|^2$.

For the VV interaction, we have
\be
Q_{V\cal N}&=&Z (2b_u+b_d)+(A-Z) (2b_d+b_u).
\en
For the AA interaction, we have
\be
Q_{A\cal N}&=&\sum_{q=u,d,s}d_q (\Delta_q^p \lambda_p+\Delta_q^n \lambda_n),\quad \lambda_{p,n}=
\frac{\la S_{p,n}\ra_{\rm eff}}{J_{\cal N}},
\en
where $\Delta^{p(n)}_q$ is the quark spin component in proton (neutron),
$\la S_{p(n)}\ra_{\rm eff}$ 
is the proton (neutron) spin expectation value 
(including the contributions of two-body current~\cite{Menendez}) and $J_{\cal N}$ is the total angular momentum of the nucleus 
$\cal N$. 
Note that we have $\la S_{p(n)}\ra_{\rm eff}\equiv \la S_{p(n)}\ra \pm \delta a_1(\la S_p\ra -\la S_n\ra)/2$~\cite{ChuaWong,Menendez}.
For the effective scalar interaction, we have
\be
f_{s\cal N}=(Z f_{sp}+(A-Z) f_{sn}),
\en
and~\cite{HINT}
\be
\frac{f_{s p}}{m_{p}}
&=&\sum_{q=u,d,s} \frac{a_q}{m_q} f^{p}_{Tq}
+\sum_{q=u,d,s,c,b}\frac{3}{4}(q(2)+\bar q(2))(g_q^{(1)}+g_q^{(2)})
-\frac{8\pi}{9\alpha_s}f_{TG}(f_G^{(a)}+f_G^{(b)}+f_G^{(c)})
\non\\
&=&\sum_{q=u,d,s}\frac{a_q}{m_q} f^{p}_{Tq}
+\sum_{q=u,d,s,c,b}\frac{3}{4}(q(2)+\bar q(2))(g_q^{(1)}+g_q^{(2)})+
\sum_{Q=c,b,t} \frac{2}{27}\frac{a_Q}{m_Q} f_{TG}
\non\\
&-&\frac{8\pi}{9\alpha_s}f_{TG}(f_G^{(b)}+f_G^{(c)}),
\label{eq:MAq=0}
\en
with
\be
\la p|m_q \bar q q|p\ra &=&m_p f^{p}_{Tq}
\non\\
\la p|m_Q {\bar Q}Q|p\ra &\simeq&\frac{2}{27} m_pf_{TG},
\quad f_{TG}=1-\sum_{q=u,d,s} f^{p}_{Tq}
\non\\
\la p|{\cal O}^q_{\mu\nu}|p\ra&=&\frac{1}{m_p}(p_\mu p_\nu-\frac{1}{4}m_p^2 g_{\mu\nu})(q(2)+\bar q(2)).
\label{eq:one-loop}
\en
In the above, $\alpha_s=g_s^2/4\pi$ ($g_s$ is the coupling constant of ($SU(3)_C$), $q(2)$ and $\bar q(2)$ are the second moments of the parton distribution functions (PDFs). The matrix elements of the light-quark currents in nucleon are obtained in chiral perturbation theory from measurements of the pion-nucleon sigma term~\cite{Cheng1,Cheng2,GLS,Alarcon2011,Alarcon2012,Cheng3}. 
Similarly, $f_{s n}$ can be written by replacing $p$ with $n$ in Eqs.~(\ref{eq:MAq=0}) and (\ref{eq:one-loop}).

In the center of mass (CM) frame, the differential cross section is 
\be 
\frac{d\sigma(\vec q=0)}{d|{\bf q}|^2}=\frac{1}{64\pi s\mu^2_{\cal N}v^2}{\sum}|M_{fi}|^2(q^2=0)
\en
Here $v$ is the DM velocity relative to the target, $\sqrt{s}\approx m_\chi+m_{\cal N}$ is the total energy, and $\mu_{\cal N}$ is the reduced mass of DM and the target nucleus $\cal N$. 
The ``standard" total cross section at zero momentum transfer defined in~\cite{JKG} is  
\be 
\sigma_0=\int^{4 \mu^2_{\cal N}v^2}_0d|{\bf q}|^2\frac{d\sigma(\vec q=0)}{d|{\bf q}|^2}=\frac{\mu_{\cal N}^2}{\pi} \bigg[\bigg(f^2_{s\cal N}
+Q_{V\cal N}^2\bigg)
+4 Q_{A\cal N}^2 J_{\cal N}(J_{\cal N}+1)\bigg]\equiv\sigma^{SI}_0+\sigma^{SD}_0,
\en
The spin-dependent cross section at zero momentum transfer can be further decomposed as   
\be 
\sigma_0^{SD}=\sigma_{0,pp}^{SD}+\sigma_{0,nn}^{SD}+\sigma_{0,pn}^{SD},
\en
where
\be
\sigma^{SD}_{0,pp(nn)}&=&\frac{4\mu_{A_i}^2}{\pi} 
\left[ (\sum d_q \Delta_q^{p(n)})^2 \lambda^2_{p(n)}
J_{A_i}(J_{A_i}+1)
\right]
\non\\
\sigma^{SD}_{0,pn}&=&\frac{4\mu_{A_i}^2}{\pi}
\left[ (\sum d_q d_{q'} \Delta_q^p \Delta_{q'}^n) \lambda_p\lambda_n
J_{A_i}(J_{A_i}+1)
\right].
\label{eq: dsigma}
\en
Hence the total cross section of DM-nucleus ${\cal N}$ scattering can be written as~\cite{ChuaWong}
\be
\sigma_{\cal N}=\int d|{\bf q}|^2 \frac{d\sigma}{d|{\bf q}|^2}=
 (\sigma^{SI}_0 r_{SI} +\sigma^{SD}_{0,pp} r_{pp}
+\sigma^{SD}_{0,nn} r_{nn}
+\sigma^{SD}_{0,pn} r_{pn}),
\label{eq: sigmaT}
\en
where
\be
r_j\equiv \int_0^{4\mu_{A_i}^2 v^2} \frac{d|{\bf q}|^2}{4\mu_{A_i}^2 v^2} F^2_{j}(|{\bf q}|),
\en
with $j=SI, pp, nn, pn$ and
\be
F^2_{pp(nn)}(|{\bf q}|)&\equiv&\frac{S_{00}(|{\bf q}|)+S_{11}(|{\bf q}|)\pm S_{01}(|{\bf q}|)}{S_{00}(0)+S_{11}(0)\pm S_{01}(0)},
\quad
F^2_{pn}(|{\bf q}|)\equiv\frac{S_{00}(|{\bf q}|)-S_{11}(|{\bf q}|)}{S_{00}(0)-S_{11}(0)}.
\en
Note that we have
$
S_{00}(0)+S_{11}(0)\pm S_{01}(0)=({(2 J_{A_i}+1)(J_{A_i}+1)}/{\pi J_{A_i}})\la S_{p,n}\ra_{\rm eff}^2$
and
$S_{00}(0)-S_{11}(0)=({(2 J_{A_i}+1)(J_{A_i}+1)}/{\pi J_{A_i}})\la S_p\ra_{\rm eff} \la S_n\ra_{\rm eff}$~\cite{ChuaWong}.

The Eq.~(\ref{eq: sigmaT}) can be shown~\cite{ChuaWong} to be equivalent to the usual expression~\cite{JKG,Ressell} 
\be 
\sigma_{{\cal N}}=\frac{\sigma^{SI}_0}{4\mu^2_{{\cal N}} v^2}
\int^{4\mu^2_{{\cal N}}v^2}_0d|{\bf q}|^2 F^2_{SI}(|{\bf q}|)+
\frac{\sigma^{SD}_0}{4\mu^2_{\cal N} v^2}
\int^{4\mu^2_{{\cal N}}v^2}_0d|{\bf q}|^2 F^2_{SD}(|{\bf q}|)\equiv\sigma^{SI}_{\cal N}+\sigma^{SD}_{\cal N},
\en
where $F^2_{SD}(|{\bf q}|)$ is the spin-dependent form factor, which involves both short and long distance physics, a feature we try to avoid in this work. 

To compare with the experimental results, we define the scaled SI and SD cross sections, respectively, for the nucleus with atomic mass number $A_i$ and isotope abundance $\eta_i$ as the following~\footnote{The terminology of spin-(in)dependent cross section is somewhat misleading. There are, in fact, two different normalizations, where 
both spin-dependent and spin-independent interactions are involved in $\sigma^{SI}$ and $\sigma^{SD}_{p,n}$.}
\be
\sigma^{SI}&\equiv&\frac{\sum_i \eta_i\sigma_{A_i}}
{\sum_j \eta_j  A^2_j\frac{\mu^2_{A_j}}{\mu^2_p}},
\label{eq: sigma Z N}
\en
and 
\be
\sigma^{SD}_{p,n}\equiv (\sum_i\eta_i\sigma_{A_i})
\left(
\sum_j \eta_j\frac{4\mu_{A_j}^2 \la S_{p,n}\ra^2_{\rm eff}  (J_{A_j}+1)}{3\mu_{p,n}^2J_{A_j}}\right)^{-1},
\label{eq: SD p n}
\en
where $\mu_{A_i}$ and $\mu_{p,n}$ are the reduced masses of the DM with the target nucleus and the DM with proton or neutron, respectively.  

\subsection{Numerical Results for Direct Search}

Now we are ready to do the numerical calculation. 
PandaX-II, XENON1T and LUX experiments ~\cite{Xenon1T2017-SI,LUX2016-SI,LUX2016-SD,PandaX-II} use the Xenon nuclei as the target material. 
They provide 
the most stringent upper limits on $\sigma^{SI}$ and $\sigma^{SD}$. 
We will compare our calculation results with these experimental data. 

We shall specify the inputs of our numerical analysis.
For the calculation of $\sigma^{SI}$, we adopt the Helm form factor~\cite{LS,VKMHS} used in the XENON1T and LUX experiments:
\be 
F^2_{SI}(|{\bf q}|)=\bigg( \frac{3j_1(qR_{\cal N})}{qR_{\cal N}}\bigg)^2e^{(qs)^2},
\en
where the nuclear radius $R^2_{\cal N}=c^2+\frac{7}{3}\pi^2a^2-5s^2$ with $c=(1.23A^{1/3}-0.6)$ fm, $a=0.52$ fm and the nuclear surface thickness $s=1$ fm. 
We use the following updated data of nucleon mass fractions: 
$f^p_{Tu} = 0.017$,
$f^p_{Td} = 0.023$,
$f^n_{Tu} = 0.012$,
$f^n_{Td} = 0.033$,
$f^{p,n}_{Ts} = 0.053$ from Ref.~\cite{Cheng3}.
We follow Ref.~\cite{Hisano2015} and use 
$u(2)=0.223$,
$\bar u(2)=0.036$,
$d(2)=0.118$,
$\bar d(2)=0.037$,
$s(2)=\bar s(2)=0.0258$,
$c(2)=\bar c(2)=0.0187$,
$b(2)=\bar b(2)=0.0117$, 
for the second moments of PDFs of quarks and antiquarks.
These values are evaluated at the scale $\mu=M_Z$ using the CJ12 next-to-leading order PDFs given by the CTEQ-Jefferson Lab collaboration~\cite{Owens2012}.
For the calculation of $\sigma_{n,p}^{SD}$, we adopt the structure factors $S_{00,01,11}(|{\bf q}|)$ for $^{129, 131}{\rm Xe}$ nucleus in Ref.~\cite{LUX2016-SD,Menendez}, and use
the following data of the quark spin components:
$\Delta^p_u=\Delta^n_d=0.85$, $\Delta^p_d=\Delta^n_u=-0.42$, 
$\Delta^p_s=\Delta^n_s=-0.08$ from Ref.~\cite{Cheng3}, which are slightly different from those in Refs.~\cite{EFO,Mallot}. 
For $^{129, 131}{\rm Xe}$ nuclei, we use the nuclear total angular momentum $J$ and the predicted spin expectation values $\la S_{p,n}\ra$ from Refs.~\cite{Menendez,XENON100SD} for $\la S_{p,n,z}\ra_{\rm eff}$ and the isotope abundance of $^{129,131}$Xe ($\eta_i$) from Refs.~\cite{LUX2016-SD}.

\begin{figure}[t!]
\centering
\captionsetup{justification=raggedright}
\subfigure[]{
 \includegraphics[width=0.48\textwidth,height=0.23\textheight]{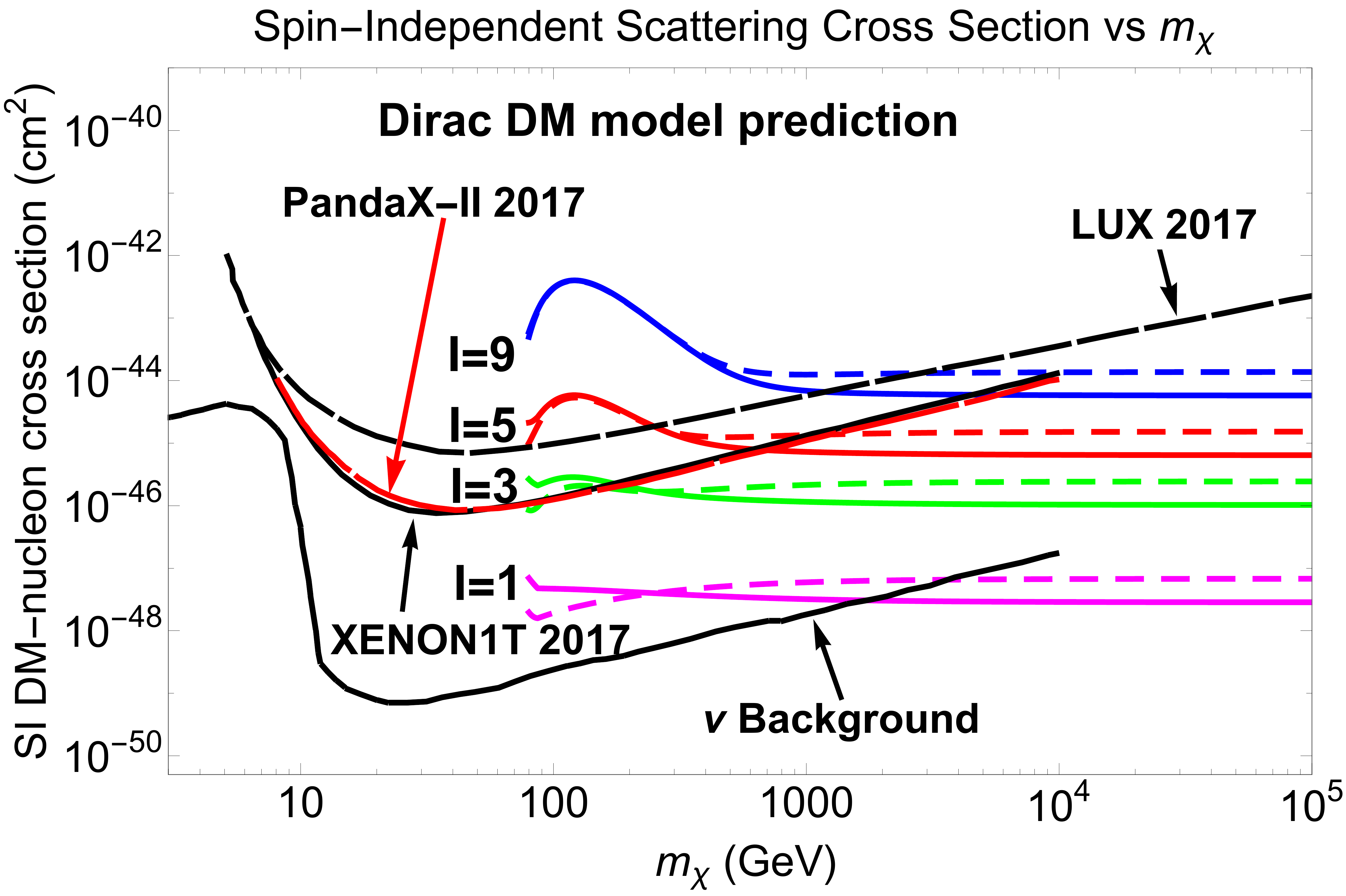}
}
\subfigure[]{
  \includegraphics[width=0.48\textwidth,height=0.23\textheight]{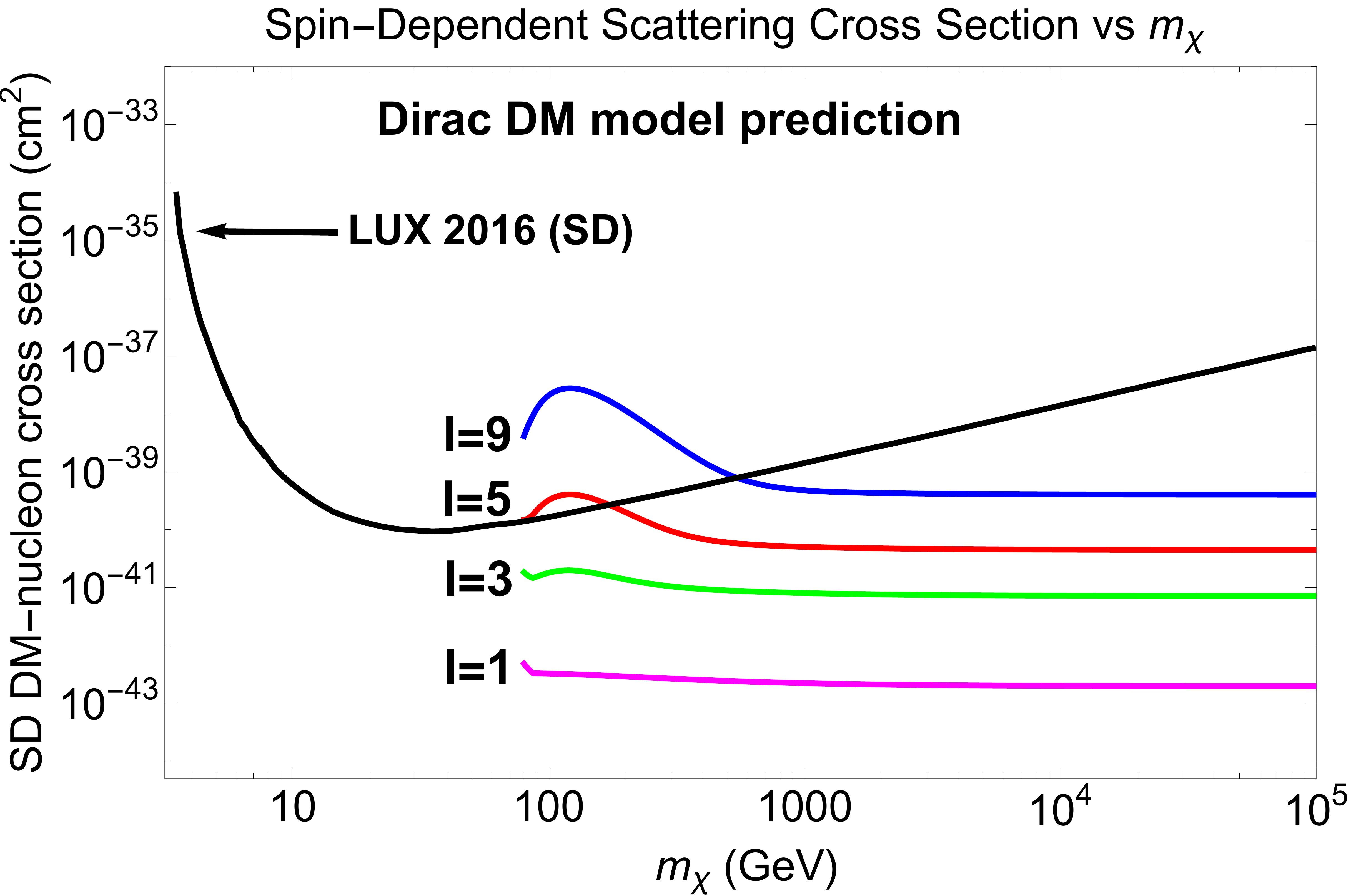}
}
\caption{Dirac DM model prediction on SI and SD cross sections of DM scattering off the nuclei $\rm{^{129,131}Xe}$ in the direct search for $I=1,3,5,9$. The dashed lines in~Fig. \ref{fig:Direct}(a) denote the results without considering the contributions of quark twist-2 operator and the two-loop diagrams from Figs.~\ref{fig:gluon}(b) and (c).}
\label{fig:Direct}
\end{figure}

For $^{129,131}{\rm Xe}$ nucleus with odd number of neutrons and even number of protons, the nuclear spin is dominant by the neutron from the odd-group model~\cite{Engel:1992bf}. 
Hence the constraint on $\sigma^{SD}_n$ is more stringent than that on $\sigma^{SD}_p$ in experiments using these nuclei.
Consequently, we only give predictions on $\sigma^{SI}$ and $\sigma^{SD}_n$. 
Furthermore we only focus on the discussion of the plot of $\sigma^{SI}$ versus $m_\chi$ in Fig.~\ref{fig:Direct}(a) since the constraint on $\sigma^{SI}$ is found to be more stringent than that on $\sigma^{SD}_n$. 

In Fig.~\ref{fig:Direct} we show our predictions on $\sigma^{SI}$ and $\sigma^{SD}_n$ for isospin $I=1,3,5,9$.
These are the main results of this work on the DM-nuclei elastic scattering cross sections.
We concentrate on the parameter space where the DM mass are greater than the $W$-boson mass and below 100 TeV. 
In Fig.~\ref{fig:Direct}(a), the solid (dashed) lines denote the prediction with (without) the contributions of quark twist-2 operator and the two-loop diagrams in Figs.~\ref{fig:gluon}(b) and (c). 
After considering these twist-2 and two-loop contributions, we find that the predicted values are roughly reduced by factors of 1.85, 2.32, 2.36, 2.37 and 2.38 for $m_\chi=1, 10, 30$, $60$ and $100 \lesssim m_\chi\lesssim 500$ TeV, respectively, regardless of the isospin $I$.
Hence, the order of magnitude of $\sigma^{SI}$ obtained in both calculations are roughly the same. 
From Fig.~\ref{fig:Direct}(a) we see that the dependence of $\sigma^{SI}$ on $m_{\chi}$ becomes mild for $m_{\chi}\gtrsim$ 1 TeV. 
In fact,  the following equation,
\be 
\sigma^{SI}\simeq I^2(I+1)^2\times7.2\times10^{-49}~{\rm cm}^2,
\en
can can nicely approximate on the SI elastic scattering cross section for $m_\chi$ in the range of few to few tens TeV.

\begin{table}[b!]
\centering
\captionsetup{justification=raggedright}
\caption{The lower limits $m_\chi^{LL}$ on Dirac DM mass are obtained from recent direct dark matter search experiments. The corresponding $\sigma^{SI}$ are also shown. 
}
\label{tab:Dmlimits}
\begin{tabular}{|c|c|c|c|c|c|c|c|}
\hline
$I$
    & $3$ 
    & $4$ 
    & $5$ 
    & $6$
    & $7$
    & $8$
    & $9$
    \\ 
    \hline
Direct $m_{\chi}^{LL}$ (TeV) 
    & 0.19 
    & 0.37 
    & 0.72
    & 1.18
    & 2.28
    & 3.55
    & 5.94 
   \\ 
   \hline
$\sigma^{SI}(10^{-46} {\rm cm}^2)$
    & 2.08 
    & 4.14  
    & 7.64
    & 13.96
    & 23.57
    & 38.21
    & 58.86
    \\
   \hline
\end{tabular}
\end{table}

We now compare our results with the data. 
Since the constraint from $\sigma^{SI}$ is much severe than that from $\sigma^{SD}$, we will concentrate on the SI part.
From Fig.~\ref{fig:Direct}(a), we see that there are plenty of parameter space for $\sigma^{SI}$ to satisfy the upper limits from the LUX, PandaX-II and XENON1T 
SI-experiments~\cite{LUX2016-SI,PandaX-II,Xenon1T2017-SI}.
Note that the cancellation in the large component 
of the VV interaction (see the previous section) is at work, 
and it significantly reduces the cross section of DM-nucleus elastic scattering and makes the Dirac DM model viable in the direct search.

For $I= 1$ and $2$, all Dirac DM masses in this surveyed region are allowed in principle. 
However, for $I=1$ we cannot distinguish the DM event from neutrino event when $m_\chi$ is greater than $1.7$ TeV, as the predicted cross section is below the curve of the neutrino background~\cite{CKW,MF,BSF}. 
Similarly, if we extrapolate the curve of neutrino background up to $m_\chi=$100 TeV, 
we find that we cannot distinguish the DM event from neutrino event when $m_\chi >$14.6 and 52.6 TeV
for $I=$2 and 3, respectively. 
We show the lower mass bound $m_\chi^{LL}$ and corresponding $\sigma^{SI}$ for $I=3\sim 9$ in Table~\ref{tab:Dmlimits}.
The lower bounds on these Dirac DM masses are extracted by comparing to the PandaX-II $\sigma^{SI}$ data [see Fig.~\ref{fig:Direct}(a)].

\section{Dirac Fermionic Dark Matter in the Indirect Search}

\subsection{Thermal Relic Dark Matter Densisty}

The DM particles are thought to have been created thermally during the big bang and frozen out of  thermal equilibrium in the early Universe contributing to the relic density.
The evolution of DM abundance obeys the following Boltzmann equation:
\begin{equation}\label{Boltzmann eq}
\frac{dn_{\chi^0}}{dt} + 3Hn_{\chi^0} = -\langle\sigma_\mathrm{ann} v_{\rm M\phi l}\rangle_{{\chi^0}\bar\chi^0}
[n_{\chi^0} n_{\bar\chi^0} - n_{\chi^0}^\mathrm{eq}
n_{\bar\chi^0}^\mathrm{eq}],
\label{eq: Boltzmann0}
\end{equation}
where $n_{\chi^0}$ ($n_{\bar\chi^0}$) is the number density of $\chi^0$ ($\bar\chi^0$), 
$H\equiv\dot{a}/a=\sqrt{4\pi^3g_*(T)T^4/(45M_{\mathrm{PL}}{}^2)}$ is the Hubble parameter, 
$M_{\mathrm{PL}}$ is the Planck mass, 
$g_*$ is the total effective numbers of relativistic degrees of freedom~\cite{Kolb,CR:03} and $n^\mathrm{eq}$ the equilibrium number density. 
Note that Eq.~(\ref{eq: Boltzmann0}) is measured in the cosmic comoving frame~\cite{GG} and  $\la\sigma_{\rm ann} v_{\rm M\phi l}\ra$ is the thermal averaged annihilation cross section times M$\phi$ller velocity which is defined by $v_{\rm M\phi l}\equiv \sqrt{(p_1\cdot p_2)^2-m_1^2m_2^2}/(E_1E_2)=\sqrt{|{\bf v}_1-{\bf  v}_2|^2-|{\bf v}_1\times {\bf v}_2|^2}$ with subscripts 1 and 2 labeling the two initial DM particles and velocities ${\bf v}_i\equiv {\bf p}_i/E_i (i=1,2)$.
For the Dirac fermionic DM particle $\chi^0$ and antiparticle $\bar\chi^0$, we define the total number density of DM as $n_{DM}\equiv n_{\chi^0}+n_{\bar\chi^0}=2n_{\chi^0({\bar \chi^0})}$ and the above equation can be written as  
 \be
 \frac{dn_{DM}}{dt}+3Hn_{DM}=-\frac{1}{2}\la\sigma_{\rm ann} v_{\rm M\phi l}\ra_{\chi^0{\bar\chi^0}}[n^2_{DM}-(n^{\rm eq}_{DM})^2].
 \label{eq: Boltzmann}
 \en

Following the standard procedure~\cite{Kolb} to solve Eq.(\ref{eq: Boltzmann}), 
the relic DM density $\Omega_{\rm DM}\equiv \rho_\chi / \rho_{\rm crit}$  can be approximately related to the velocity averaged annihilation cross section $\la\sigma_{\rm ann}v\ra$ as
\begin{equation}\label{abundance}
\Omega _{\text{DM}}h^2 \approx 1.04\times
10^9\frac{{\rm GeV}^{-1}}{M_{\mathrm{PL}} \sqrt{g_*\left(T_f\right)}J(x_f)},
\end{equation}
where
\begin{equation}\label{xf}
x_f \approx \ln\left[\frac{2\times0.038 m_\chi
M_{\mathrm{PL}}\langle\sigma_{\mathrm{ann}}v\rangle
}{\sqrt{g_*\left(T_f\right)}{}x_f^{1/2}}\right],
\end{equation}
and 
\begin{equation}\label{Jfactor}
J\left(x_f\right) 
\equiv \int_{x_f}^{\infty }
\frac{\langle\sigma_{\mathrm{ann}}v\rangle}{x^2} dx
\equiv \int_{x_f}^{\infty }
\frac{\langle\sigma_{\mathrm{ann}}v_{\rm M\phi l}\rangle_{{\chi^0}\bar{\chi^0}}}{2x^2} dx.
\end{equation} 
In the above, $x_f\equiv m_\chi/T_f$ and $T_f$ is the freeze-out temperature. For convenient, 
the velocity averaged annihilation cross section $\langle\sigma_{\mathrm{ann}}v\rangle$
with  $v$ the ``relative velocity'' is defined as \footnote{In general, the collision is not collinear in the comoving frame. We have to use the M$\phi$ller velocity. The M$\phi$ller velocity is not equal to the relative velocity $v\equiv |{\bf v}_1-{\bf v_2}|$. Nevertheless, it has been shown~\cite{GG} that  $\la\sigma_{\rm ann} v_{\rm M\phi l}\ra=\la\sigma_{\rm ann}v_{\rm lab}\ra^{\rm lab}$ where $v_{\rm lab}\equiv |{\bf v}_{1,{\rm lab}}- {\bf v}_{2,{\rm lab}}|$ is calculated in the lab frame with one of two initial particles being at rest.}
\begin{eqnarray}\label{T_average}
\langle\sigma_{\mathrm{ann}}v\rangle
\equiv\frac{\langle\sigma_{\mathrm{ann}}v_{\rm M\phi l}\rangle_{\chi^0\bar\chi^0}}{2}
&\equiv&\frac{3\sqrt 6}{\sqrt{\pi}v_0^3}\int_0^\infty dv\, v^2
\frac{(\sigma_{\mathrm{ann}}v)_{\chi^0\bar\chi^0}}{2}
e^{-3v^2/2v_0^2}
\non\\
&=&
\frac{x^{3/2}}{2\sqrt{\pi}}\int_0^\infty dv\, v^2
\frac{(\sigma_{\mathrm{ann}}v)_{\chi^0\bar\chi^0}}{2}e^{-xv^2/4},
\end{eqnarray}
where we define $v_0\equiv\la v^2\ra^{1/2}$ and $v_0=\sqrt{6/x_f}$ has been used in the last expression.
It is straightforward to obtain
\be
J\left(x_f\right) = \int_{x_f}^{\infty }
\frac{\langle\sigma_{\mathrm{ann}}v\rangle}{x^2} \, dx
=\int_0^\infty dv\frac{(\sigma_{\rm ann} v)_{\chi^0\bar\chi^0}}{2} v\left[1-{\rm erf}\left(v\sqrt x_f/2\right)\right].
\en

\subsection{Dirac Dark Matter Annihilation}

\begin{figure}[t] \centering
\includegraphics[width=0.8\textwidth]{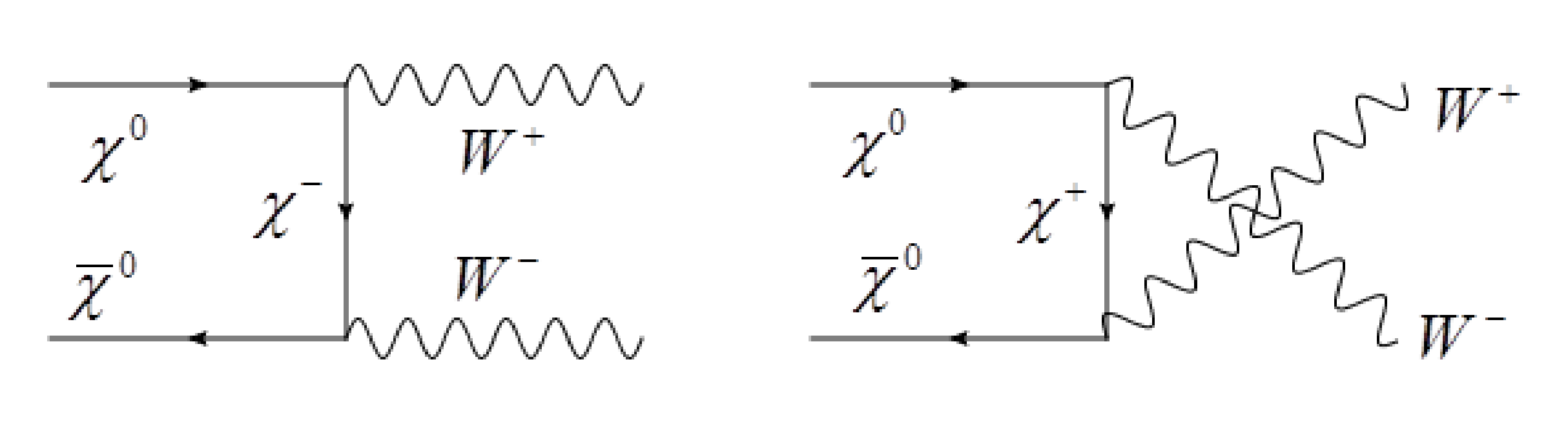}
\caption{The Feynman diagrams of DM annihilation for $W^+W^-$
channel.} \label{fig:WW}
\end{figure}

DM relic density is determined by the velocity averaged cross section $\langle \sigma_{\rm ann} v\rangle$ of DM annihilation processes, 
which have been ceased after the freeze-out stage in the cosmological scale. 
Nevertheless, the DM annihilation to the SM particles can still occur today in regions of high DM density. 
These annihilations result in the end products as excesses relative to products from the SM astrophysical processes, 
where the excesses are actively search for in the indirect search experiments. 
As the DM particles became non-relativistic when they froze out of thermal equilibrium in the early universe, 
non-perturbative effects, such as the so-called Sommerfeld enhancement effect, can be important~\cite{Hisano:2004,Hisano:2005,Hisano:2006nn,AFSW,Chun1,Chun2,Chun3}.

\begin{figure}[b!]
	\centering
	\captionsetup{justification=raggedright}
	\subfigure[]{
		\includegraphics[width=0.3\textwidth,height=0.16\textheight]{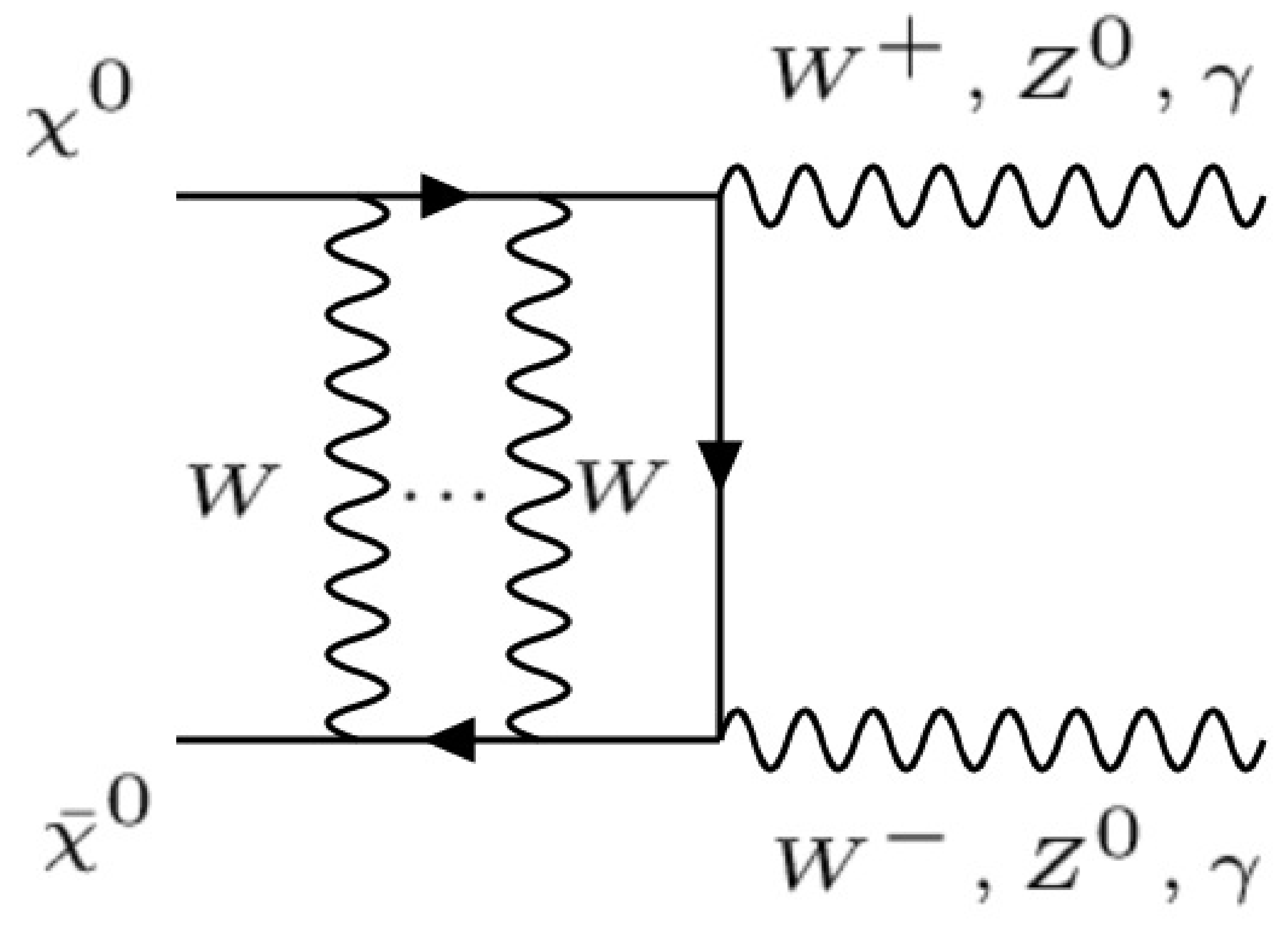}
	}
	\subfigure[]{
    	\includegraphics[width=0.3\textwidth,height=0.16\textheight]{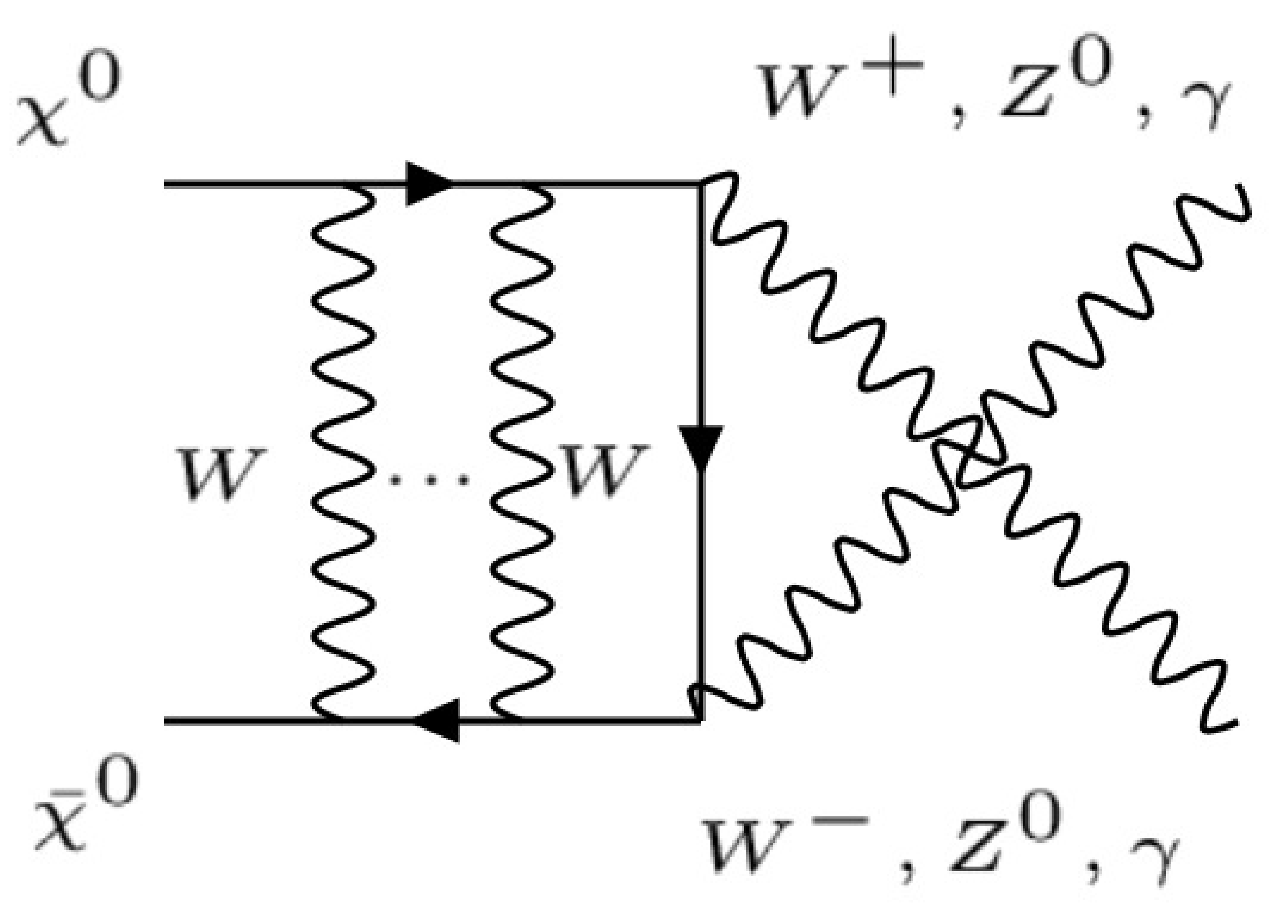}
    }
	\subfigure[]{
		\includegraphics[width=0.3\textwidth,height=0.16\textheight]{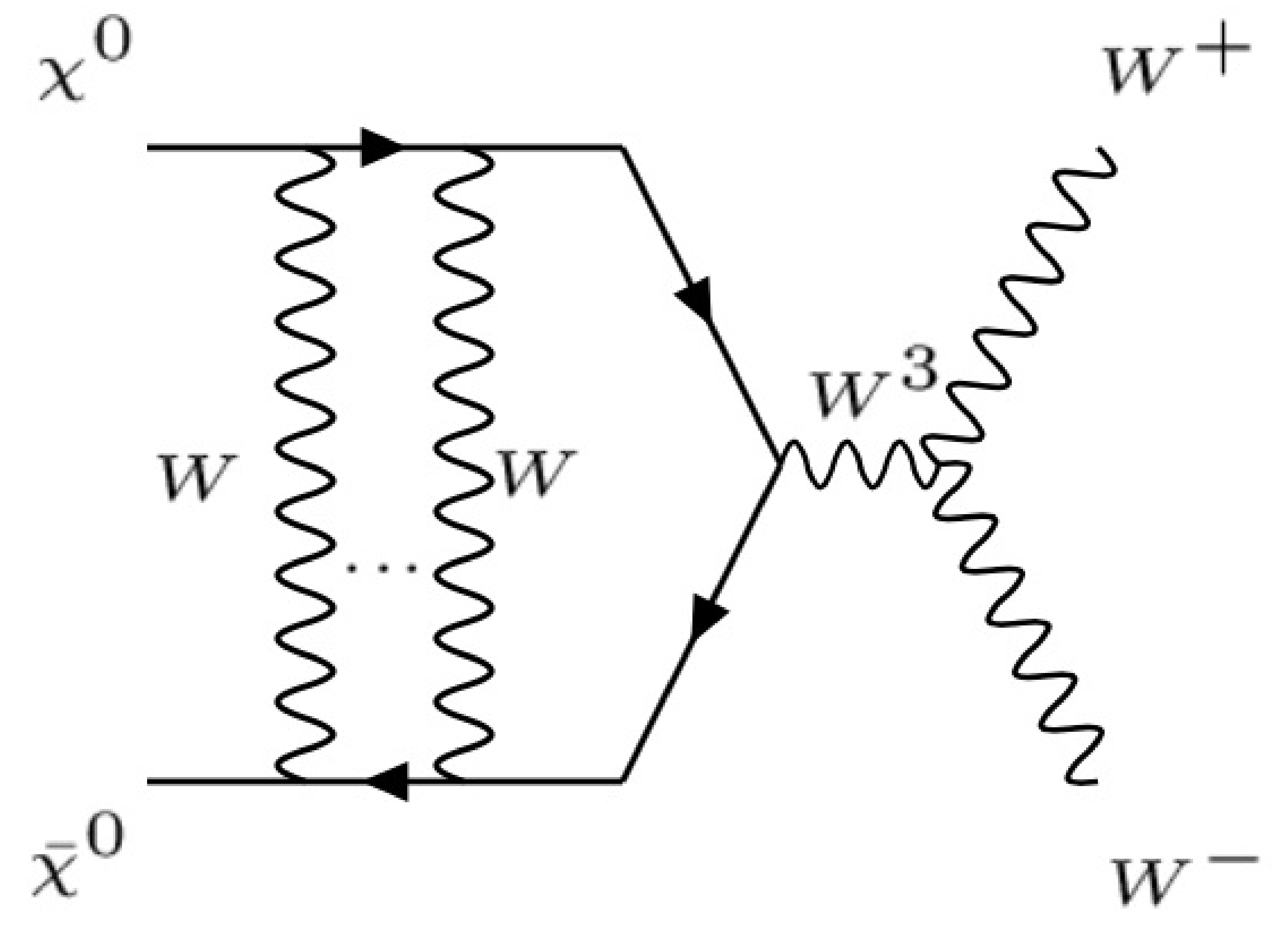}
	}
	\caption{$\chi^0\bar\chi^0\rightarrow VV$ annihilation diagrams with the Sommerfeld effect.}
	\label{fig:Som}
\end{figure}

In this work, we only consider the $I\neq 0, I_3=Y=0$ case. 
The interactions of DM with $Z$-boson and $\gamma$ are vanishing at tree level 
and, hence, the DM particles cannot annihilate into a pair of neutral gauge bosons, 
such as $ZZ, Z\gamma$ and $\gamma\gamma$. 
On the other hand, the DM particles can annihilate into a pair of $W^+W^-$ at tree level as shown in Fig.~\ref{fig:WW}, 
as the tree-level interaction of DM with $W$-boson is allowed. 
It is interesting that through diagrams shown in Fig.~\ref{fig:Som}, the Sommerfeld effect opens the possibilities for $\chi^0\bar\chi^0$ to annihilate into a pair of neutral gauge bosons.

The $\chi^0\overline{\chi^0}\to W^+W^-$ annihilation process can occur through diagrams shown in Fig.~\ref{fig:WW}. 
The corresponding cross section can be calculated to be~\cite{chua}

\begin{eqnarray}
&&(\sigma_{\mathrm{ann}}v)_{\chi^0\bar\chi^0} 
\non\\
&&\quad= [I(I+1)]^2\frac{g^4}
{32\pi s^{1/2} \left(s-2 m_\chi^2\right)}
\left\{-\frac{\sqrt{s-4m^2_W}\left(s m_{\chi }^2+4 m_{\chi }^4+2
m_W^4\right)}{\left(m_{\chi }^2 \left(s-4
m_W^2\right)+m_W^4\right)}\right.
\non\\ 
&&\qquad+ \left.
\frac{ \left(4
m_{\chi }^2 \left(s-2 m_W^2\right)-8 m_{\chi }^4+4
m_W^4+s^2\right)}{\sqrt{\left(s-4 m_{\chi }^2\right)} \left(s-2
m_W^2\right)} \log \left[-\frac{\sqrt{\left(s-4 m_{\chi }^2\right)
\left(s-4 m_W^2\right)}-2 m_W^2+s}{\sqrt{\left(s-4 m_{\chi
}^2\right) \left(s-4 m_W^2\right)}+2 m_W^2-s}\right]\right\}.\non\\
\label{eq:sigmav xx->WW}
\end{eqnarray}
After substituting 
$s=2m_\chi^2(1+1/\sqrt{1-v^2})$
into the above equation and expanding around $v^2$, one obtains:
\begin{eqnarray}
\langle\sigma_{\mathrm{ann}}v\rangle = \langle a^{+-} + b^{+-} v^2 + {\cal
O}(v^4)\rangle,
\label{eq:sigmavann}
\end{eqnarray}
where we have
\begin{eqnarray}
a^{+-} &\equiv& [I(I+1)]^2 \frac{g^4(m_\chi^2-m_W^2)^{3/2}}
{16\pi m_\chi (2m_\chi^2-m_W^2)^2},
\non\\
b^{+-} &\equiv&  [I(I+1)]^2\frac{g^4 (m_{\chi }^2-m_W^2)^{1/2}
\left(76 m_{\chi }^4m_W^2
-66 m_{\chi }^2 m_W^4+23 m_W^6\right)}
{384 \pi m_{\chi }  \left(2m_{\chi }^2 -
m_W^2\right)^4},
\label{eq:ab}
\end{eqnarray}
with $g=e/\sin\theta_W$.
For non-relativistic annihilation of DM, neglecting $v^4$ and higher order terms is a good approximation.
From Eq.~(\ref{T_average}), we have $\langle v^2\rangle=6
x^{-1}_f$, and hence the velocity averaged cross section becomes 
\be
\langle\sigma_{\mathrm{ann}}v\rangle \simeq  a^{+-} +6 \frac{b^{+-}}{ x_f},
\qquad
J(x_f)\simeq \frac{a^{+-} +3 b^{+-}/ x_f}{x_f}.
\label{eq:sigmav}
\en

For indirect search, we will compare our calculation result with the Fermi-LAT result  \cite{Fermi-LAT2015} which is  
from a combined analysis of 15 dwarf spheroidal satellite galaxies (dSphs) of the Milky Way and the H.E.S.S result~\cite{HESS2016} using $\gamma$-ray observation towards the inner 300 parsecs of the Milky Way.
As we know that the DM halo is immersed in the Galaxy. The speed of the sun moving around the Galactic center is about 220 km/s at the local distance $r\approx$ 8.5 kpc and the Galactic circular rotation speed is about 230 km/s at radii $\approx$ 100 kpc~\cite{JKG,Kochanek:1995xv}. On the other hand, the shortest and longest distance of these 15 dSphs from the sun are $\approx$ 23 and 233 kpc, 
respectively~\cite{Fermi-LAT2015}. Hence we will use a typical DM velocity $v_0\simeq 300$ km/s in the indirect-detection calculation.
In $\la\sigma v\ra$ the $b$-term is sub-leading. 
For $m_\chi\gtrsim 100$ GeV, it can be shown that we have $bv^2/(a+b v^2)<5\times 10^{-7}$ and $15\%$ for $v=v_0$ and $c/2$, respectively.
Hence, neglecting $b$ is a good approximation for indirect-detection calculations, 
while it introduces a $15\%$ error in the relic abundance calculation, as the DM velocity is about half the speed of light in the latter case. 
To simplify the calculation, we only keep the first term (the $a$ term) in Eq.~(\ref{eq:sigmavann}) for both relic density calculation and the indirect annihilation processes, namely the $S$-wave contribution.

\begin{figure}[t] \centering
\includegraphics[width=0.7\textwidth]{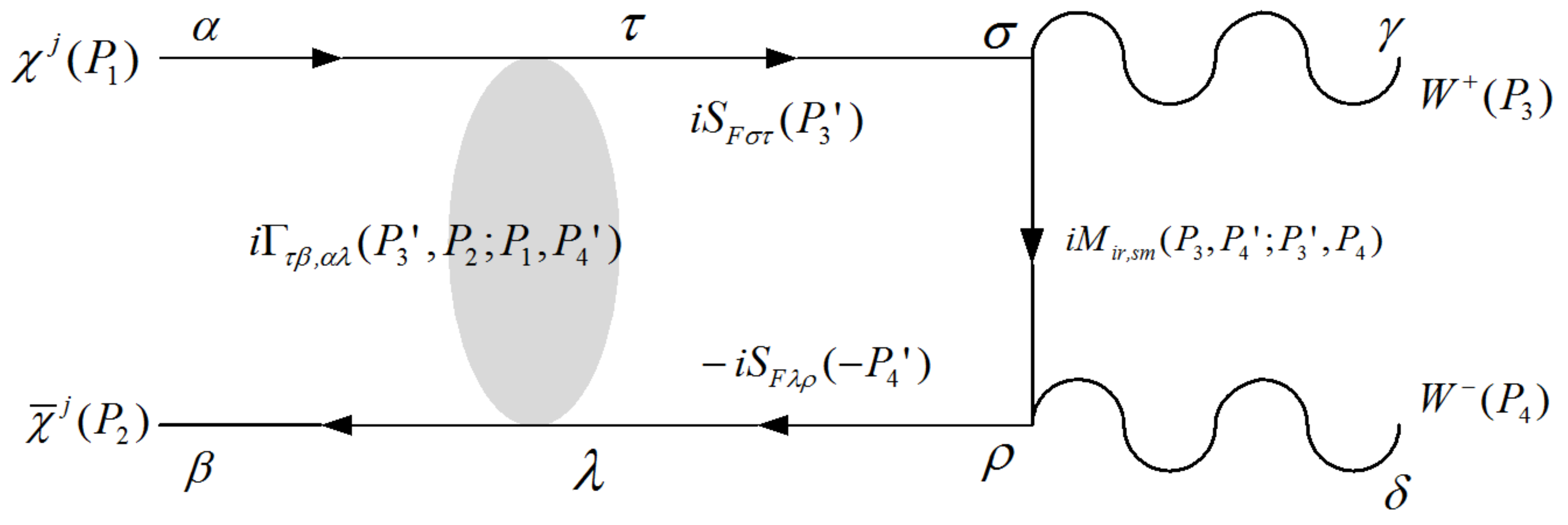}
\caption{The Feynman diagram for WIMPs annihilation in $W^+W^-$
channel.}
\label{fig:Som4}
\end{figure}

Now we consider the Sommerfeld enhancement effect through diagrams shown in Fig.~\ref{fig:Som}~\cite{chua}.
The Sommerfeld enhancement is rather complicated here,
since the $\chi^0\overline{\chi^0}$ state can rescatter into other
states, such as $\chi^\pm\overline{\chi^{\pm}}$ and so on, through
$t$-channel diagrams by exchanging $W$ and $Z$ with the rescattered
state annihilated into $W^+W^-$. 
Hence we need consider the scattering processes in a general form:
$\chi^j\overline{\chi^j}\to \chi^i\overline{\chi^i}\to VV$ (with
$i,j=-I,-I+1,\dots,I-1,I$) for a generic isospin $I$. In other words, the non-perturbative scattering of 
$\chi^j\overline{\chi^j}\to \chi^i\overline{\chi^i}$ follows the main perturbative scattering of $\chi^i\overline{\chi^i}\to VV$.
To simplify the calculation we
follow \cite{Strumia,Cirelli2009,chua} to consider the SU(2)
symmetric limit, which is expected to be a good one when the DM mass $m_\chi$ is much greater than $m_{W,Z}$.

\begin{figure}[t!] \centering
\captionsetup{justification=raggedright}
\includegraphics[width=0.7\textwidth]{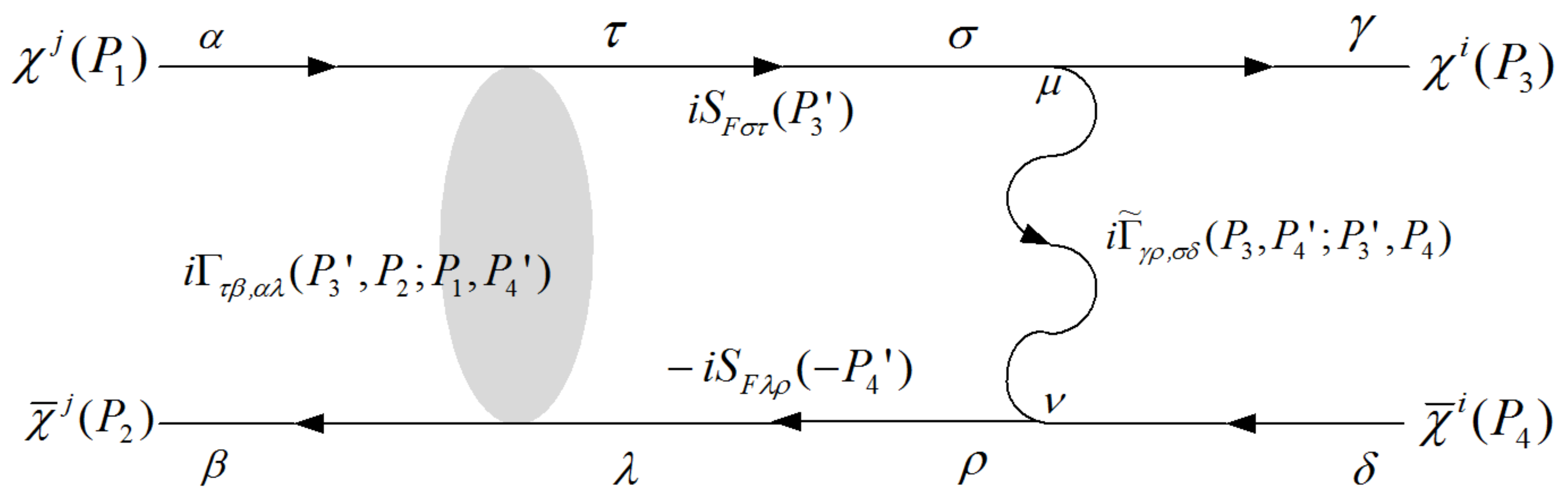}
\caption{The diagram corresponds to the second term in Eq.~(\ref{eq:master0 main text}) of the $\chi^j\bar\chi^j\to \chi^i\bar\chi^i$
process. Note that $i\Gamma^{ij}$ contains an infinite series of the ladder diagrams. 
} 
\label{fig:Som5}
\end{figure}

We consider the non-perturbative scattering process $\chi^j(p_1)\bar\chi^j(p_2)\to W^+(p_3)W^-(p_4)$ as shown in Fig.~\ref{fig:Som4}. 
The $\chi^j\bar\chi^j$ annihilation amplitude $iM^{jS}$ can be expressed as 
\be
iM^{jS}_{\beta,\alpha}(p_3, p_2, p_1, p_4) \bar v_\beta(p_2) u_\alpha (p_1)
&=&iM^j_{\beta,\alpha}(p_3, p_2, p_1, p_4) \bar v_\beta(p_2) u_\alpha (p_1)
\non\\
&&+\int \frac{d^4p'_3}{(2\pi)^4}
iM^i_{\rho,\sigma}(p_3,p'_4;p'_3,p_4)
\non\\
&& (iS_{F \sigma\tau}(p'_3))(-iS_{F \lambda\rho}(-p'_4)) 
\non\\
&&
i\Gamma^{ij}_{\tau\beta,\alpha\lambda}(p'_3,p_2;p_1,p'_4) \bar v_\beta(p_2) u_\alpha(p_1),
\label{eq:wholeM}
\en
where $i\Gamma^{ij}$ is 
the amputated non-perturbative 4-point vertex function for the $\chi^j(p_1)\bar\chi^j(p_2)\to \chi^i(p_3)\bar\chi^i(p_4)$ scattering as shown in Fig.~\ref{fig:Som5}. The vertex function satisfies the following equation, 
\be
&&\hspace{-0.5cm}
i\Gamma^{ij}_{\gamma\beta,\alpha\delta}(p_3, p_2, p_1, p_4) 
\non\\
&&=i\tilde\Gamma^{ij}_{\gamma\beta,\alpha\delta}(p_3, p_2, p_1, p_4)
\non\\
&&\quad
+\int \frac{d^4p'_3}{(2\pi)^4}
i\tilde\Gamma^{ik}_{\gamma \rho,\sigma\delta}(p_3,p'_4;p'_3,p_4)
(iS_{F \sigma\tau}(p'_3)) 
i\Gamma^{kj}_{\tau\beta,\alpha \lambda}(p'_3,p_2;p_1,p'_4)(-iS_{F \lambda\rho}(-p'_4)),
\label{eq:master0 main text}
\en
where we have $p'_4=-p'_3+p_3+p_4$, $S_{F\sigma\tau}$ and $S_{F\lambda\rho}$ are the fermion propagators,
and the relevant lowest order perturbative 4-point vertex function in the SU(2) symmetric limit is given by
\be
i\tilde\Gamma^{ij}_{\gamma\beta,\alpha\delta}(p_3, p_2, p_1, p_4)=-ig^2\sum_{a=1,2,3}T^a_{ij}T^a_{ji}(\gamma^\mu)_{\gamma\alpha}(\gamma^\nu)_{\beta\delta}\frac{g_{\mu\nu}}{(p_1-p_3)^2-M^2_W}.
\en 
Recall that the hypercharge $Y$ of $\chi^{i}$ is vanishing, hence, the $U(1)$ field $B^\mu$ does not contribute to $\tilde\Gamma$ in the SU(2) symmetric limit.
Note that through iteration $i\Gamma^{ij}$ contains an infinite series of the ladder diagrams [see Fig.~\ref{fig:Som5} and 
Eq.~(\ref{eq:master0 main text})].
In the NR limit, Eq.~(\ref{eq:master0 main text}) can lead to the Lippmann-Schwinger equation (see Appendix B for details). 

For the case of $S$-wave rescattering,
Eq.~(\ref{eq:wholeM}) can be expressed as (see Appendix B)
\be
iM^{jS}_{\beta,\alpha}(p_3, p_2, p_1, p_4) \bar v_\beta(p_2) u_\alpha (p_1)&=&
\sum_i iM^i_{\rho,\sigma}(\vec p_3,\vec p_2;\vec p_1,\vec p_4) \bar v_\rho(p_2) u_\sigma(p_1)
{\cal Q}_{ij}
\label{eq: s wave scattering main text} 
\en
where we have
\be
{\cal Q}_{ij}\equiv\sum_{{\cal I}=0}^{2I} (U^T)_{i{\cal I}}\psi_{\cal I}(\vec r=0)U_{{\cal I}j},
\en
with
\be
U_{{\cal I}j}=(-1)^j\la I j I(-j)|{\cal I}0\ra,
\en
the matrix that diagonalizes $\sum_c T^a_{ij}T^a_{ji}$ [see Eq. (\ref{eq: diagonal})]~\cite{chua},
and $\la I j I(-j)|{\cal I}0\ra$ is the Clebsch-Gordan coefficient (in the $\la j_1 m_1 j_2 m_2|JM\ra$ notation).
Note that ${\cal I}$ is the total isospin of the $\chi\bar\chi$ pair. 
The wave function satisfies the Schr\"odinger equation,
\be
-\frac{1}{2\mu} \nabla^2\psi_{\cal I}(\vec r)+V_{\cal I}(\vec r)\psi_{\cal I}(\vec r)=E\psi_{\cal I}(\vec r)=\frac{1}{2} \mu v^2\psi_{\cal I}(\vec r),
\label{eq: Schrodinger eq}
\en
where $V_{\cal I}(r)$ is a Yukawa-type potential
\be
V_{\cal I}(r)=-\alpha_{\rm w}\{I(I+1)-{\cal I}({\cal I}+1)/2\}\frac{e^{-M_Wr}}{r},
\label{eq: Yukawa}
\en
with $\alpha_{\rm w}$ the weak fine structure constant,
$E=|\vec p|^2/2\mu\equiv \mu v^2/2$, and the wave function goes to $e^{i \vec p\cdot \vec r}$ asymptotically.
Consequently, the rate is modified as
\be
|\bar v(p_2) M^{jS}(p_3, p_2, p_1, p_4) u(p_1)|^2
&=&
\sum_{i,i'}
(\bar v M^{i'}(\vec p_3,\vec p_2;\vec p_1,\vec p_4)u)^*
(\bar v M^i_{\rho,\sigma}(\vec p_3,\vec p_2;\vec p_1,\vec p_4) u)
{\cal Q}_{ij}{\cal Q}^\dagger_{ji'} ,
\label{eq: Sommerfeld S-wave}
\en
where the additional term (${\cal Q}_{ij}{\cal Q}^\dagger_{ji'}$) is the Sommerfeld enhancement factor, which cannot be factorized in this work,
and will be dealt with later. As a cross check we note that in the $V_{\cal I}=0$ limit, we have $\psi_{\cal I}(\vec r=0)=1$ giving ${\cal Q}_{ij}=\delta_{ij}$, and, consequently, Eq. (\ref{eq: s wave scattering main text}) and the above equation are just trivial identities.

Using the Hulth\'{e}n potential to approximate the Yukawa potential,~\cite{Cassel} 
\be
V_{\cal I}(\vec r)\simeq-\alpha_{\cal I} \frac{(\pi^2 m_W/6) e^{-\pi^2 m_W r/6}}{1-e^{-\pi^2 m_W r/6}},
\label{eq: Huthen}
\en
one obtains (see Appendix B.3)
\be
\psi_{\cal I}(\vec r=0)&=&
i\frac{\pi^2\epsilon_W/6}{2\epsilon_v} \,
\Gamma\left(1-i \frac{\epsilon_v}{\pi^2\epsilon_W/6}\left(1+\sqrt{1-\frac{\pi^2\epsilon_W/6}{\epsilon^2_v}}\right)\right)
\non\\
&&\times
\Gamma\left(1-i \frac{\epsilon_v}{\pi^2\epsilon_W/6}\left(1-\sqrt{1-\frac{\pi^2\epsilon_W/6}{\epsilon^2_v}}\right)\right)
\bigg/\Gamma\left(\frac{-2i\epsilon_v}{\pi^2\epsilon_W/6}\right),
\label{eq: psi0Hulthen}
\en
with~\footnote{Note that the $\beta$ in the formula of \cite{Cassel} is in fact $v/2$ in this work.} 
\be 
\epsilon_v\equiv\frac{v}{2\alpha_{\cal I}}, 
\quad
\epsilon_W\equiv\frac{m_W}{\alpha_{\cal I} m_\chi},
\quad
\alpha_{\cal I}\equiv[I(I+1)-{\cal I}({\cal I}+1)/2]\alpha_W,
\en 
and $|\psi_{\cal I}(\vec r=0)|^2$ is just the Sommerfeld factor shown in~\cite{Feng:2010zp}, 
\be
|\psi_{\cal I}(\vec r=0)|^2
=
\frac{\sinh\left(\frac{2\pi\epsilon_v}{\pi^2\epsilon_W/6}\right)}
{\cosh\left(\frac{2\pi\epsilon_v}{\pi^2\epsilon_W/6}\right)
-\cos\left(2\pi \sqrt{\frac{1}{\pi^2\epsilon_W/6}-\frac{\epsilon^2_v}{(\pi^2\epsilon_W/6)^2}}\right)}.
\label{eq:Sold} 
\en 
Note that the phase convention of the above $\psi_{\cal I}(\vec r=0)$ is fixed by its asymptotic behavior (see Appendix B.3) and 
$\psi_{\cal I}(\vec r=0)$ indeed goes to 1 in the $\alpha_{\cal I}=0$ limit.
Also note that the phase of $\psi_{\cal I}(\vec r=0)$ is simple the $S$-wave phase shift, $\delta_{l=0}$ (see Appendix B.3).
We find that the above $\psi_{\cal I}(\vec r=0)$, including its size and phase, agree well 
with the results obtained by  numerically solving the Schr\"{o}dinger equation with the Yukawa potential (see below and Appendix B.4).

It is usually advocated that in the $m_\chi\gg m_W$ limit, the Yukawa potential can be approximated by a coulomb potential:
\be
V_{\cal I}(r)
\simeq-\{I(I+1)-{\cal I}({\cal I}+1)/2\}\frac{\alpha_{\rm w}}{r}.
\label{eq: Coulomb}
\en
The corresponding wave function is given by~\cite{Coulomb}
\be
\psi^{(coul)}_{\cal I}(\vec r)=\Gamma(1+i\gamma_{\cal I}) e^{-\pi\gamma_{\cal I}/2} e^{i\vec p\cdot\vec r}
{}_1F_1(-i\gamma_{\cal I}, 1,ipr-i\vec p\cdot \vec r),
\en
where
\be
\gamma_{\cal I}=-\{I(I+1)-{\cal I}({\cal I}+1)/2\}\frac{\alpha_{\rm w} }{v}=-\{I(I+1)-{\cal I}({\cal I}+1)/2\}\frac{\alpha_{\rm w} \mu_\chi}{|\vec p|}.
\en
In this approximation we have
\be
\psi^{(coul)}_{\cal I}(\vec r=0)=\Gamma(1+i\gamma_{\cal I}) e^{-\pi\gamma_{\cal I}/2}.
\label{eq: psi coulomb}
\en
Note that as a cross check, we have
\be
|\psi^{(coul)}_{\cal I}(\vec r=0)|^2&=&\Gamma(1+i\gamma_{\cal I})\Gamma(1-i\gamma_{\cal I}) e^{-\pi\gamma_{\cal I}}
=\frac{2\pi\gamma_{\cal I}}{e^{2\pi\gamma_{\cal I}}-1}=S^{(coul)},
\label{eq: SQED}
\en
which is the usual Sommerfeld enhancement factor, $S$, in the Coulomb potential case.
In fact the Sommerfeld factor in Eq.~(\ref{eq:Sold}) does reduce to $S^{(coul)}$ in the large $m_\chi$ region, as one can check it directly, or by considering the limiting behavior of $|\psi_{\cal I}(\vec r=0)|$, which goes to $|\psi^{coul}_{\cal I}(\vec r=0)|$ in the large $m_\chi$ limit~\cite{Cassel,Lebedev0}.
We expect the Coulomb approximation to be a good one in the large $m_\chi$ region,
but as we will see shortly that this is not really the case.

The $\chi^0\overline{\chi^0}\to W^+ W^-$ amplitude with Sommerfeld enhancement, $A_S$, is now given by
\be
A_S(\chi^0\overline{\chi^0}\to W^+W^-)&=&\sum_i A(\chi^i\overline{\chi^i}\to W^+W^-) {\cal Q}_{i0},
\en
where $i$ is summed over all $\chi^i\overline{\chi^i}$ states. Therefore, the Sommerfeld enhanced $S$-wave contribution of $\la\sigma_{\rm ann} v\ra$ in Eq.(~\ref{eq:sigmav}) is given by
\be
a_S^{+-}=\sum_{i,j} {\cal Q}^{\dagger}_{0i} a^{+-}_{ij}{\cal Q}_{j 0}.
\en
In the above, $i$ and $j$ are summed over $\chi^i\overline{\chi^i}$ and $\chi^j\overline{\chi^j}$ states, respectively, and $a_{ij}^{+-}$ corresponds to the $S$-wave contribution from $A^*(\chi^i\overline{\chi^i}\to W^+W^-) A(\chi^j\overline{\chi^j}\to W^+W^-)$.

It is straightforward to obtain
\be
a^{+-}_{ij}&=&a^{+-}
\left\{\frac{[I(I+1)-i^2][I(I+1)-j^2]}{[I(I+1)]^2}
+ij\frac{(4m^4_\chi+20m^2_\chi m^2_W+3m^4_W)}{2(4m^2_\chi-m_W^2)^2(I(I+1))^2}\right\}
\en
and, consequently,~\footnote{For a generic $I$, we have the following property:
$\sum_{i,j=-I}^I{\cal Q}^\dagger_{0i}(i,j,i j, i j^2,i^2 j){\cal Q}_{j0}=(0,0,0,0,0)$.\label{fn:QQ=0}} 
\be
a^{+-}_S=a^{+-}\,
S_{WW}(v),
\en
with
\be
S_{WW}(v)=\frac{1}{9}\left|2 \psi_{{\cal I}=0}(\vec r=0)+\psi_{{\cal I}=2}(\vec r=0)\right|^2.
\label{eq: SWW}
\en
Finally, 
we obtain 
\be
\la \sigma^{+-} v\ra=
\left
\la a^{+-} 
S_{WW}(v)
\right\ra.
\label{eq: sigma v +-}
\en
In the $\psi_{\cal I}(\vec r=0)=1$ limit, $\la\sigma^{+-} v\ra$ reduces $\la(\sigma^{+-} v)_0\ra$. 
It is reasonable that the total isospin of the $\chi\bar\chi$ pair, ${\cal I}$, cannot be greater than 2 in the $\chi\bar\chi\to W^+W^-$ process, since the total isospin of $WW$ can at most be 2 and we are considering the SU(2) symmetric limit.

Note that through rescattering we can also have $\chi^0\overline{\chi^0}\to Z^0Z^0, Z^0\gamma,\gamma\gamma$ annihilations, with
\be
A_S(\chi^0\overline{\chi^0}\to Z^0Z^0)&=&\sum_i A(\chi^i\overline{\chi^i}\to Z^0Z^0) {\cal Q}_{i0}
\non\\
A_S(\chi^0\overline{\chi^0}\to Z^0\gamma)&=&\sum_i A(\chi^i\overline{\chi^i}\to Z^0\gamma) {\cal Q}_{i0},
\non\\
A_S(\chi^0\overline{\chi^0}\to \gamma\gamma)&=&\sum_i A(\chi^i\overline{\chi^i}\to \gamma\gamma) {\cal Q}_{i0},
\en
and, consequently,
\be
a^{00,0\gamma,\gamma\gamma}_S=\sum_{i,j} {\cal Q}^{\dagger}_{0i} a^{00,0\gamma,\gamma\gamma}_{ij} {\cal Q}_{j 0},
\en
with
\be
a^{00}_{ij}=\frac{g^4\cos^4\theta_W(m_\chi^2-m^2_Z)^{3/2}i^2j^2}{8\pi m_\chi(2m^2_\chi-m_Z^2)^2},
\quad
a^{0\gamma}_{ij}=\frac{e^2g^2\cos^2\theta_W(4m_\chi^2-m^2_Z) i^2j^2}
{64\pi m^4_\chi},
\quad
a^{\gamma\gamma}_{ij}=\frac{e^4 i^2j^2}{32\pi m^2_\chi}.
\en

\begin{figure}[t!]
\centering
\captionsetup{justification=raggedright}
\subfigure[]{
 \includegraphics[width=0.44\textwidth]  {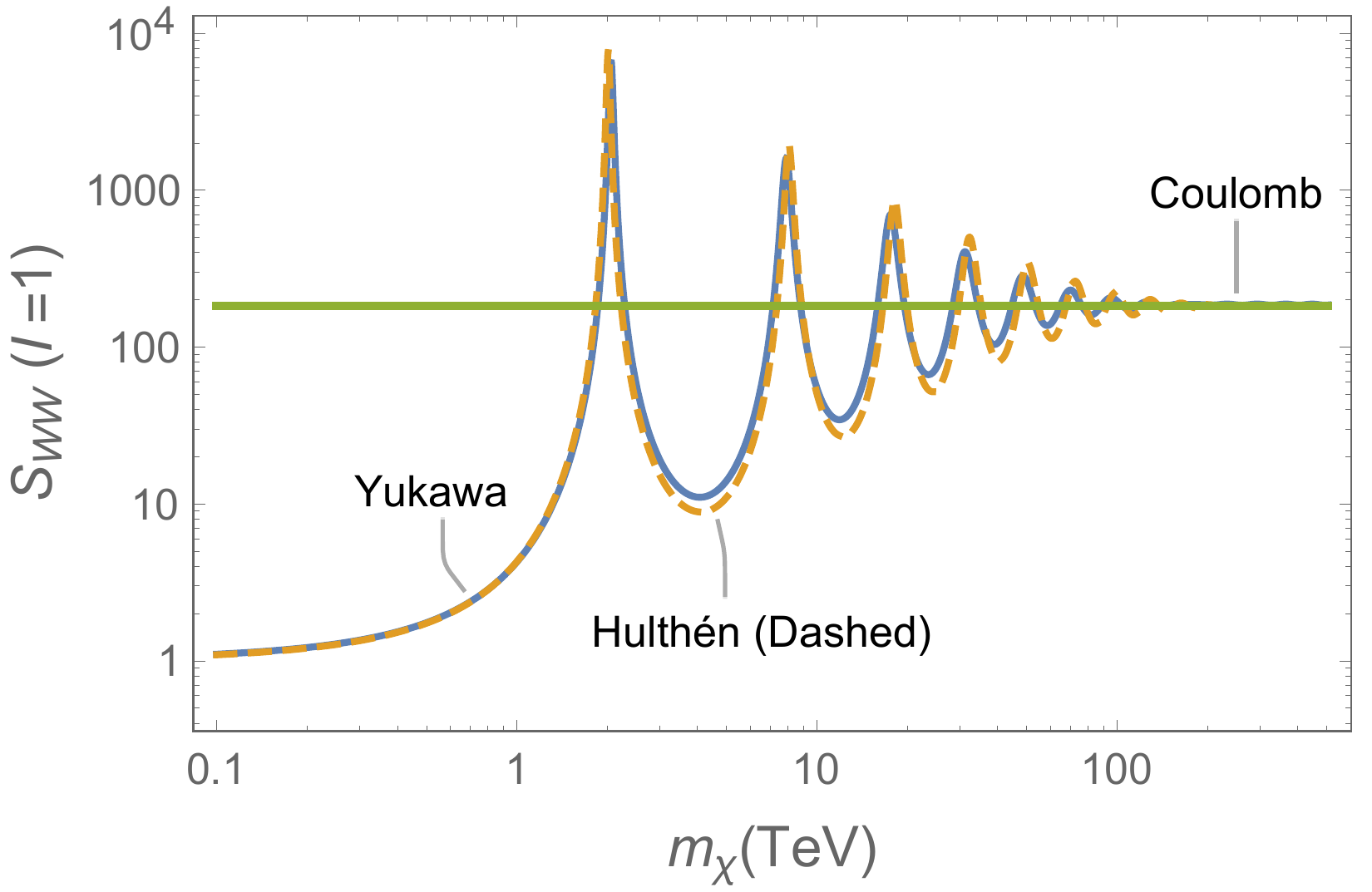}
}
\subfigure[]{
  \includegraphics[width=0.44\textwidth]  {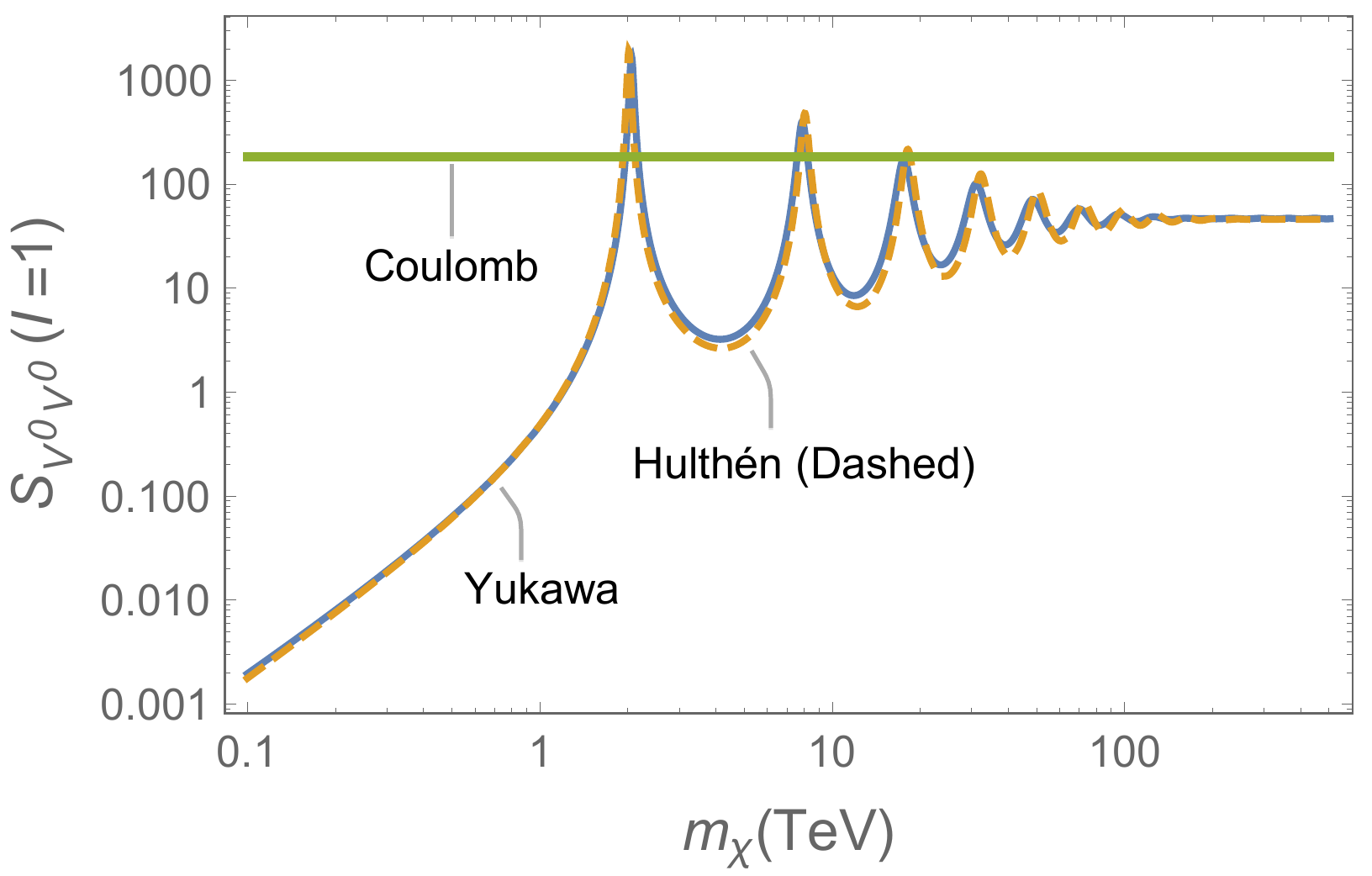}
}
\subfigure[]{
 \includegraphics[width=0.44\textwidth]  {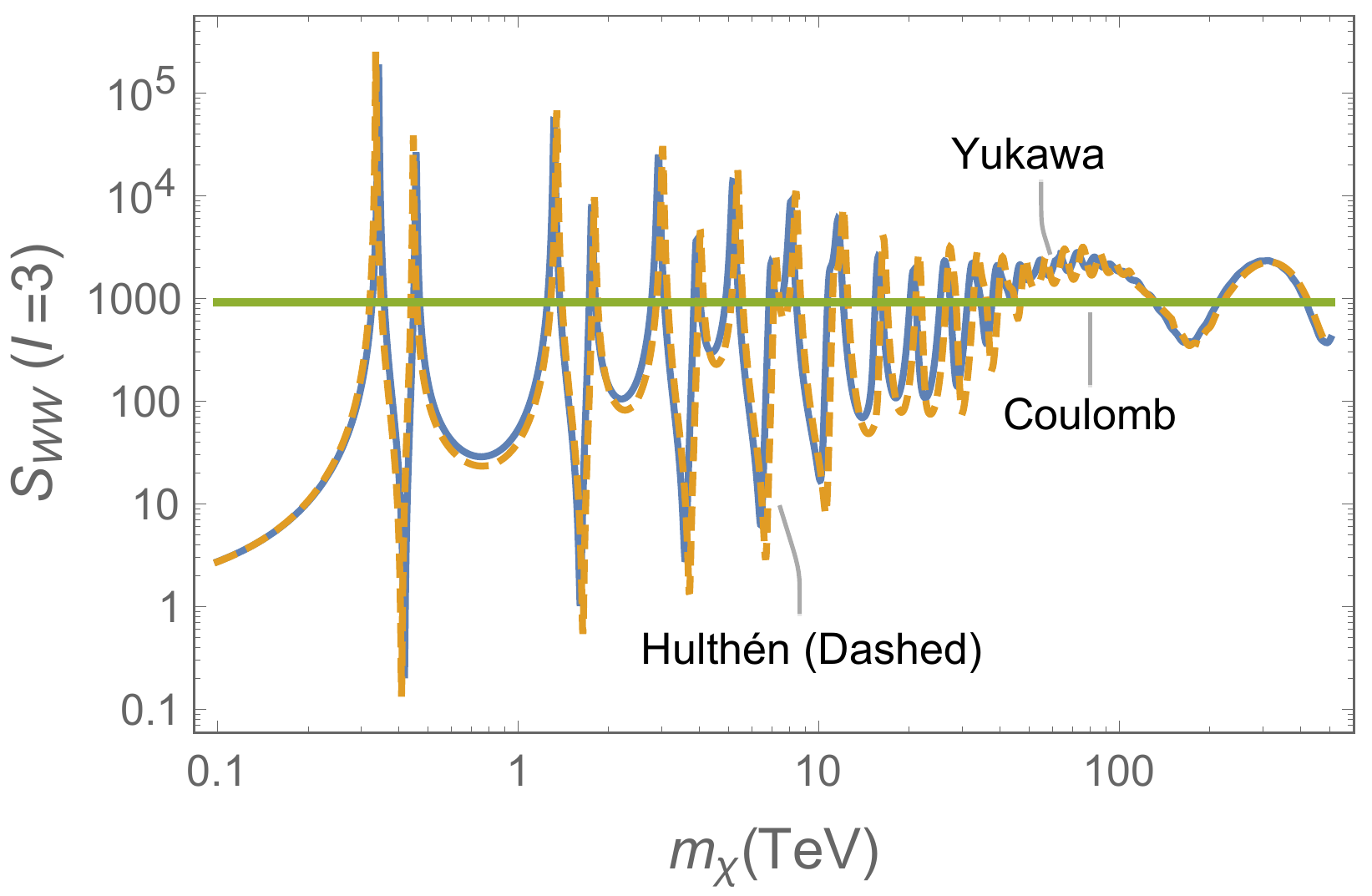}
}
\subfigure[]{
  \includegraphics[width=0.44\textwidth] {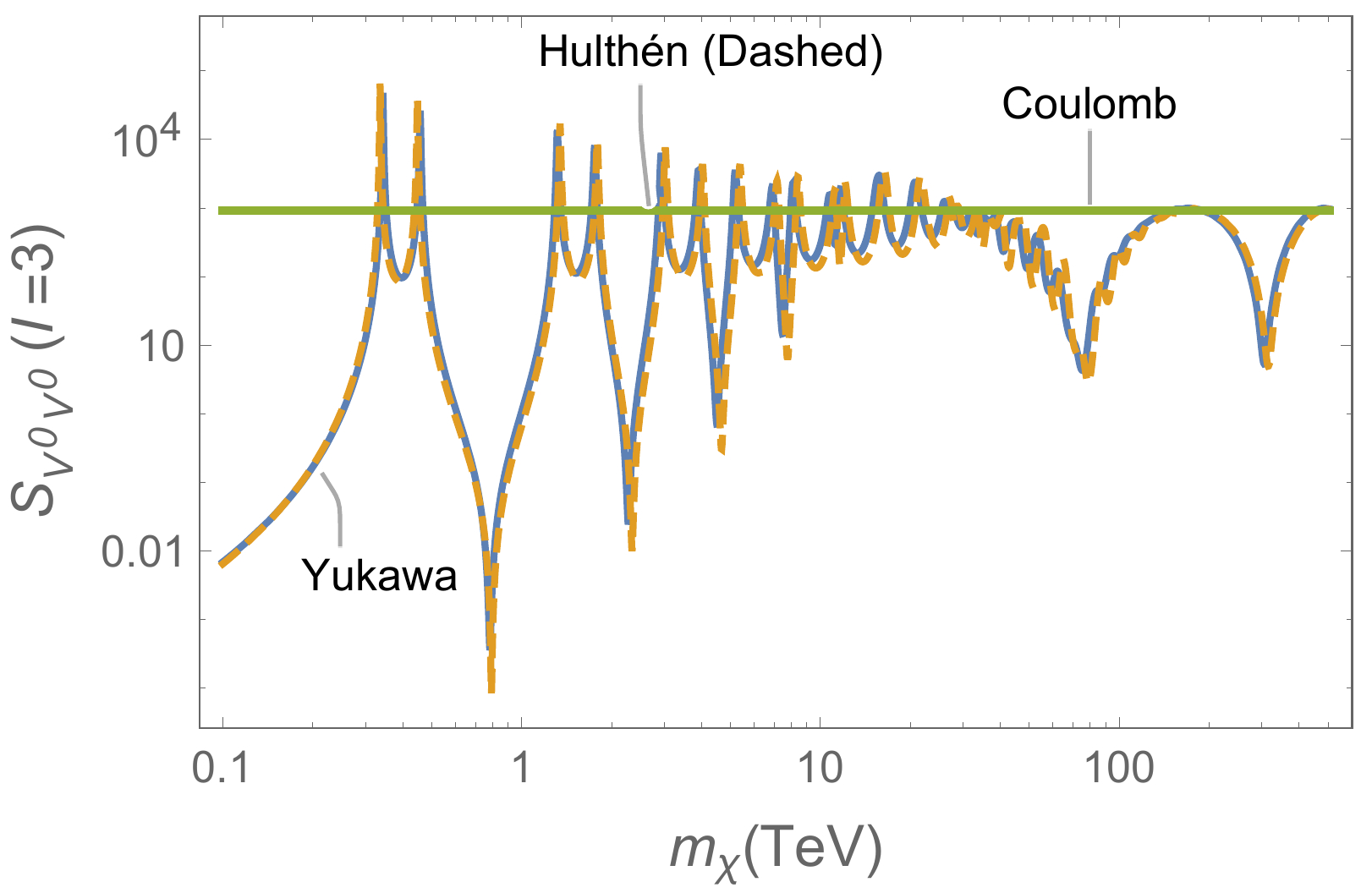}
}
\subfigure[]{
 \includegraphics[width=0.44\textwidth]  {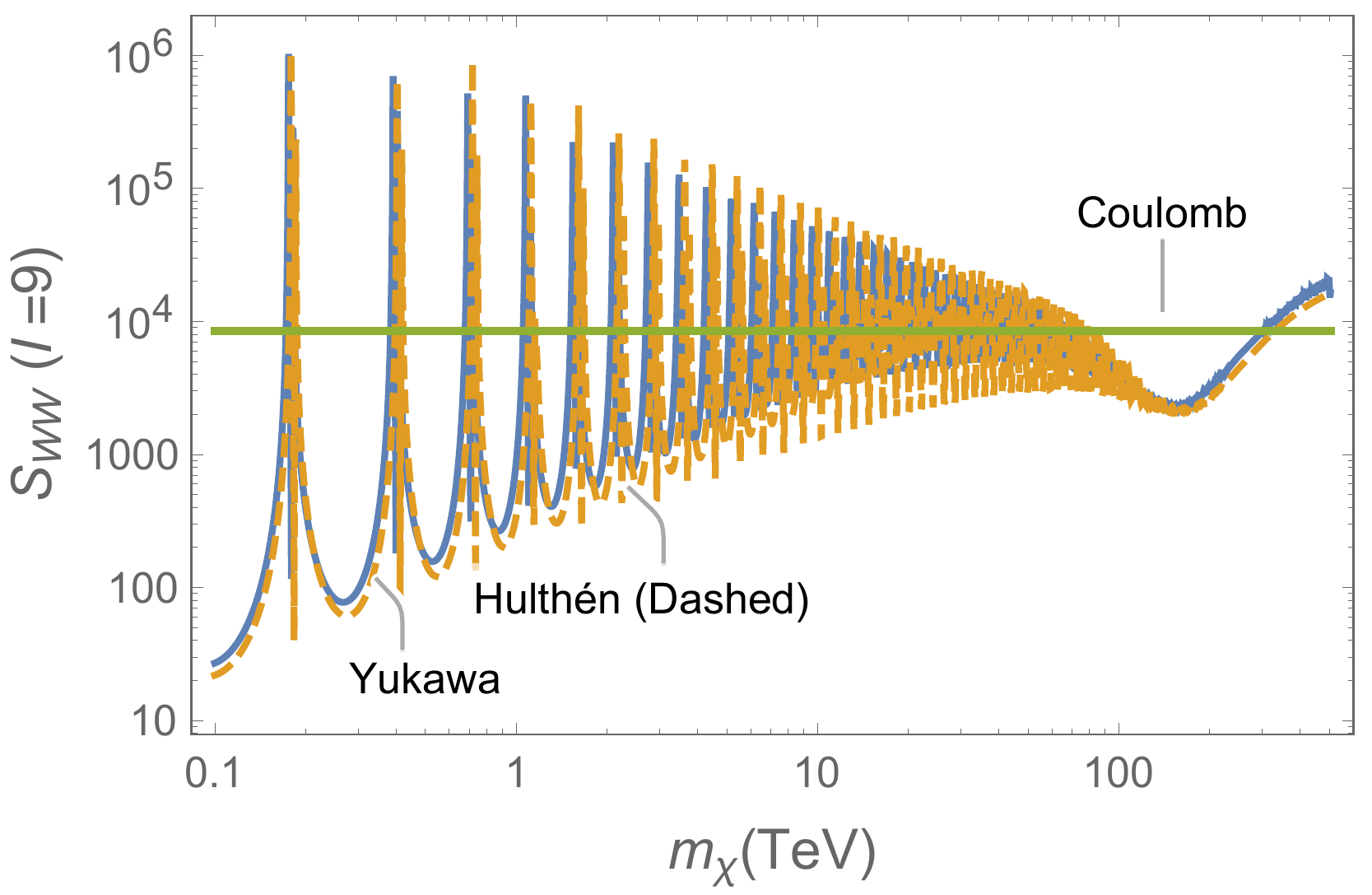}
}
\subfigure[]{
  \includegraphics[width=0.44\textwidth]   {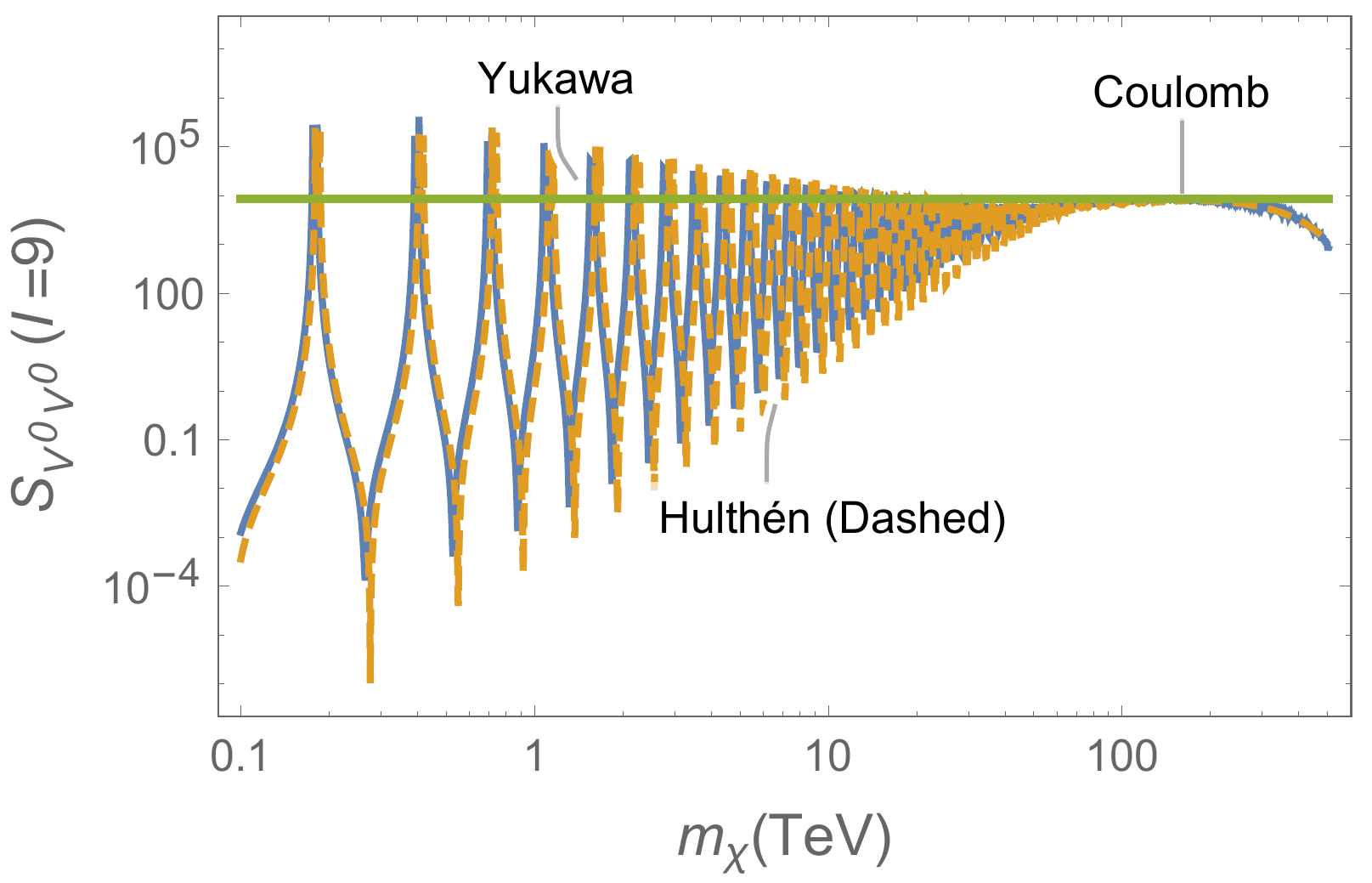}
}
\caption{The Sommerfeld factors $S_{WW}$ and $S_{V^0V^0}$ in the cases of Yukawa potential [Eq. (\ref{eq: Yukawa})], Hulth\'en potential [Eq. (\ref{eq: Huthen})] and the Coulomb potential [Eq. (\ref{eq: Coulomb})] for $I=1,3,9$ with $v=10^{-3}$ are plotted. The $\psi_{\cal I}(\vec r=0)$ in the Yukawa case are obtained by solving the differential equation numerically (see Appendix B.4).}
\label{fig:SWWSV0V0}
\end{figure}

\begin{figure}[t!]
\centering
\captionsetup{justification=raggedright}
\subfigure[]{
 \includegraphics[width=0.44\textwidth]{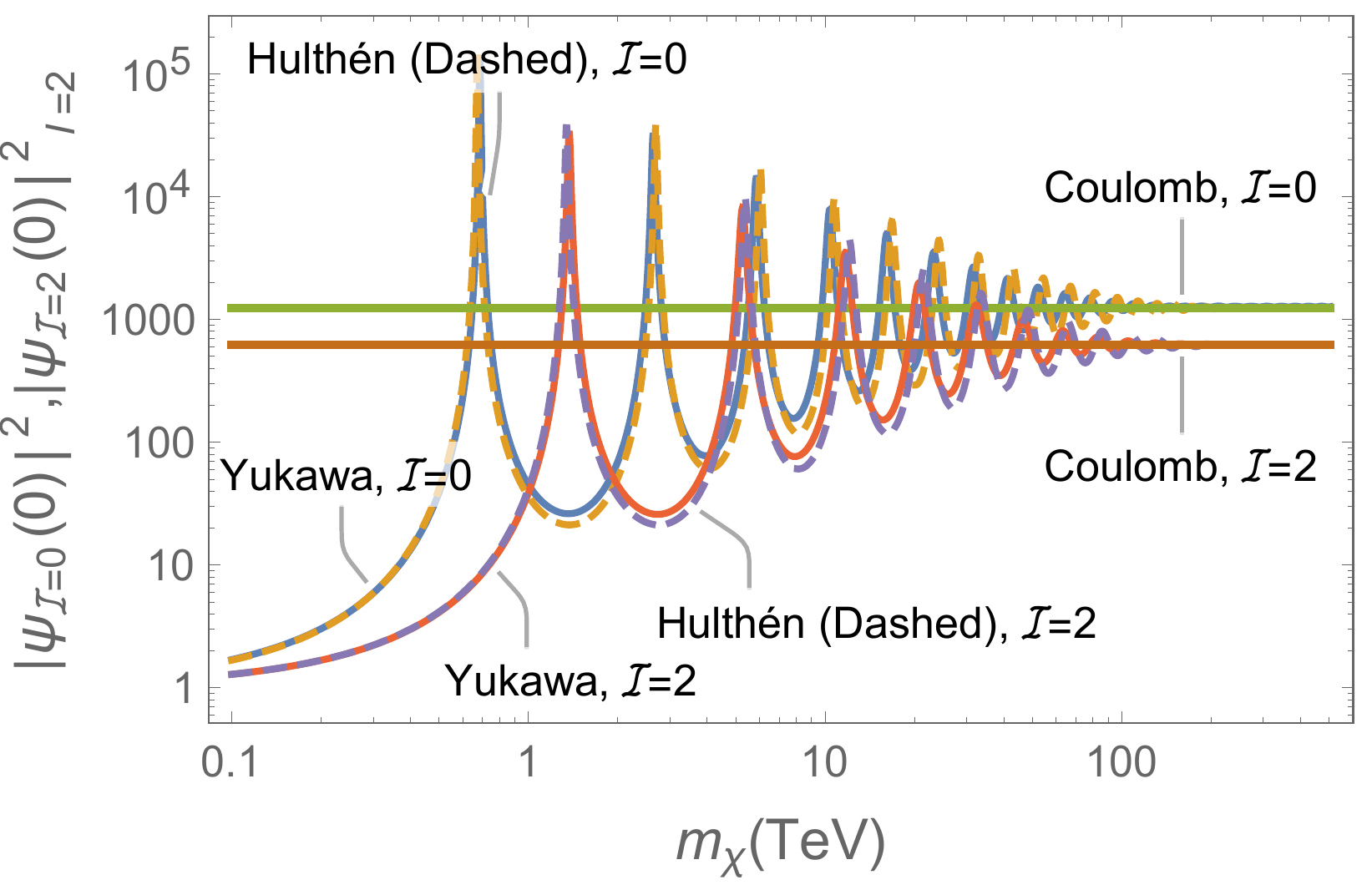}
}
\subfigure[]{
 \includegraphics[width=0.44\textwidth]{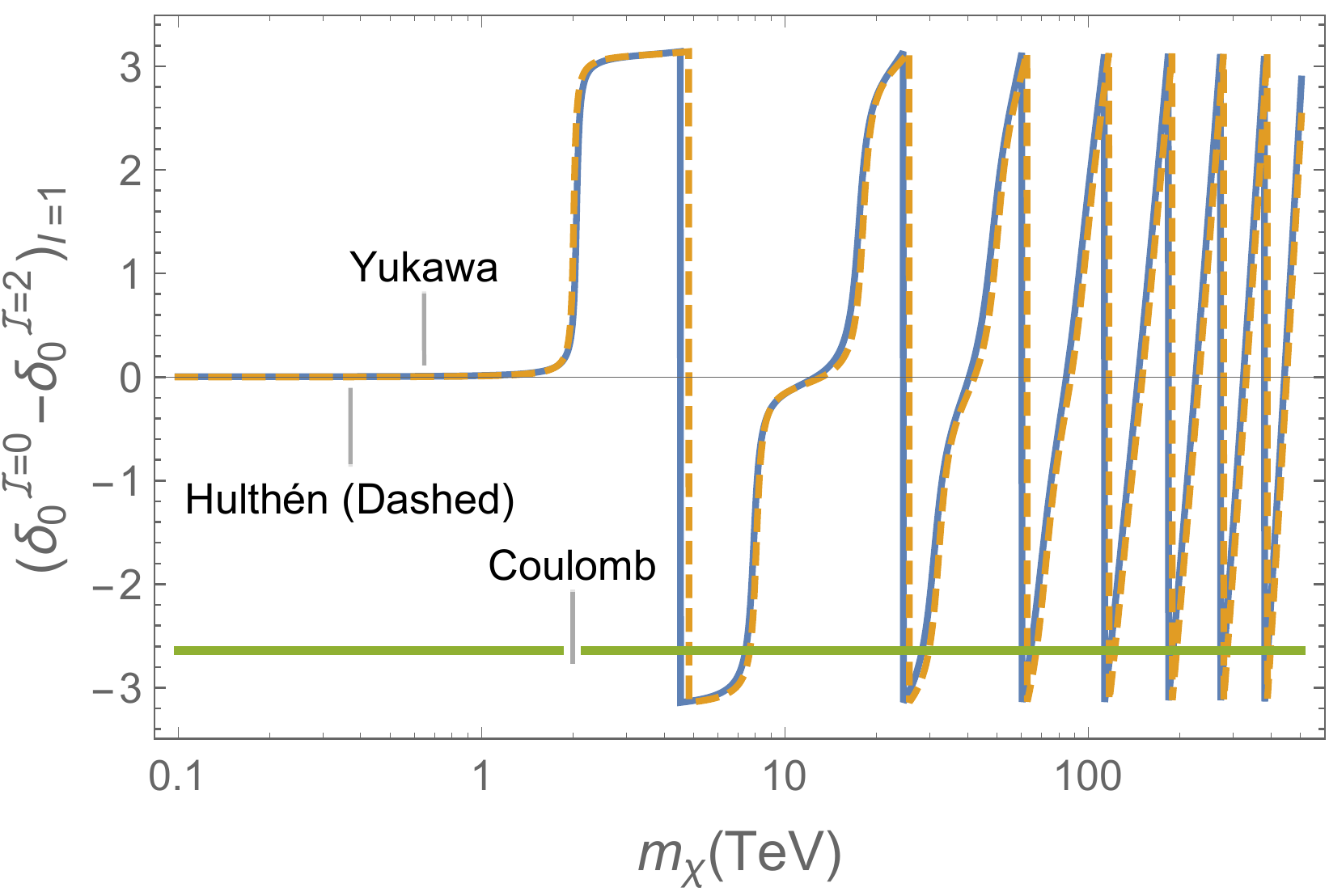}
}
\subfigure[]{
 \includegraphics[width=0.44\textwidth]{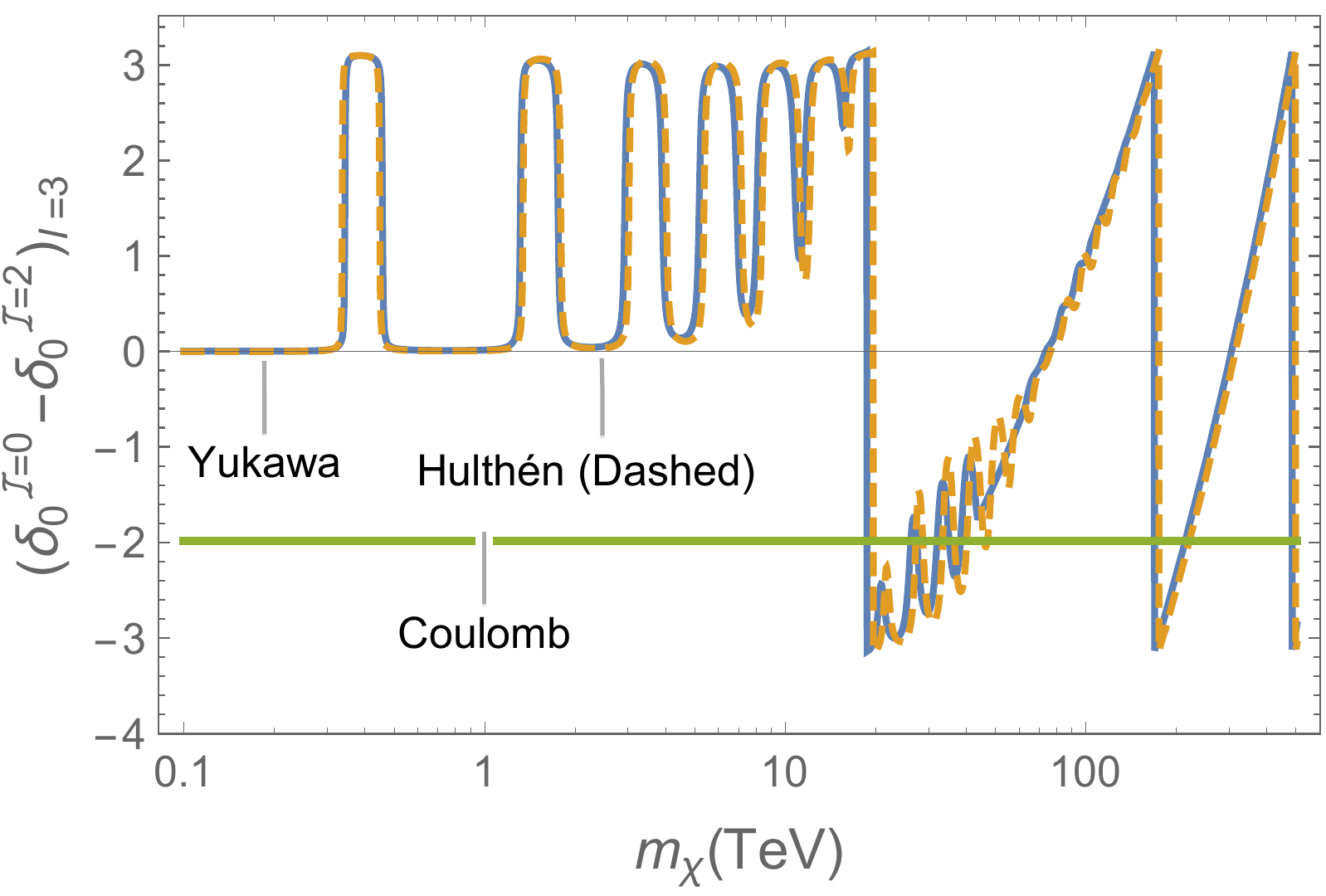}
}
\subfigure[]{
 \includegraphics[width=0.44\textwidth]{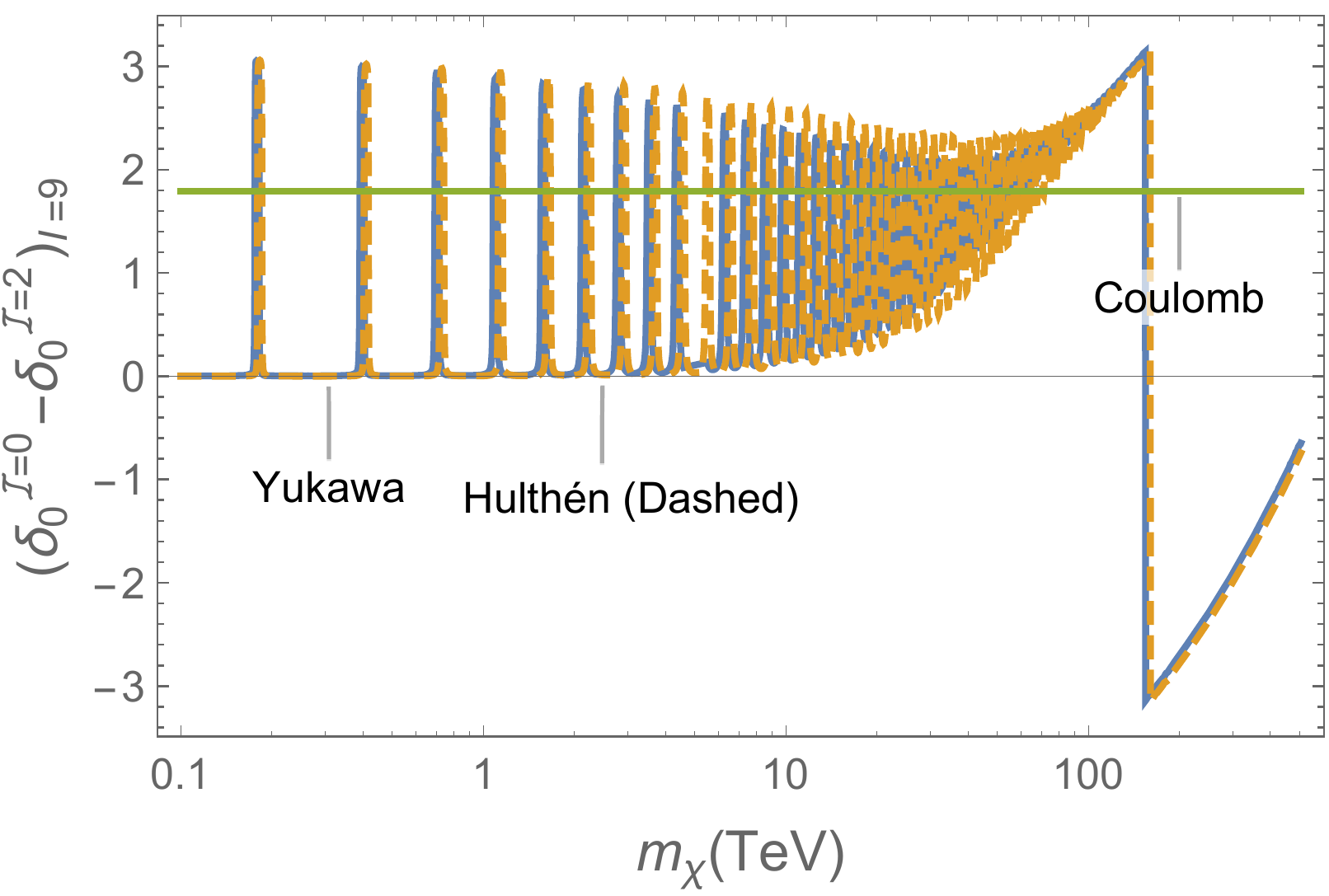}
}
\caption{$|\psi_{{\cal I}=0,2}(\vec r=0)|^2$ for $I=2$ with $v=10^{-3}$ are plotted in (a). The phase shift differences, $\delta_0^{{\cal I}=0}-\delta_0^{{\cal I}=2}$, in the cases of Yukawa potential, Hulth\'en potential and the Coulomb potential for 
(b) $I=1$, (c) $I=3$ and (d) $I=9$ with $v=10^{-3}$ are plotted. }
\label{fig:Deltadelta0}
\end{figure}

We obtain the annihilation cross sections for $\chi^0\overline{\chi^0}\to Z^0Z^0,Z^0\gamma,\gamma\gamma$ as
\be
\la \sigma^{\alpha} v\ra=
\left
\la a^{\alpha}
S_{V^0V^0}(v)
\right\ra,
\label{eq: sigma v alpha}
\en
with $\alpha=00,0\gamma,\gamma\gamma$,
\be
S_{V^0V^0}(v)=
\frac{1}{9}\left|\psi_{{\cal I}=0}(\vec r=0)-\psi_{{\cal I}=2}(\vec r=0)\right|^2,
\label{eq: SV0V0}
\en
and
\be
a^{00}&=&2a^{+-}\big|_{g\to g\cos\theta_W,m_W\to m_Z},
\non\\
a^{\gamma\gamma}&=&2a^{+-}\big|_{g\to e, m_W\to 0},
\non\\
a^{0\gamma}&=&[I(I+1)]^2\frac{e^2g^2\cos^2\theta_W(4m_\chi^2-m^2_Z) }
{64\pi m^4_\chi}.
\en
It is clear that these $\la\sigma^\alpha v\ra$s go to zero in the $\psi_{\cal I}(\vec r=0)=1$ limit.

In this model, $\chi^0\bar\chi^0$ cannot directly annihilate into a fermion pair, but it may be produced through loop diagrams by diagrams similar to Fig. \ref{fig:Som}(c) but with the final state replaced. Through rescattering $\chi^0\bar\chi^0$ can go to $\chi^i\bar \chi^i$, which can decay to a fermion pair $f\bar f$. The amplitude for $\chi^i\bar\chi^i\to f\bar f$ is proportional to $T_3(=i)$.
Hence the Sommerfeld enhanced rate is proportional to ${\cal Q}^\dagger_{0j} ji{\cal Q}_{i0}$, see footnote~\ref{fn:QQ=0}, which is in fact vanishing.

In Fig.~\ref{fig:SWWSV0V0} we compare results of the Sommerfeld factors $S_{WW}$ and $S_{V^0V^0}$ obtained from the Hulth\'en approximation, the Coulomb approximation results with the numerical ones, which are obtained by solving the equation with the Yukawa potential numerically (see Appendix B.4), for $I=1,3$ and $9$. We see that the Hulth\'en results can successfully mimic the numerical ones. 
In general the Sommerfeld factors $S_{WW}$ are enhancements,
but occasionally the interference is a destructive one leading to a suppression ($S_{WW}<1$) instead~(see, also~\cite{Chun1}). 
For example, for $I=3$ the factor $S_{WW}$ is smaller than one for $m_\chi\simeq 0.4$~TeV.
One can also see the behaviors of these Sommerfeld factors for different $I$.
These Sommerfeld factors oscillates as $m_\chi$ changes.
When $I$ is increasing the oscillations get faster in the low $m_\chi$ region, 
while simple oscillation patterns at high $m_\chi$ region are occurring.

The shapes of these Sommerfeld factors are governed by the norms of $\psi_{{\cal I}=0,2}$ and the relative phases between them. 
The former exhibit resonant-like behavior, see Eq.~(\ref{eq:Sold}) and, for example, Fig.~\ref{fig:Deltadelta0} (a). 
They approach $|\psi^{coul}_{\cal I}(\vec r=0)|$ in the large $m_\chi$ region [see Fig.~\ref{fig:Deltadelta0} (a)].
The phase differences between $\psi_{{\cal I}=0}(\vec r=0)$ and $\psi_{{\cal I}=2}(\vec r=2)$, 
which are also the phase shift differences, $\delta_0^{{\cal I}=0}-\delta_0^{{\cal I}=2}$, for $I=1,3,9$ are shown in Fig.~\ref{fig:Deltadelta0} (b), (c) and (d).
We see indeed for $I=3$, $\psi_{{\cal I}=0}(\vec r=0)$ and $\psi_{{\cal I}=2}(\vec r=2)$ are out of phase around $m_\chi\simeq 0.4$~TeV, giving the above mentioned destructive interference in $S_{WW}$.
Note that when $S_{WW}$ have destructive interference, $S_{V^0V^0}$ are enjoying constructive interference, instead, as one can see from Eqs. (\ref{eq: SWW}) and (\ref{eq: SV0V0}). 
From Fig.~\ref{fig:Deltadelta0}(a), (c) and (d), we see that 
the norms of $\psi_{\cal I}(\vec r=0)$ and
the phase differences oscillate rapidly with $m_\chi$ in the low mass region, but in large $m_\chi$ region the norms approaching constants, while 
the phase differences increase monotonically (with modulus $2\pi$). 
The above features are more prominent when $I$ increases. 
This explains the simple oscillation patterns of $S_{WW}$ and $S_{V^0V^0}$ in the large mass region. 

From Figs.~\ref{fig:SWWSV0V0} and \ref{fig:Deltadelta0}, we see that although $|\psi_{\cal I}(\vec r=0)|$ can go to $|\psi^{coul}_{\cal I}(\vec r=0)|$ in the large $m_\chi$ limit [see Fig.~\ref{fig:Deltadelta0}(a), in particular], the Coulomb approximation of the Yukawa potential does not reproduce the interference effects in the large $m_\chi$ region. 
Nevertheless the numerical and Hulth\'en results seems to hover around the Coulomb ones (see Fig.~\ref{fig:SWWSV0V0}). 
The latter seems to provide the ``averaged" (about the neighbouring $m_\chi$) behaviors of the formers. 
In this work we shall use the Hulth\'en approximation for $\psi_{\cal I}(\vec r=0)$ in the later numerical analysis.

\subsection{Numerical Results for Relic Density and Indirect Search}

\begin{figure}[t!] \centering
\includegraphics[width=0.7\textwidth]{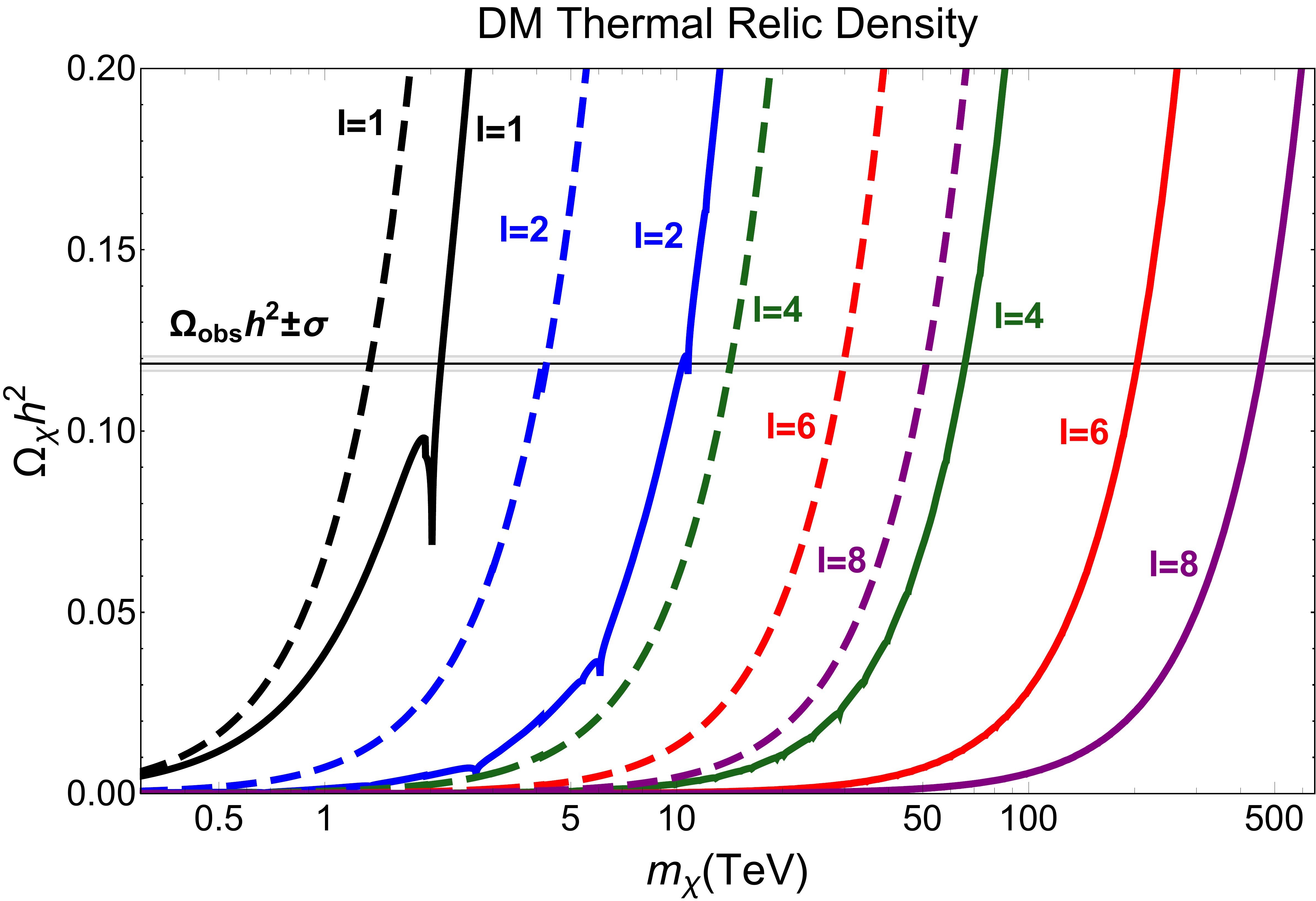}
\caption{Predicted DM thermal relic density for $I=1,2,4,6,8.$} 
\label{fig:relic}
\end{figure}

We present our numerical results of relic density and indirect search here. 
Fig.~\ref{fig:relic} shows the predicted relic density using $x_f=24$ for $I=1,2,4,6,8$ cases. The shadow area shows the observed dark matter relic density $\Omega_{\rm obs} h^2=0.1186\pm 0.0020$~\cite{PDG}. The solid (dashed) lines denote the results with (without) considering the Sommerfeld enhancement effect. 
Note that the relic density $\Omega_{\chi}$ is inversely proportional to $\la\sigma v\ra$, which is enhanced by the Sommerfeld effect. Hence, $\Omega_{\chi}$ is reduced and the DM mass satisfying the relic constraint is shifted to a larger value.
We consider the possibility that $\chi$ may not saturate the DM relic density, i.e., $\Omega_{\chi}\leqslant\Omega_{\rm obs}$. 
Hence, the observed relic density provides the upper mass bounds for DM particles with different isospin $I$. 
We see that the upper limit on Dirac DM mass becomes larger for a larger isospin $I$.

\begin{table}[b!]
\centering
\captionsetup{justification=raggedright}
\caption{The lower limits $m_\chi^{LL}$ obtained from recent direct and indirect dark matter search experiments, the upper limits $m_\chi^{UL}$ obtained from the observed relic density are shown. Note that $m_\chi$ for $I=1, 2$ cases are not constrained by direct search. The upper values (lower values within the parentheses) for indirect $m_{\chi}^{LL}$ and relic $m_{\chi}^{UL}$ denote the results with (without) considering the Sommerfeld effect. Dirac DM mass is given in the unit of TeV.
\label{tab:mlimits}}
\label{tab:summary}
\begin{tabular}{|c|c|c|c|c|c|c|c|}
\hline
$I$
    & $1$ 
    & $2$ 
    & $3$ 
    & $4$ 
    & $5$ 
    & $6$
    & $7$
    \\ 
    \hline
Direct $m_{\chi}^{LL}$ 
    & --- 
    & ---
    & 0.19
    & 0.37 
    & 0.72 
    & 1.18 
    & 2.28 
   \\ 
 \hline
    \multirow{2}{*}{Indirect $m_{\chi}^{LL}$}
    & 3.00
    & 1.09 
    & 6.64 
    & 41.16
    & 67.33
    & $>$ 70
    & $>$ 70
   \\ 
    & $(0.60)$ 
    & (2.38)
    & (4.29)
    & (6.38)  
    & (8.55)
    & (10.86)
    & (13.34)
    \\
    \hline
     \multirow{2}{*}{Relic $m_{\chi}^{UL}$}
     & 2.15
    & 10.58
    & 30.31
    & 66.54
    & 123.20
    & 205.23
    & 316.94
   \\ 
    & (1.36)
    & (4.30)
    & (8.60)
    & (14.34)     
    & (21.51)
    & (30.11)
    & (41.14)
   \\ 
\hline
\end{tabular}
\end{table}

\begin{figure}[t!]
\centering
\captionsetup{justification=raggedright}
 \subfigure[without Sommerfeld enhancement effect]{
  \includegraphics[width=0.44\textwidth]{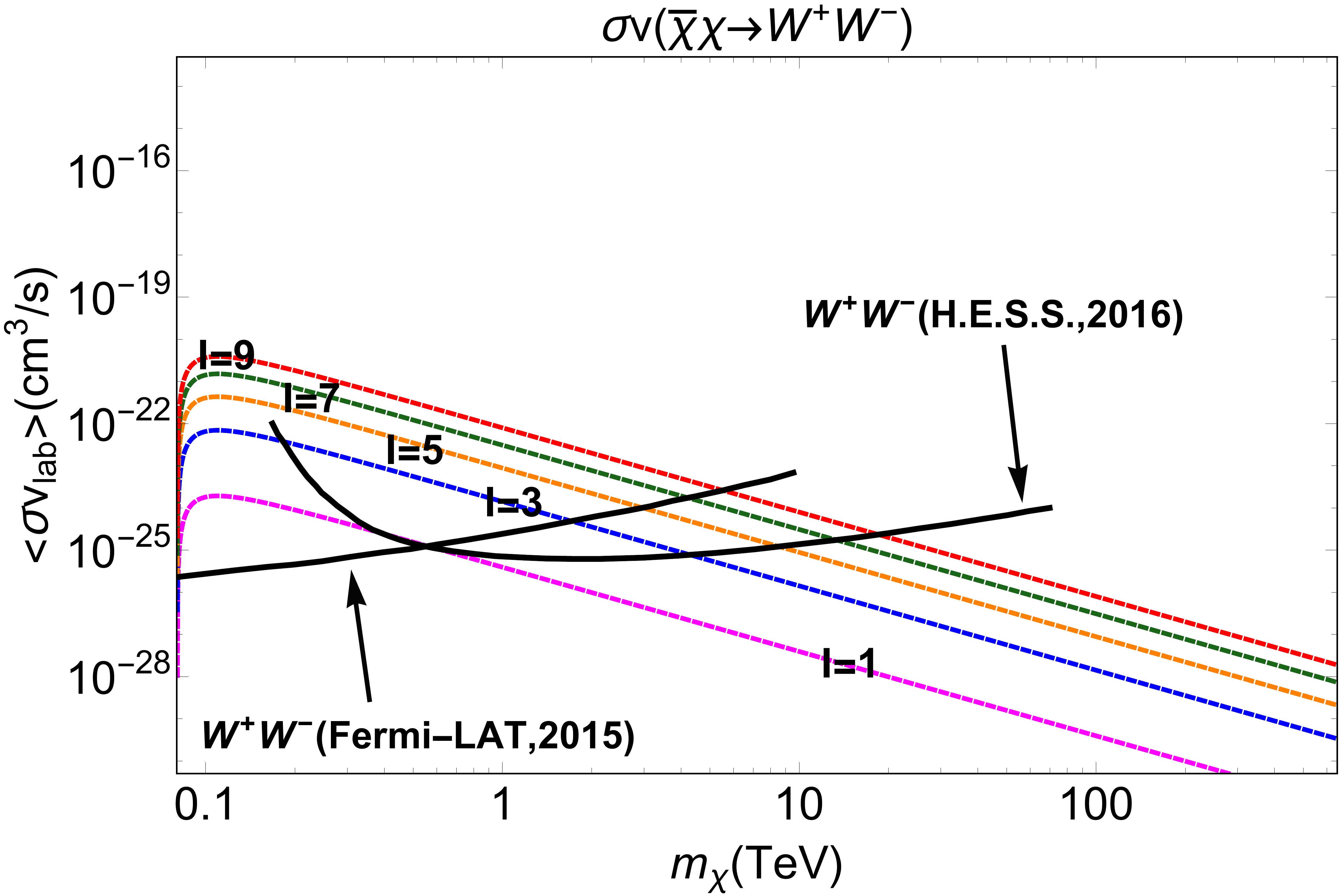}
}\subfigure[with Sommerfeld enhancement effect]{
  \includegraphics[width=0.44\textwidth]{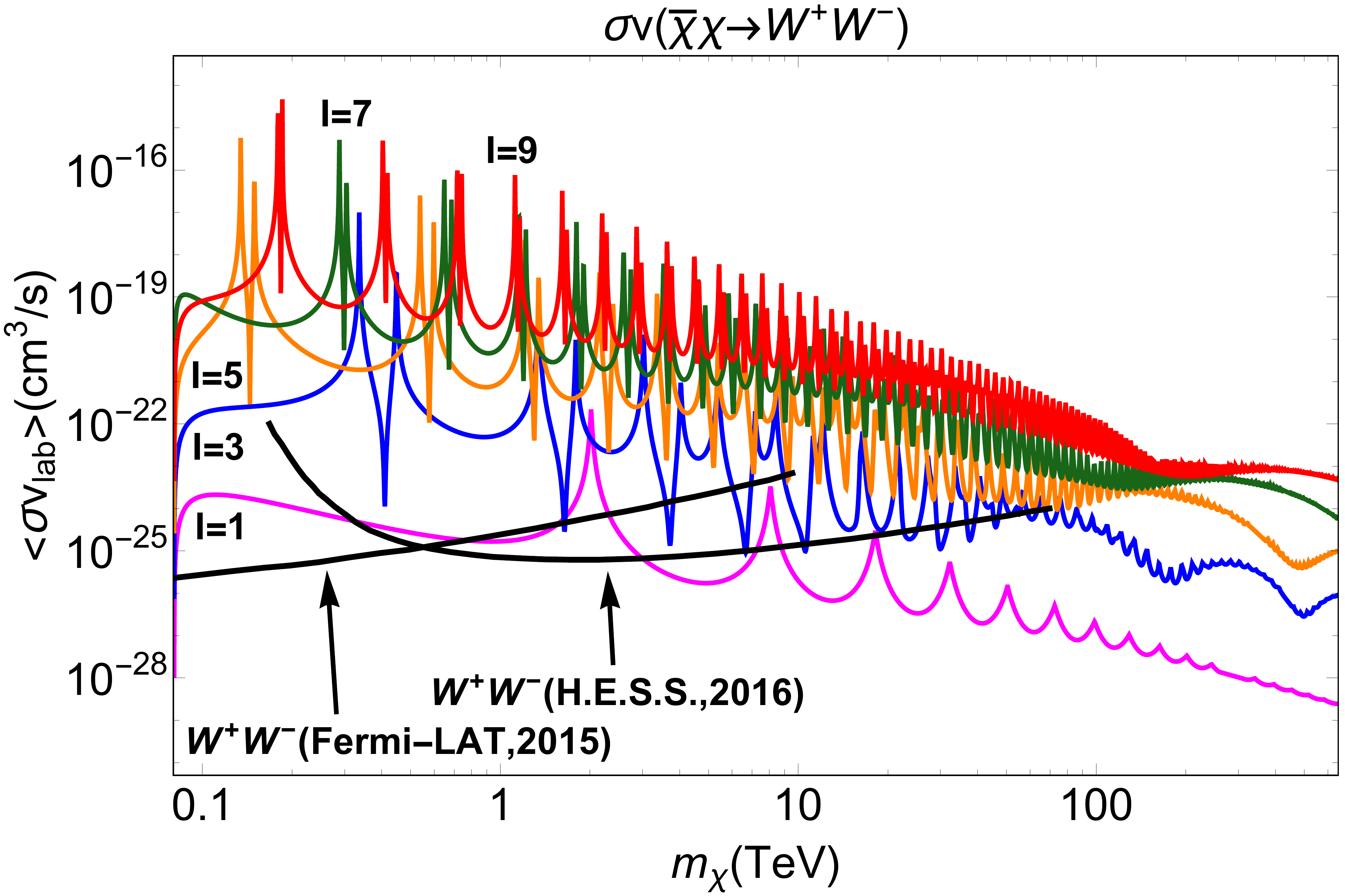}
}
\caption{Predicted velocity averaged cross sections of galactic DM annihilation to $W^+W^-$ processes for $I=1,3,5,7$ and $9$ cases.}
\label{fig:DMtoWW}
\end{figure}

Figs.~\ref{fig:DMtoWW}(a)  and \ref{fig:DMtoWW}(b) show the plots of velocity averaged cross section $\la \sigma v\ra$ of DM annihilation in $W^+W^-$ channel without and with considering the Sommerfeld enhancement effect for $I=1,3,5,7$ and $9$, respectively. 
We see that the pattern of $\la\sigma(\chi\bar\chi\rightarrow W^+W^-)v\ra$ versus $m_\chi$ in Fig.~\ref{fig:DMtoWW}(b) is determined the shape of the original  $\la\sigma(\chi\bar\chi\rightarrow W^+W^-)v\ra$ in Fig.~\ref{fig:DMtoWW}(a) and the Sommerfeld factor $S_{WW}(v)$ (see Fig.~\ref{fig:SWWSV0V0}). 
Note that the Sommerfeld enhancement in Galactic DM annihilations is much stronger than cosmological DM annihilations, as the velocity of the Galactic DM is much slower.
In each plot, we compare our results with $W^+W^-$ data from H.E.S.S.~\cite{HESS2016} and Fermi-LAT~\cite{Fermi-LAT2015} astrophysical observations.  Fermi-LAT provides upper limits on $\la \sigma v\ra$ for DM annihilating into $W^+W^-$ and various SM fermion pairs at $95\%$ confidence level with WIMPs masses between 2 GeV to 10 TeV, while H.E.S.S. gives the corresponding upper limits with masses from 160 GeV to 70 TeV. 
In the $W^+W^-$ channel, for $m_{\chi} \lesssim 0.55$ TeV the Feremi-LAT limit dominates, while for $0.55 \lesssim m_{\chi} \leqslant 70$ TeV the H.E.S.S limit dominates.
We see that the low mass region with $m_{\chi}\leqslant 0.55$ TeV are ruled out by the Fermi-LAT constraint and the H.E.S.S data is use to constrain the mass region with $0.55 < m_{\chi}\leqslant 70$ TeV.
The combining $W^+W^-$ data basically excludes light DM and provides us the lower mass bounds and allowed mass regions for the Dirac DM particles with different isospin $I$. 
The Sommerfeld enhancement increases the cross sections by 1 to 4 order of magnitude, and hence the lower limits on the Dirac DM masses are in principle shifted to larger values. 

\begin{figure}[t!]
\centering
\captionsetup{justification=raggedright}
\subfigure[]{
  \includegraphics[width=0.44\textwidth]{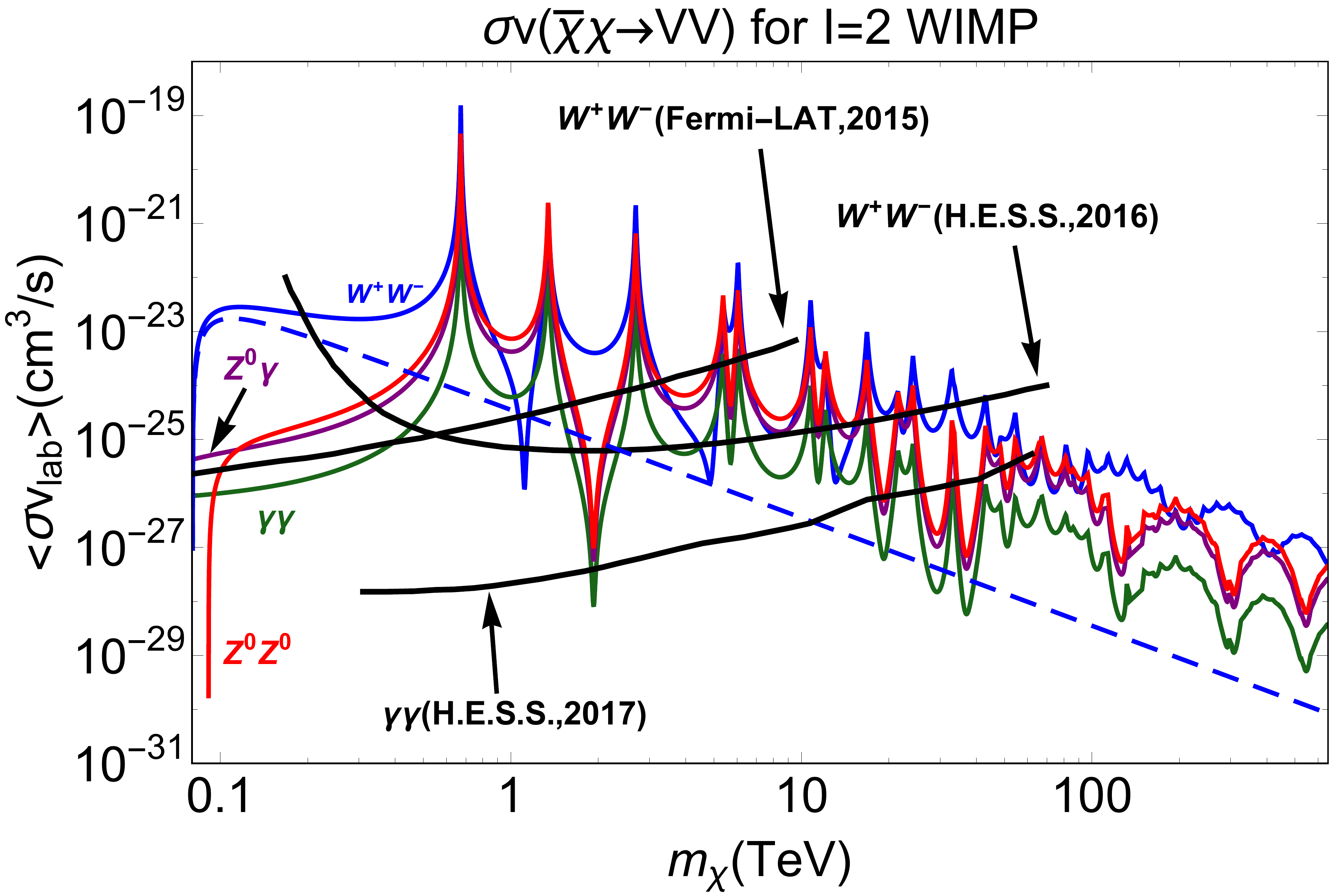}
}\subfigure[]{
  \includegraphics[width=0.44\textwidth]{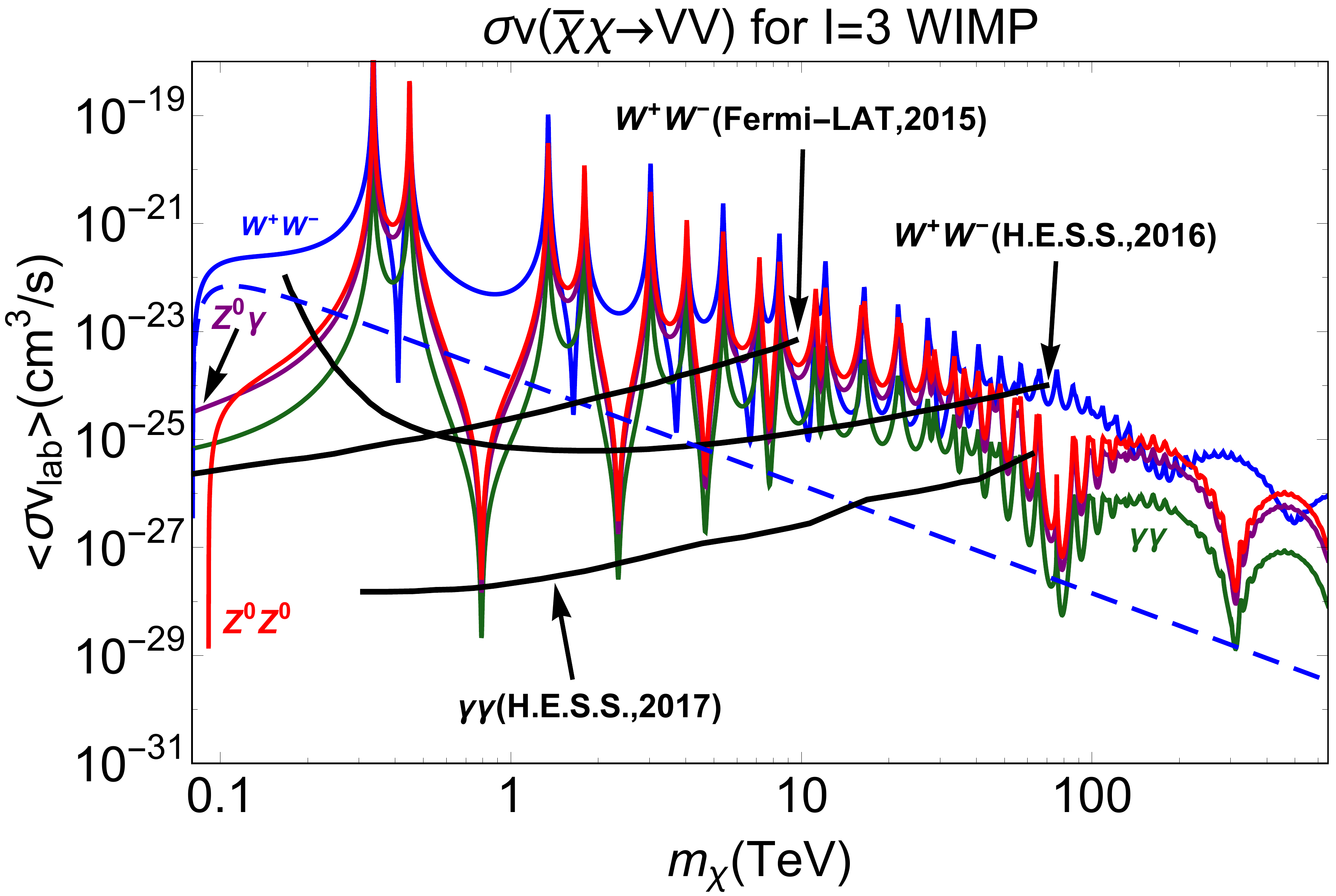}
}\\\subfigure[]{
  \includegraphics[width=0.44\textwidth]{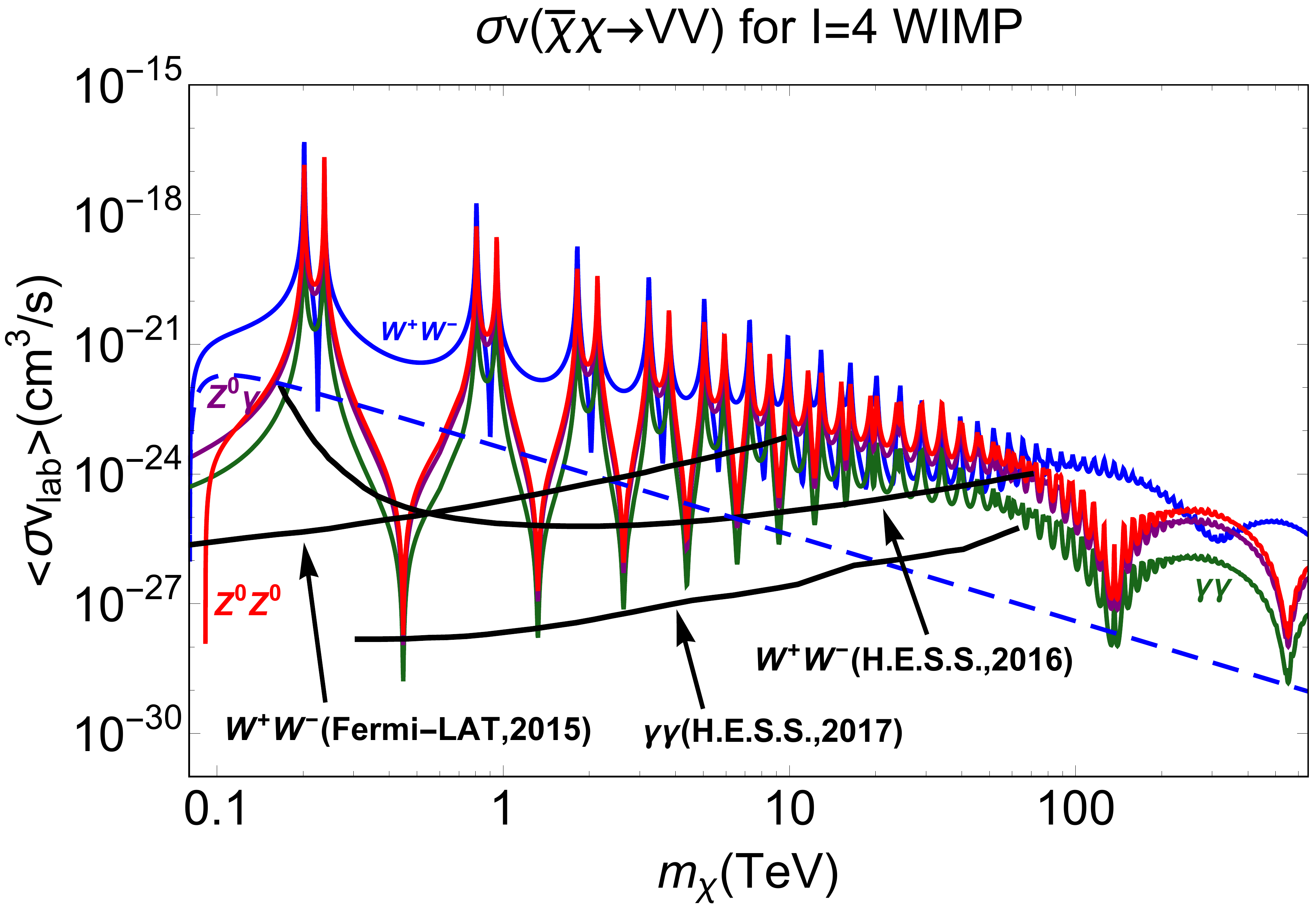}
}\subfigure[]{
  \includegraphics[width=0.44\textwidth]{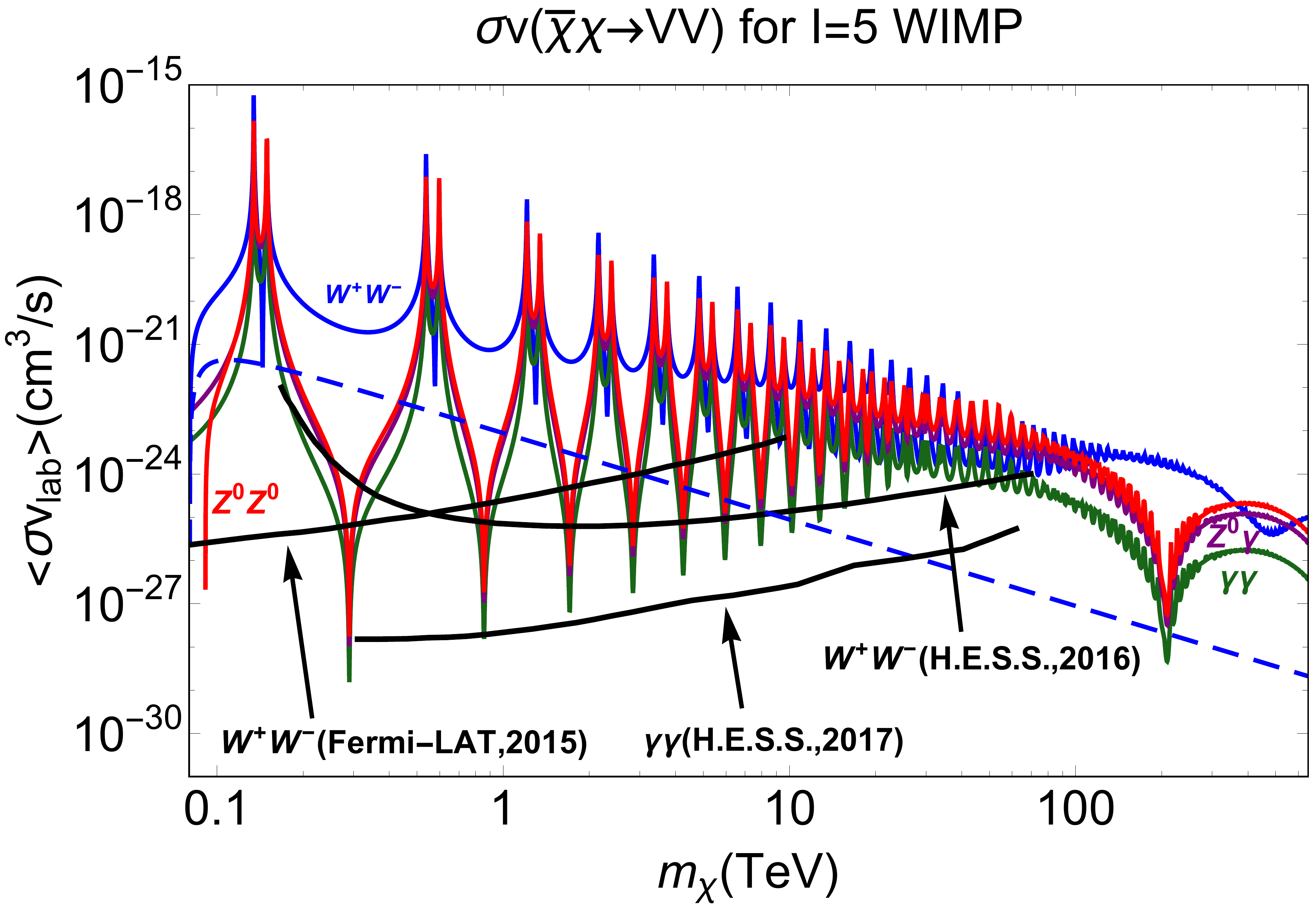}
}\\\subfigure[]{
  \includegraphics[width=0.44\textwidth]{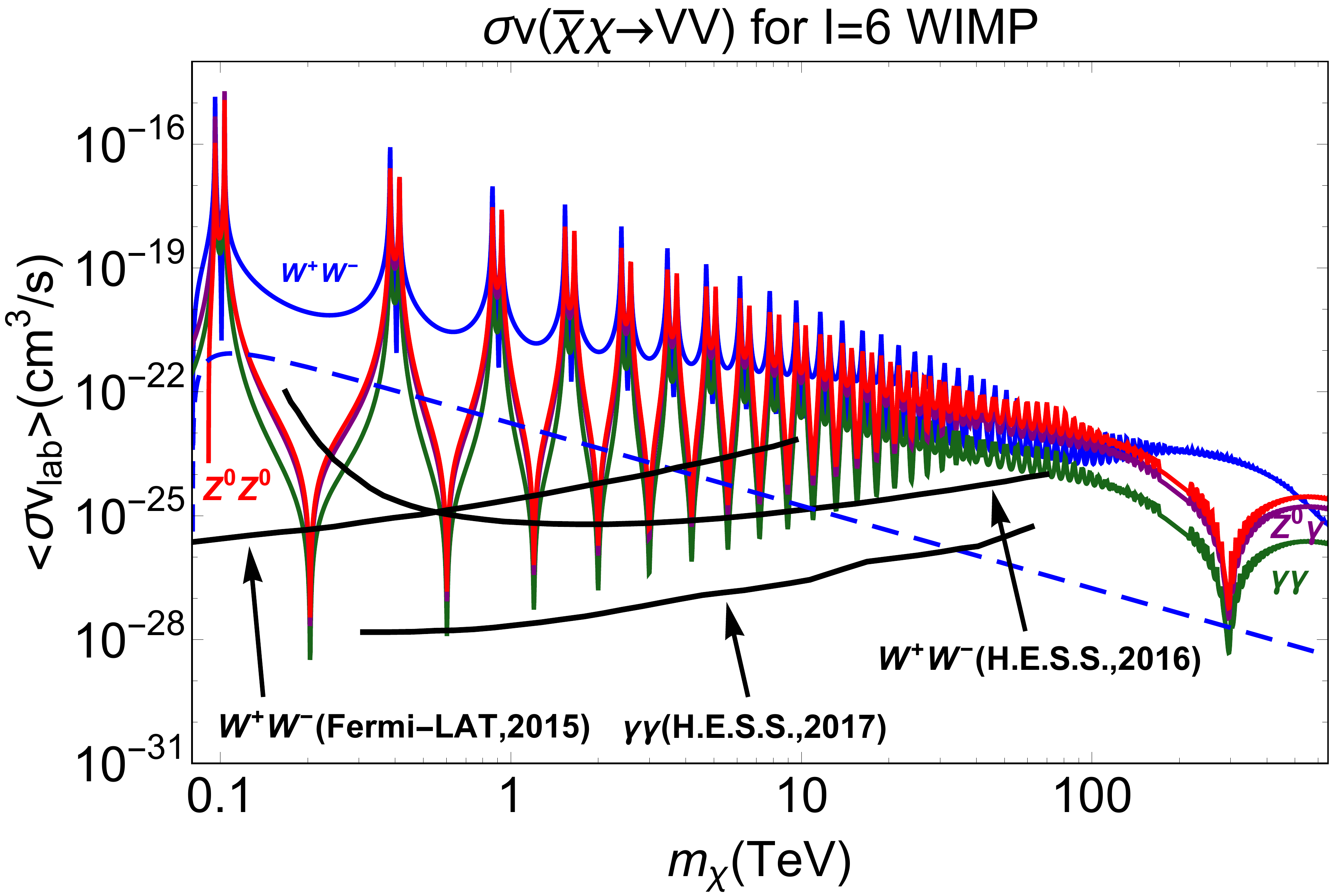}
}\subfigure[]{
  \includegraphics[width=0.44\textwidth]{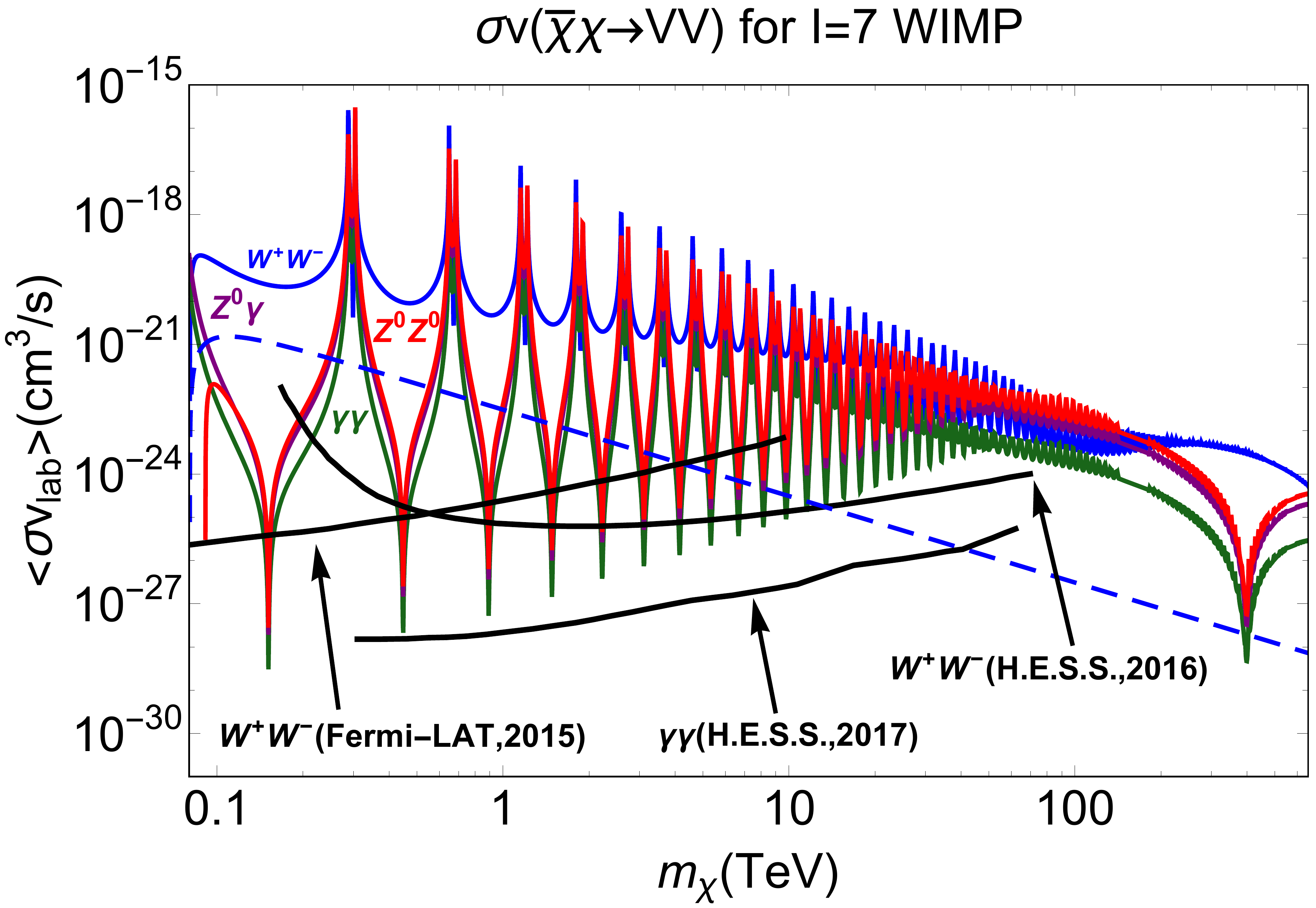}
}
\caption{Predicted velocity averaged cross sections of galactic DM annihilation to $W^+W^-$, $Z^0Z^0$, $Z^0\gamma$ and $\gamma\gamma$ processes for $I=2,3,4,5,6,7$ cases.}
\label{fig:ind}
\end{figure}

In Table.~\ref{tab:mlimits}, we show the lower mass limits 
$m_\chi^{LL}$ from the direct search of PandaX-II and XENON1T constraints, the indirect search of Fermi-LAT and H.E.S.S. constraints and the upper mass limits $m_\chi^{UL}$ from the observed DM relic density. Note that we require the contribution to relic density does not excess the upper $1$ $\sigma$ range of the observed dark matter relic density, $\Omega_{\rm obs} h^2$.
The upper values (lower values within the parentheses) for indirect $m_{\chi}^{LL}$ and relic $m_{\chi}^{UL}$ in the table denote the results with (without) considering the Sommerfeld effect. 
The XENON1T constraint does not provide lower limits on DM masses for $I\leqslant 2$. For the larger $I$ case, the lower mass limit constrained from the indirect search is more stringent than that from the direct search. Without considering the Sommerfeld effect, all cases (with different isospin $I$) are allowed, while after turning on the Sommerfeld effect,
the $I=1$ case is ruled out, since the lower mass limit $m_{\chi}^{LL}=3.00$ TeV is greater than the upper mass limit $m_{\chi}^{UL}=2.15$ TeV.
 
Fig.~\ref{fig:ind} shows the plots of velocity averaged cross section of DM annihilating to $W^+W^-$, $Z^0Z^0$, $Z^0\gamma$ and $\gamma\gamma$ channels for $I=2\sim7$, separately. 
Besides the constraints from H.E.S.S. $\la\sigma v\ra(W^+W^-)$ and Fermi-LAT $\la\sigma v\ra(W^+W^-)$, we also show the constraint from H.E.S.S. $\la\sigma v\ra(\gamma\gamma)$, which provides $95\%$ confidence level with WIMPs masses from 300 GeV to 
60 TeV~\cite{HESS2017rr}.
The solid (dashed) lines are our numerical results with (without) considering the Sommerfeld enhancement effect. DM can annihilate into a pair of neutral gauge bosons only through the Sommerfeld enhancement effect. 
In general, we see that $\la\sigma (\chi^0\bar\chi^0\to W^+W^-)v\ra \gtrsim \la\sigma (\chi^0\bar\chi^0\to V^0V^0)v\ra$, but occasionally it is the other way around.
As mentioned in previous subsection, the Sommerfeld factor can be occasionally suppressed due to the destructive interference between $\psi_{{\cal I}=0} (\vec r=0)$ and $\psi_{{\cal I}=2} (\vec r=0)$ resulting in smaller lower mass bound. This can be seen that the lower mass bound originally from 2.38 TeV is shifted to a smaller value of 1.09 TeV for $I=2$ case (see Fig.~\ref{fig:ind}(a)) after considering the Sommerfeld effect.

From Figs.~\ref{fig:ind}(a) and (b), we see that the H.E.S.S constraint on $\la\sigma v\ra(\gamma\gamma)$ rules out $m_\chi\leq$ 17.8 and 42.3 TeV in the cases of $I=2$ and 3, respectively.
Note that there are a few dips in the predicted $\la\sigma v\ra(\gamma\gamma)$ for $m_\chi$ below the above mentioned lower limits that can satisfy the H.E.S.S. $\gamma\gamma$ constraint, but they are ruled out by the H.E.S.S. $W^+W^-$ constraint.
Furthermore, since the upper bounds on $m_\chi$ from the relic density are 10.6 and 30.3 TeV for $I=2$ and 3, respectively (see Table~\ref{tab:mlimits}), 
these two cases are ruled out by the H.E.S.S. and relic density constraints.

\begin{table}[t!]
\centering
\captionsetup{justification=raggedright}
\caption{The allowed region of $m_\chi$ in the $I=4\sim7$ cases after imposing Fermi-LAT $\la\sigma v\ra(W^+W^-)$ and H.E.S.S. $\la\sigma v\ra(W^+W^-,\gamma\gamma)$ constraints and the relic density constraint are given. 
Predictions on $\la\sigma v\ra$ for Galactic $\chi\bar\chi\to W^+W^-, ZZ, Z\gamma,\gamma\gamma$ annihilations, $\Omega_\chi h^2$ and $\sigma^{SI}$ are also shown.
DM mass are given in the unit of  TeV, while $\la\sigma v\ra $ in the unit of $10^{-25}$ cm$^3$/s and $\sigma^{SI}$ in $10^{-46}$ cm$^2$. 
The values in the parentheses for $\sigma^{SI}$ denote the prediction value without considering the contributions of quark twist-2 operator and the two-loop diagrams.
\label{tab: I2}}
\label{tab:summary}
\begin{tabular}{|c|c|c|c|c|c|c|c|c|}
\hline
$I$
    &Allowed $m_\chi$ 
    & $\la\sigma v\ra(W^+W^-)$ 
    & $\la\sigma v\ra(Z^0Z^0)$ 
    & $\la\sigma v\ra(Z^0\gamma)$ 
    & $\la\sigma v\ra(\gamma\gamma)$ 
    & $\Omega_\chi h^2(\%)$
    & $\sigma^{SI}$
    \\ 
    \hline
 4   
         & $(61.45, 63.49)$ 
    & $(5.6, 9.1)$ 
    & $(14.6, 21.9)$ 
    & $(8.4, 12.6)$ 
    & $(1.2, 1.8)$ 
    & $(10.30, 11.00)$
    & $\sim2.85\ (6.76)$ 
    \\
  \hline
\multirow{2}{*}{5} 
    & $(67.33, 68.04)$ 
    & $(9.4, 9.8)$  
    & $(43.7, 54.9)$  
    & $(25.1, 31.6)$  
    & $(3.6, 4.5)$  
    & $(3.60, 3.68)$ 
    & $\sim 6.41\ (15.21)$ 
   \\ 
   &  $(70, 123.25)$ 
   &  $(9.4, 144.6)$  
   &  $(6.5, 101.2)$ 
    & $(3.7, 59.5)$  
    & $(0.54, 8.5)$ 
    & $(3.89, 12.06)$
    & $\sim 6.41\ (15.2)$ 
   \\
 \hline
6
     & $(70, 205.23)$ 
    & $(18.2, 42.8)$  
    & $(5.2,550.8 )$  
    & $(3.0, 316.7)$  
    & $(0.4, 45.6)$  
    & $(1.38, 12.06)$ 
    & $\sim 12.6\ (29.9)$ 
   \\ 
\hline
7  
    & $(70, 317.94)$ 
    & $(29.2, 734.2)$  
    & $(2.4, 720.6)$  
    & $(1.4, 410.7)$  
    & $(0.2, 57.0)$  
    & $(0.59, 12.06)$ 
    & $\sim22.3\ (53)$ 
   \\ 
\hline
\end{tabular}
\end{table}

We summarize our results for the $I=4\sim7$ cases in Table~\ref{tab:summary}.
We show the allowed DM mass regions after imposing the Fermi-LAT $\la\sigma v\ra(W^+W^-)$, H.E.S.S. $\la\sigma v\ra(W^+W^-,\gamma\gamma)$ constraints and the relic density constraint (requiring the contribution to relic density does not excess the upper $1$ $\sigma$ range of the observed dark matter relic density, $\Omega_{\rm obs} h^2$).
Predictions on $\la\sigma v\ra$ for Galactic $\chi\bar\chi\to W^+W^-, ZZ, Z\gamma,\gamma\gamma$ annihilations, $\Omega_\chi h^2$ and $\sigma^{SI}$ are also shown.
 
From Table~\ref{tab: I2}, we see that the allowed mass regions are greater than $\sim 60$ or $\sim 70$ TeV for these cases.  
For $I=5,6,7$, the allowed Dirac DM masses are mostly in the range of $70\, {\rm TeV} $ to few hundreds TeV.
For $I=4,5$ with $m_{\chi} \lesssim 70$ TeV, the allowed values for $\la\sigma v\ra({W^+W^-})$ are less than the corresponding $\la\sigma v\ra({Z^0Z^0})$ and $\la\sigma v\ra({Z^0\gamma})$, due to the destructive interference effect in the former channel. 
In addition to the $W^+W^-$ and $\gamma\gamma$ channels, the $\chi\bar\chi\to 
Z^0Z^0$ and $Z^0\gamma$ annihilations are also useful to search for DM. 
In most cases, the predicted relic density does not saturate the observed relic density. 
We see that $\sigma^{SI}\simeq I^2(I+1)^2\times7\times10^{-49}~{\rm cm}^2$ for $m_\chi$ in the range of few tens to one hundred TeV, 
giving
$\sigma^{SI}\simeq 10^{-46}\sim10^{-45}$ cm$^2$. 
In conclusion, the predicted $\la\sigma (\chi^0\bar\chi^0\to Z^0Z^0, Z^0\gamma, \gamma\gamma)v\ra$ and $\sigma^{SI}$ are sizable and they will be useful to search for DM in astrophysical observation and in direct search in near future.

\section{Conclusion}

In this paper, we work on a Dirac DM model by adding on SM a Dirac fermionic DM multiplet with quantum numbers $I_3=Y=0$ to evade the dangerous tree-level $Z$-exchange diagram in elastic DM-nucleus scattering. Nevertheless, there are loop diagrams contributing to the cross section $\sigma^{SI}$. We consider loop diagram contributions and find that there are some cancellations in $W$-exchange diagrams that makes the model viable. 
We find that the dependence of $\sigma^{SI}$ on $m_{\chi}$ is mild 
for $m_{\chi}\gtrsim$ 1 TeV, and $\sigma^{SI}\simeq I^2(I+1)^2\times7.1\times10^{-49}~{\rm cm}^2$ for $m_\chi$ in the range of few to few tens TeV.
By comparing to PandaX-II and XENON1T constraints on SI cross section $\sigma^{SI}$, we find that the constraints do not give lower bound on DM mass for $I =1$ and $2$. 
For $I=3\sim 5$, the lower bounds on DM mass are sub-TeV, while for $I=6\sim 9$, the lower bounds are as large as few TeV.

For indirect search, since the Galactic DM particles are nonrelativistic, the Sommerfeld enhancement effect in DM annihilation processes should be included.
In this model, the DM can only annihilate into $W^+W^-$ at the tree level, but can annihilate into $ZZ, Z\gamma$ and $\gamma\gamma$ through the Sommerfeld enhancement effect, while fermionic final states are still prohibited. 
Analytic formulas of the Sommerfeld enhancement factors for arbitrary $I$ are obtained using the SU(2) symmetric limit and the Hulth\'en approximation. 
We find that the Galactic DM annihilation cross sections are in general significantly enhanced by the Sommerfeld effect, 
but it can occasionally suppressed by some destructive interference. 

The calculated velocity averaged cross sections $\la \sigma v\ra(W^+W^-,\gamma\gamma)$ are compared with the data from Fermi-LAT and H.E.S.S astrophysical observations. 
We consider the possibility that $\chi$ may not saturate the DM relic density, i.e., $\Omega_{\chi}\leqslant\Omega_{\rm obs}$. 
The observed relic density provides the upper mass bounds for DM particles. 
The Fermi-LAT and H.E.S.S. upper limits on $\la \sigma v\ra(W^+W^-,\gamma\gamma)$ provide lower bounds on $m_{\chi}$, while the relic density provides upper bounds. 
Working together they exclude the $I=1\sim 3$ cases and force the Dirac DM mass to be within the ranges of few tens to few hundreds TeV  for $I \geqslant 4$. 
In the allowed parameter space, the corresponding DM-nucleus cross section $\sigma^{SI}$ is in the range of $10^{-46}\sim10^{-45}$ cm$^2$. 
In conclusion, the predicted $\la\sigma (\chi^0\bar\chi^0\to Z^0Z^0, Z^0\gamma, \gamma\gamma)v\ra$ and $\sigma^{SI}$ are sizable and they will be useful to search for DM in astrophysical observation and in direct search in near future. 




\section*{Acknowledgments}

This research was supported by the Ministry of Science and Technology of R.O.C. under Grant Nos. 105-2811-M-033-007 and in part by 103-2112-M-033-002-MY3 and 106-2112-M-033 -004 -MY3.

\appendix

\section{Loop Integral Functions in the Loop Induced Effective Couplings}

The loop integral function $F^S_h(q,x,y)$ with $x\equiv {M_W^2}/{m_\chi^2}$ and $y\equiv {m_t^2}/{m_\chi^2}$ from the SS interaction of Higgs-exchange are the same for all flavors of quarks since there is no quark propagator in the loop. Hence $F^S_h(q,x,y)=F^S_h(x)$ is given by 
\be
F^S_h(x)&=&
\frac{1}{2}x^{1/2}
-\frac{2+2x-x^2}{2\sqrt{4-x}}\bigg[\tan^{-1}(\frac{\sqrt{x}}{4-x})+\tan^{-1}(\frac{2-x} {\sqrt{(4-x)x}})\bigg]
-\frac{1}{4}x^{3/2}\ln(x).
\en
For the effect couplings of quarks $q=u,d,s,c$, all these quarks is much lighter than $W$-boson mass, we only keep the effective coupling up to the leading order of in $m_q$. Hence all these quarks have the same loop integral functions (independent on $y$) as follows:
\be
\non\\
F^S_{W}(u,x,y)&=&
\frac{1}{4}x^{1/2}
-\frac{2x-x^2}{4\sqrt{4-x}} \bigg[\tan^{-1}(\frac{\sqrt{x}}{4-x})+\tan^{-1}(\frac{2-x} {\sqrt{(4-x)x}})\bigg]
-\frac{1}{8}x^{3/2}\ln(x),
\non\\
F^V_{W_1}(u,x,y)&=&
-x^{1/2}\frac{\sec^{-1}(2\sqrt{1/x})}{\sqrt{(4-x)x}}
+\frac{1}{4}x^{1/2}
-\frac{2x-x^2}{4\sqrt{4-x}}\bigg[\tan^{-1}(\frac{\sqrt{x}}{4-x})
+\tan^{-1}(\frac{2-x} {\sqrt{(4-x)x}})\bigg]
\non\\
&&-\frac{1}{8}x^{3/2}\ln(x),
\non\\
F^V_{W_2}(u,x,y)&=&0,
\non\\
F^A_{W_1}(u,x,y)&=&
\frac{1}{3}x
+\frac{2x-7x^2+2x^3}{6\sqrt{(4-x)x}}\bigg[\tan^{-1}(\frac{\sqrt{x}}{4-x})+\tan^{-1}(\frac{2-x} {\sqrt{(4-x)x}})\bigg]
+(\frac{1}{4}x-\frac{1}{6}x^2)\ln(x),
\non\\
F^A_{W_2}(u,x,y)&=&
-\frac{1}{4}
-\frac{(8-5x+2x^2)\pi
}{8x}\sqrt{\frac{x}{4-x}},
\non\\
&&-(1-x)\sqrt{\frac{4-x}{x}}
\bigg[\tan^{-1}(\sqrt{\frac{4-x}{x}})+\tan^{-1}(\frac{\sqrt{(4-x)x}}{2-x})\bigg]
-\frac{1}{8}(3-x)\ln(x).
\en
For the effective couplings of bottom quark, the top quark is in the loop of the box diagrams. Hence the top quark mass in the loop calculation can not be ignored. With the assumption that $m_\chi\gg m_t$, the corresponding loop integral functions are given by 
\be
F^S_{W}(b,x,y)&=&
+\frac{x^{3/2}}{8(x-y)^{1/2}}
 \bigg\{2xy\ln(x)-2y\ln(x/y)-x^2\ln(x)-y2\ln(y)-2y+2x
 \non\\
 & &+2(x^2-2x-2xy+6y)\sqrt{\frac{x}{4-x}}\big[\tan^{-1}(\sqrt{\frac{x}{4-x}})+ \tan^{-1}(\frac{2-x}{\sqrt{(4-x)x}})\big]
 \non\\
& &-2y\sqrt{y(4-y)}\big[\tan^{-1}(\sqrt{\frac{y}{4-y}})+\tan^{-1}(\frac{2-y}{\sqrt{(4-y)y}})\big]\bigg\},
\non\\
 F^V_{W_1}(b,x,y)&=&
-\frac{x^{3/2}}{2(x-y)^2}\bigg\{\frac{2(x+y-xy)}{\sqrt{x(4-x)}}+\frac{1}{2}(y-x)
\non\\
& & +(\frac{1}{2}x^2-x-xy+3y)\sqrt{\frac{x}{4-x}}
\big[\tan^{-1}(\sqrt{\frac{x}{4-x}})+\tan^{-1}(\frac{2-x}{\sqrt{(4-x)x}})\big]
 \non\\
& & -(1+\frac{y}{2})\sqrt{y(4-y)}\big[\tan^{-1}(\sqrt{\frac{y}{4-y}})+\tan^{-1}(\frac{2-y}{\sqrt{(4-y)y}})\big]
\non\\
& & +y\ln(\frac{x}{y})+(\frac{x^2}{4}-\frac{xy}{2})\ln(x)+\frac{y^2}{4}\ln(y)\bigg\},
\non\\
F^V_{W_2}(b,x,y)&=&0,
\non\\
F^A_{W_1}(b,x,y)&=&
\frac{x}{2(x-y)^2}\bigg\{2(2x^2-3xy+y^2)+\big[3x^3(1+y)-2x^3-6xy\big]\ln(x)+y^2(3-y)\ln(y)
\non\\
& &+2\big[2(x-3y)+x(12y-7x)+x^2(2x-3y)\big]\sqrt{\frac{x}{4-x}}
\non\\
&&\quad
\times
\big[\tan^{-1}(\sqrt{\frac{x}{4-x}})+\tan^{-1}(\frac{2-x}{\sqrt{(4-x)x}})\big]
\non\\
& &+2y(1-y)\sqrt{y(4-y)}\big[\tan^{-1}(\sqrt{\frac{y}{4-y}})+\tan^{-1}(\frac{2-y}{\sqrt{(4-y)y}})\big]\bigg\},
\non\\
F^A_{W_2}(b,x,y)&=&
-\frac{1}{16(x-y)^2}\bigg\{2(2x^2-3xy+y^2)+(6x^2-12xy-2x^3+3x^2y)\ln(x)+y^2(6-y)\ln(y)
\non\\
& &+2\sqrt{x(4-x)}(2x-6y-2x^2+3xy)\big[\tan^{-1}(\sqrt{\frac{x}{4-x}})+\tan^{-1}(\frac{2-x}{\sqrt{(4-x)x}})\big]
\non\\
& &+2y(4-y)\sqrt{y(4-y)}\big[\tan^{-1}(\sqrt{\frac{y}{4-y}})+\tan^{-1}(\frac{2-y}{\sqrt{(4-y)y}})\big]\bigg\}.
\en
For the effective couplings of top quark, the top quark is in the external line of the box diagrams, and it still involves the top quark mass. In this case, we can not have an analytical form  for each loop integral function. With the assumption that $m_\chi\gg m_t$, the corresponding loop integral functions can be numerically calculated
from the following expressions:

\be
F^S_{W}(t,x,y)&=&
2x^{3/2}\int^1_0du\int^{1-u}_0dv\frac{u(1-u-v)}{[(1-u-v)^2+xu-yv-i\epsilon]^2},
 \non\\
F^V_{W_1}(t,x,y)&=&
-x^{3/2}\bigg\{
\int^1_0du\int^{1-u}_0dv
\frac{u}{[(1-u-v)^2+xu-yv-i\epsilon]^2}
\non\\
& & \qquad -\int^1_0du\int^{1-u}_0dv
\frac{u(1-u-v)}{[(1-u-v)^2+xu-yv-i\epsilon]^2}\bigg\},
\non\\
F^V_{W_2}(t,x,y)&=&0,
\non\\
F^A_{W_1}(t,x,y)&=&
\frac{1}{2}x\int^1_0du\int^{1-u}_0dv\frac{u(1-u-v)^2}{[
(1-u-v)^2+xu-yv-i\epsilon]^2},
\non\\
F^A_{W_2}(t,x,y)&=&
-\frac{3}{4}\int^1_0du\int^{1-u}_0dv\frac{u}{(1-u-v)^2+xu-yv-i\epsilon}.
\en
The loop integral functions with $x\equiv {M_W^2}/{m_\chi^2}$ and $y\equiv {m_t^2}/{m_\chi^2}$ derived from the quark twist-2 operator are~\cite{HINT} (see footnote~2) 
\be
 g_{T1}(x)&=&\frac{\sqrt{4-x}}{6}(2+x^2)
\tan^{-1}(\sqrt{\frac{4-x}{x}})+\frac{\sqrt{x}}{12}[1-2x-x(2-x)\ln(x)],
\non\\
 g_{T2}(x)&=&\frac{1}{2\sqrt{4-x}}x(2-4x+x^2)
\tan^{-1}(\sqrt{\frac{4-x}{x}})-\frac{\sqrt{x}}{4}[1-2x-x(2-x)\ln(x)].
\en
The loop integral functions with $x\equiv {M_W^2}/{m_\chi^2}$ and $y\equiv {m_t^2}/{m_\chi^2}$ derived from two-loop diagrams in Fig.~\ref{fig:gluon} are~\cite{HINT} (see footnote~3)
\be
g_H(x)&=&-\frac{4}{\sqrt{4-x}}(2+2x+x^2)\tan^{-1}(\sqrt{\frac{4-x}{x}})+2\sqrt{x}(2-x\ln(x)),
\\
g_W(x,y)&=&2g_{B1}(x)+g_{B3}^{(1)}(x,y)+c_bg_{B3}^{(2)}(x,y),
\en
where
\be
g_{B1}(x)=\frac{1}{24}\sqrt{x}(x\ln(x)-2)
+\frac{(x^2-2x+4)\tan^{-1}(\sqrt{\frac{4-x}{x}})}{12\sqrt{4-x}},
\en
and
\be
g_{B3}^{(1)}(x,y)&=&\frac{-x^{3/2}}{12(y-x)}
+\frac{-x^{3/2}y^2}{24(y-x)^2}\ln(y)
-\frac{-x^{5/2}(x-2y)}{24(y-x)^2}\ln(x)\non\\
& &-\frac{x^{3/2}\sqrt{y}(y+2)\sqrt{4-y}}{12(y-x)^2}
\tan^{-1}(\sqrt{\frac{4-y}{y}})\non\\
& &+\frac{x[x^3-2(y+1)x^2+4(y+1)x+4y]}{12(y-x)^2\sqrt{4-y}}
\tan^{-1}(\sqrt{\frac{4-x}{x}}),\non\\
g_{B3}^{(2)}(x,y)&=&\frac{-x^{3/2}y}{12(y-x)^2}
+\frac{-x^{5/2}y}{24(y-x)^3}\ln(y)
-\frac{-x^{5/2}(x-2y)}{24(y-x)^3}\ln(x)\non\\
& &+\frac{x^{3/2}\sqrt{y}(-6y+xy^2-2xy-2x)}{12(y-x)^3\sqrt{4-y}}
\tan^{-1}(\sqrt{\frac{4-y}{y}})\non\\
& &+\frac{-xy(x^2y-2xy-6x-2y)}{12(y-x)^3\sqrt{4-x}}
\tan^{-1}(\sqrt{\frac{4-x}{x}}).
\en

\section{Sommerfeld enhancement in $\chi\bar\chi$ annihilations}

\subsection{Lippmann-Schwinger equation in $\chi^j\bar\chi^j\to \chi^i\bar\chi^i$ process}

The Feynman diagram of non-perturbative scattering $\chi^j(p_1)\bar\chi^j(p_2)\to \chi^i(p_3)\bar\chi^i(p_4)$ is shown in Fig.~\ref{fig:Som5}.
Note that $p_3$ and $p_4$ are not necessary on-shell as these two lines will be connected to $\chi^i\bar\chi^i$ annihilation diagrams later. 
We will basically following \cite{Landau} but will depart from it and obtain the Lippmann-Schwinger equation at the end.
The amputated non-perturbative 4-point vertex function can be written as
\be
&&\hspace{-0.5cm}
i\Gamma^{ij}_{\gamma\beta,\alpha\delta}(p_3, p_2, p_1, p_4) 
\non\\
&&=i\tilde\Gamma^{ij}_{\gamma\beta,\alpha\delta}(p_3, p_2, p_1, p_4)
\non\\
&&\quad
+\int \frac{d^4p'_3}{(2\pi)^4}
i\tilde\Gamma^{ik}_{\gamma \rho,\sigma\delta}(p_3,p'_4;p'_3,p_4)
(iS_{F \sigma\tau}(p'_3)) 
i\Gamma^{kj}_{\tau\beta,\alpha \lambda}(p'_3,p_2;p_1,p'_4)(-iS_{F \lambda\rho}(-p'_4)),
\label{eq:master0}
\en
where we have $p'_4=-p'_3+p_3+p_4$, $S_{F}$ is the fermion propagator
and the relevant lowest order perturbative 4-point vertex function in the SU(2) symmetric limit is given by~\footnote{Note an additional minus sign from different pairing for particle and antiparticle lines. We will return to this later. Also note that in the SU(2) symmetric limit the $U(1)$ field $B^\mu$ does not contribute to $\tilde\Gamma$ as the hypercharge of $\chi$ is vanishing.}
\be
i\tilde\Gamma^{ij}_{\gamma\beta,\alpha\delta}(p_3, p_2, p_1, p_4)=-ig^2\sum_{a=1,2,3}T^a_{ij}T^a_{ji}(\gamma^\mu)_{\gamma\alpha}(\gamma^\nu)_{\beta\delta}\frac{g_{\mu\nu}}{(p_1-p_3)^2-M^2_W}.
\en 
We make use of the  instantaneous approximation,
where the time component of the momentum transfer is neglected, and obtain 
\be
i\tilde\Gamma^{ij}_{\gamma\beta,\alpha\delta}(p_3, p_2, p_1, p_4)
= -ig^2\sum_{a=1,2,3}T^a_{ij}T^a_{ji}\gamma^\mu_{\gamma\alpha}\gamma_{\mu\,\beta\delta}\frac{-1}{(\vec p_1-\vec p_3)^2+M^2_W}.
\label{eq:vertex}
\en

Now using~\cite{chua}~\footnote{The expression is obtained with the help of $\sum_c T^c_{ij} T^c_{ji}=-\sum_c T^c_{ij} T^c_{-i-j}(-)^{i-j}$ and the standard method of addition of angular momentum.}
\be
\sum_{a=1,2,3}T^a_{ij}T^a_{ji}=\sum_{{\cal I}=0}^{2I}(U^T)_{i{\cal I}}\{I(I+1)-{\cal I}({\cal I}+1)/2\}U_{{\cal I}j},
\label{eq: diagonal}
\en
with
\be
U_{{\cal I}j}=(-1)^j\la I j I(-j)|{\cal I}0\ra,
\en
where $\la I j I(-j)|{\cal I}0\ra$ is the Clebsch-Gordan coefficient (in the $\la j_1 m_1 j_2 m_2|JM\ra$ notation),
we can diagonal $\tilde\Gamma^{ij}$ in the flavor space as
\be
\tilde\Gamma^{ij}_{\gamma\beta,\alpha\delta}(p_3, p_2, p_1, p_4)
=\sum_{{\cal I}=0}^{2I}(U^T)_{i{\cal I}} \tilde \Gamma^{\cal I}_{\gamma\beta,\alpha\delta}(p_3, p_2, p_1, p_4) U_{{\cal I}j},
\en
where, we have
\be
i\tilde\Gamma^{{\cal I}}_{\gamma\beta,\alpha\delta}(p_3, p_2, p_1, p_4)
= -i \gamma^\mu_{\gamma\alpha}\gamma_{\mu\,\beta\delta}V_{\cal I}(\vec q),
\label{eq:vertexI}
\en
with $V_{\cal I}(\vec p_1-\vec p_3)$ the potential given by 
\be
V_{\cal I}(\vec p_1-\vec p_3)\equiv -4\pi\alpha_{\rm w}\{I(I+1)-{\cal I}({\cal I}+1)/2\}\frac{1}{(\vec p_1-\vec p_3)^2+M^2_W},
\label{eq:Uij}
\en
and with $\alpha_{\rm w}$ the electroweak fine-structure constant. 
Note that ${\cal I}$ is the total isospin of the $\chi\bar\chi$ pair.

It can be easily shown (by iteration) that $\Gamma^{ij}$ can be diagonalized in the flavor space similarly,
\be
\Gamma^{ij}=\sum_{{\cal I}=0}^{2I}(U^T)_{i{\cal I}} \Gamma^{\cal I} U_{{\cal I}j}.
\label{eq:UTGammaU}
\en
To proceed we define two auxiliary functions as following:
\be
i\eta^{\cal I}_{\sigma\beta,\alpha\rho}(p_3,p_2;p_1,p_4)
&\equiv&iS_{F \sigma\tau}(p_3) i\Gamma^{\cal I}_{\tau\beta,\alpha\lambda}(p_3,p_2;p_1,p_4)(-iS_{F \lambda\rho}(-p_4)),
\non\\
i\tilde\chi^{\cal I}_{\sigma\beta,\alpha\rho}(p_3,p_2;p_1,p_4)
&\equiv&iS_{F \sigma\tau}(p_3) i\tilde\Gamma^{\cal I}_{\tau\beta,\alpha\lambda}(p_3,p_2;p_1,p_4)(-iS_{F \lambda\rho}(-p_4)),
\label{eq:vertex2}
\en
and the master formula, Eq.~(\ref{eq:master0}), can be expressed as
\be
i\eta^{\cal I}_{\gamma\beta,\alpha\delta}(p_3,p_2;p_1,p_4)
&=&
i\tilde \chi^{\cal I}_{\gamma\beta,\alpha\delta}(p_3,p_2;p_1,p_4)
\non\\
&&+
\int \frac{d^4p'_3}{(2\pi)^4}
i\tilde\chi^{\cal I}_{\gamma\rho,\sigma\delta}(p_3,p^{\prime}_4;p^{\prime}_3,p_4)
i\eta^{\cal I}_{\sigma\beta,\alpha\rho}(p^{\prime}_3,p_2;p_1,p^{\prime}_4),
\label{eq: ieta}
\en
Adding $\delta_{\gamma\alpha}\delta_{\beta\delta} (2\pi)^4\delta^4(p_3-p_1)$
to both side of the above equation,~\footnote{This we depart from~\cite{Landau}.}
and defining 
\be
i\chi^{\cal I}_{\gamma\beta,\alpha\delta}(p_3,p_2;p_1,p_4)&\equiv&
\delta_{\gamma\alpha}\delta_{\beta\delta} (2\pi)^4\delta^4(p_3-p_1)+i\eta^{\cal I}_{\gamma\beta,\alpha\delta}(p_3,p_2;p_1,p_4),
\label{eq: ichidefine1}
\en
now the master formula, Eq. (\ref{eq: ieta}), becomes
\be
i\chi^{\cal I}_{\gamma\beta,\alpha\delta}(p_3,p_2;p_1,p_4)
&=&
\delta_{\gamma\alpha}\delta_{\beta\delta} (2\pi)^4\delta^4(p_3-p_1)
\non\\
&&+
\int \frac{d^4p'_3}{(2\pi)^4}
i\tilde\chi^{\cal I}_{\gamma\rho,\sigma\delta}(p_3,p^{\prime}_4;p^{\prime}_3,p_4)
i\chi^{\cal I}_{\sigma\beta,\alpha\rho}(p^{\prime}_3,p_2;p_1,p^{\prime}_4).
\label{eq: ichi11}
\en

In the NR limit, the fermion and anti-fermion propagators can be approximated as
\be
S_F(\pm k)=\frac{1}{2}(1\pm\gamma^0)\frac{1}{(k_0-m_\chi)-\vec k^2/2m_\chi+i\epsilon}\equiv \frac{1}{2}(1\pm\gamma^0)g(k),
\en
where the projection operators correspond to $u\bar u/2m_\chi$ and $-v\bar v/2m_\chi$ in the NR limit.
Using the above approximation and Eq.~(\ref{eq:vertex2}), we have
\be
i\tilde\chi^{\cal I}_{\gamma\rho,\sigma\delta}(p_3,p^{\prime}_4;p^{\prime}_3,p_4)
\simeq i\frac{1}{2}(1+\gamma_0 ) _{\gamma\gamma'}
{\gamma}^\mu_{\gamma'\sigma}{\gamma}_{\mu\,\rho\delta'}
\frac{1}{2}(\gamma_0-1)_{\delta'\delta} 
        g(p_3)g(p_4)V_{\cal I}(\vec p^{\,\prime}_3-\vec p_3).
\label{eq: tilde chi0}               
\en
On the other hand, $i \chi_{\gamma\beta,\alpha\delta}$ will eventually contract with $u(\vec p_3,s_3)$ and $\bar v(\vec p_4,s_4)$, 
and in the NR limit,
\be
\frac{1}{2}(1+\gamma_0 ) u(\vec p_3,s_3)
&\simeq& \frac{1}{2}(1+\gamma_0 ) u(\vec 0,s_3)
= u(\vec 0,s_3)
\simeq u(\vec p_3,s_3),
\non\\
\bar v(\vec p_4, s_4) \frac{1}{2}(1-\gamma_0 ) 
&\simeq&
\bar v(\vec 0, s_4) \frac{1}{2}(1-\gamma_0 ) 
=\bar v(\vec 0,s_4)
\simeq \bar v(\vec p_4,s_4),
\en
we should take
\be
\delta_{\gamma\alpha}\delta_{\beta\delta}\to \frac{1}{2}(1+\gamma_0)_{\gamma\alpha}\frac{1}{2}(1-\gamma_0 )_{\beta\delta}.
\en
Hence Eq.(\ref{eq: ichi11}) becomes 
\be
i\chi^{\cal I}_{\gamma\beta,\alpha\delta}(p_3, p_2 ;p_1,p_4)
&\simeq&
\frac{1}{2}(1+\gamma_0)_{\gamma\alpha}\frac{1}{2}(1-\gamma_0 )_{\beta\delta} (2\pi)^4\delta^4(p_3-p_1)
\non\\
&&+
g(p_3)g(p_4)\int \frac{d^4p'_3}{(2\pi)^4}\bigg\{
i\frac{1}{2}(1+\gamma_0 ) _{\gamma\gamma'}
{\gamma}^\mu_{\gamma'\sigma}{\gamma}_{\mu\,\rho\delta'}
\frac{1}{2}(\gamma_0-1)_{\delta'\delta} 
\non\\
&&
\qquad
i\chi_{\sigma\beta,\alpha\rho}(p_3,p^{\prime}_4;p^{\prime}_3,p_4) V_{\cal I}(\vec p'_3-\vec p_3)\bigg\}.
\label{eq: ichi21}
\en
The solution of the above equation, $i \chi^{\cal I}_{\gamma\beta,\alpha\delta}$, can be written as
\be
i\chi^{\cal I}_{\gamma\beta,\alpha\delta}(p_3, p_2 ;p_1,p_4)
=\frac{1}{2}(1+\gamma_0)_{\gamma\alpha}\frac{1}{2}(1-\gamma_0 )_{\beta\delta}
 i\chi_{\cal I}(p_3, p_2 ;p_1,p_4),
\label{eq: ichi form1}
\en
Substitute it into the above equation, we obtain the equation for $\chi_{\cal I}$, 
\be
i\chi_{\cal I}(p_3, p_2 ;p_1,p_4)
=(2\pi)^4\delta^4(p'_3-p_3)
-g(p_3)g(p_4)\int \frac{d^4p'_3}{(2\pi)^4}
V_{\cal I}(\vec p'_3-\vec p_3)
\chi_{\cal I}(p'_3, p_2 ;p_1,p'_4).
\label{eq: ichi31}
\en
The above equation can be readily brought into the Lippmann-Schwinger equation. 

Define
\be
p^{(\prime)}\equiv\frac{1}{2}(p_{1(3)}-p_{2(4)})=(\varepsilon^{(\prime)},\vec p{}^{(\prime)}),
\quad
P\equiv p_1+p_2=p^{(\prime)}_3+p^{(\prime)}_4=(E+2m,\vec 0),
\en
and redefine
\be
i\chi_{\cal I}(p',P,p)=i\chi_{\cal I}\left(\frac{1}{2}(p_3-p_4),p_3+p_4,\frac{1}{2}(p_1-p_2)\right)
\equiv i\chi_{\cal I}(p_3,p_2 ;p_1,p_4).
\en
We now have for Eq.~(\ref{eq: ichi31})
\be
i\chi_{\cal I}(p',P,p)
&=&(2\pi)^4\delta^4(p'-p)
\non\\
&&
-g\left(\frac{1}{2}P+p'\right)g\left(\frac{1}{2}P-p'\right)
\int \frac{d^3q'}{(2\pi)^3}
V_{\cal I}(\vec q^{\,\prime}-\vec p^{\,\prime})
\int \frac{d q'_0}{2\pi}
\chi_{\cal I}\left(q',P,p\right),
\label{eq: chi=...}
\en
where we use
$
q'\equiv p'_3-P/2$
and
$p'_3-p_3
=q'-p',
$
and note that the integration of $q'_0$ only applies to $\chi_{\cal I}$.

There are further simplification by using the center of  mass frame:
\be
\vec P=\vec 0,
\quad
\vec p_{1(3)}=-\vec p_{2(4)}=\vec p^{\,(\prime)},
\quad 
E=P^0-2m_\chi=\frac{|\vec p|^2}{m_\chi},
\quad
p^0=\epsilon=0.
\en
It is useful to recall that $p_{1,2}$ are on-shell, but $p_{3,4}$ are not necessarily so as noted in the beginning.
Therefore, at the center of mass frame, $\epsilon=(p^0_1-p^0_2)/2=[(|\vec p|^2/2m_\chi)-(|\vec p|^2/2m_\chi)]/2=0$, 
but $\epsilon'=(p^{0}_3-p^{0}_4)/2$ and $\vec p^{\,\prime}$ are not fixed as $p^0_{3,4}$ are not necessarily equal to  $\vec p^{\,\prime2}/2m$.

Define
\be
\psi_{\cal I}(\vec q,\vec p)\equiv\frac{1}{2\pi}\int_{-\infty}^\infty dq_0 i \chi_{\cal I}(q,P,p) ,
\en
and integrate both side of Eq. (\ref{eq: chi=...}) with respect to $p^{\prime 0}=\epsilon'$,
we obtain
\be
\psi_{\cal I}(\vec p^{\,\prime},\vec p)
&=&(2\pi)^3\delta^3(\vec p^{\,\prime}-\vec p)
\non\\
&&
-\frac{1}{2\pi i}\int_{-\infty}^\infty d\epsilon' 
g\left(\frac{1}{2}P+p'\right)g\left(\frac{1}{2}P-p'\right)
\int \frac{d^3q}{(2\pi)^3}
V_{\cal I}(\vec q-\vec p^{\,\prime})
\psi_{\cal I}\left(\vec q,\vec p\right),
\en
where we redefine the integration variable $\vec q^{\,\prime}$ to $\vec q$ in the last line.
After integrating over $\epsilon'$,~\footnote{Note that 
$g\left(\frac{1}{2}P\pm p'\right)
=1/({\pm \varepsilon'+\frac{1}{2} E-(\vec p^{\,\prime 2}/2m_\chi)+i\epsilon})$.}
we finally obtain
\be
\psi_{\cal I}(\vec p{\,}',\vec p)
&=&(2\pi)^3\delta^3(\vec p{\,}'-\vec p)
+\frac{1}{E-(\vec p{\,}'{}^2/2\mu)+i\epsilon}
\int \frac{d^3q}{(2\pi)^3}
V_{\cal I}(\vec q-\vec p^{\,\prime})
\psi_{\cal I}\left(\vec q,\vec p\right),
\label{eq: LSp}
\en
where $\mu=m_\chi/2$ is the reduced mass of the $\chi\bar\chi$ system.
The above equation is simply the Lippman-Schwinger equation in the momentum space representation.

Furthermore, using
\be
V_{\cal I}(\vec r)=\int \frac{d^3 q}{(2\pi)^3} e^{-i \vec q\cdot \vec r}V_{\cal I}(\vec q),
\quad
\psi_{\cal I}(\vec r)
=
\int \frac{d^3 \vec p\,'}{(2\pi)^3} 
e^{i\vec p\,'\cdot \vec r}
\psi_{\cal I}(\vec p{\,}',\vec p),
\en
the Lippmann-Schwinger equation can be written in the position space as
\be
\psi_{\cal I}(\vec r)
&=&e^{i \vec p\cdot \vec r}
+\int d^3 r'\la \vec r|\frac{1}{E-H_0+i\epsilon}|\vec r\,'\ra
V_{\cal I}(\vec r\,')\psi_{\cal I}(\vec r\,').
\label{eq: LSr}
\en
The wave function satisfies the Schr\"odinger equation,
\be
-\frac{1}{2\mu} \nabla^2\psi_{\cal I}(\vec r)+V_{\cal I}(\vec r)\psi_{\cal I}(\vec r)=E\psi_{\cal I}(\vec r)=\frac{1}{2} \mu v^2\psi_{\cal I}(\vec r),
\label{eq: Schrodinger eq}
\en
with
\be
V_{\cal I}(r)=-\alpha_{\rm w}\{I(I+1)-{\cal I}({\cal I}+1)/2\}\frac{e^{-M_Wr}}{r},
\en
$E=|\vec p|^2/2\mu\equiv \mu v^2/2$ and the boundary condition that the wave function goes to $e^{i \vec p\cdot \vec r}$ asymptotically.

\subsection{Sommerfeld enhancement}

For a $\chi^j\bar\chi^j$ annihilation process, there may be interaction among initial state as depicted in Fig.~\ref{fig:Som4} going into $\chi^i\bar \chi^i$, before the main annihilation process takes place. The  $\chi^j\bar\chi^j$ annihilation amplitude $iM^{jS}$ can be expressed as 
\be
iM^{jS}_{\beta,\alpha}(p_3, p_2, p_1, p_4) \bar v_\beta(p_2) u_\alpha (p_1)
&=&iM^j_{\beta,\alpha}(p_3, p_2, p_1, p_4) \bar v_\beta(p_2) u_\alpha (p_1)
\non\\
&&+\int \frac{d^4p'_3}{(2\pi)^4}
iM^i_{\rho,\sigma}(p_3,p'_4;p'_3,p_4)
\non\\
&& (iS_{F \sigma\tau}(p'_3))(-iS_{F \lambda\rho}(-p'_4)) 
\non\\
&&
i\Gamma^{ij}_{\tau\beta,\alpha\lambda}(p'_3,p_2;p_1,p'_4) \bar v_\beta(p_2) u_\alpha(p_1).
\en
In the above equation, $iM^i$ represents the main $\chi^i\bar\chi^i$ annihilation, while $i\Gamma^{ij}$ contains an infinite series of the ladder diagrams (see Fig.~\ref{fig:Som5})
representing the initial state interaction.
With the help of Eqs. (\ref{eq:UTGammaU}) and (\ref{eq: ichidefine1}), we have
\be
iM^{jS}_{\beta,\alpha}(p_3, p_2, p_1, p_4) \bar v_\beta(p_2) u_\alpha (p_1)&=&\int \frac{d^4 p'_3}{(2\pi)^4}
iM^i_{\rho,\sigma}(p_3,p'_4;p'_3,p_4)
\non\\
&&
\sum_{{\cal I}=0}^{2I}
(U^T)_{i{\cal I}}
i\chi^{\cal I}_{\sigma\beta,\alpha\rho}(p'_3,p_2;p_1,p'_4) 
U_{{\cal I}j}\bar v_\beta(p_2) u_\alpha(p_1),
\en
where $p'_4=p_1+p_2-p'_3$ and $p'_{3,4}$ are not necessarily on-shell.

As shown previously, in the NR limit, we have
\be
i\chi^{\cal I}_{\sigma\beta,\alpha\rho}(p'_3,p_2;p_1,p'_4) \bar v_\beta(p_2) u_\alpha(p_1)
&\simeq&\frac{1}{2}(1+\gamma_0)_{\sigma\alpha}\frac{1}{2}(1-\gamma_0 )_{\beta\rho} i\chi_{\cal I}(p'_3, p_2 ;p_1,p'_4)\bar v_\beta(p_2) u_\alpha(p_1)
\non\\
&\simeq& i\chi_{\cal I}(p'_3, p_2 ;p_1,p'_4)\bar v_\rho(p_2) u_\sigma(p_1).
\en
Furthermore, we assume the energy dependent of $iM$ in the integral can be neglected, which is called the instantaneous approximation. Therefore, the integration on $E'_3$ only applies to $\chi(p'_3,p_2;p_1,p'_4)$,
and the above equation becomes
\be
iM^{jS}_{\beta,\alpha}(p_3, p_2, p_1, p_4) \bar v_\beta(p_2) u_\alpha (p_1)
&=&\int \frac{d^3 \vec p^{\,\prime}_3}{(2\pi)^3}
iM_{\rho,\sigma}(\vec p_3,\vec p'_4;\vec p'_3,\vec p_4) \bar v_\rho(p_2) u_\sigma(p_1)
\non\\
&&\times
\sum_{{\cal I}=0}^{2I}
 (U^T)_{iK}\psi_{\cal I}\left(\frac{\vec p^{\,\prime}_3-\vec p^{\,\prime}_4}{2},\frac{\vec p_1-\vec p_2}{2}\right)U_{{\cal I}j},
\label{eq: Sommerfeld S}
\en
where we made use of 
\be
\int \frac{dE'_3}{2\pi} i\chi_{\cal I}(p'_3,p_2;p_1,p'_4)
&=&\int \frac{d(E'_3-E'_4)/2}{2\pi} i\chi_{\cal I}(p'_3,p_2;p_1,p'_4)
\non\\
&=&\psi_{\cal I}\left(\frac{\vec p^{\,\prime}_3-\vec p^{\,\prime}_4}{2},\frac{\vec p_1-\vec p_2}{2}\right)
\en
with $\psi_{\cal I}(\vec p\,',\vec p)$ the solution of the Lippmann-Schwinger equation, Eq. (\ref{eq: LSp}).
Note that in the $V_{\cal I}=0$ case, $\psi_{\cal I}(\vec p{\,}',\vec p)$ reduces to $(2\pi)^3\delta^3(\vec p{\,}'-\vec p)$ 
and Eq. (\ref{eq: Sommerfeld S}) is just a trivial identity equation.

In the case of $S$-wave rescattering, $M_{\rho,\sigma}(\vec p_3,\vec p'_4;\vec p'_3,\vec p_4)$ is independent of the momentum,
therefore, we can move the $iM^i$ out from the integration
\be
iM^{jS}_{\beta,\alpha}(p_3, p_2, p_1, p_4) \bar v_\beta(p_2) u_\alpha (p_1)
&=&
iM^i_{\gamma\rho,\sigma\delta}(\vec p_3,\vec p_2;\vec p_1,\vec p_4) \bar v_\rho(p_2) u_\sigma(p_1)
\non\\
&&\times
\int \frac{d^3 (\vec p^{\,\prime}_3-\vec p^{\,\prime}_4)/2}{(2\pi)^3}
\sum_{{\cal I}=0}^{2I}
(U^T)_{i{\cal I}}
\psi_{\cal I}\left(\frac{\vec p^{\,\prime}_3-\vec p^{\,\prime}_4}{2},\frac{\vec p_1-\vec p_2}{2}\right)
U_{{\cal I}j}
\non\\
\en
and obtain
\be
iM^{jS}_{\beta,\alpha}(p_3, p_2, p_1, p_4) \bar v_\beta(p_2) u_\alpha (p_1)&=&
iM^i_{\rho,\sigma}(\vec p_3,\vec p_2;\vec p_1,\vec p_4) \bar v_\rho(p_2) u_\sigma(p_1)
{\cal Q}_{ij}
\label{eq: s wave scattering} 
\en
with
\be
{\cal Q}_{ij}\equiv\sum_{{\cal I}=0}^{2I} (U^T)_{i{\cal I}}\psi_{\cal I}(\vec r=0)U_{{\cal I}j},
\en
where $\psi_{\cal I}(\vec r)$ is normalized to give $e^{i \vec p\cdot \vec r}$ asymptotically.
Finally, we have
\be
|\bar v M^{jS}(p_3, p_2, p_1, p_4) u|^2
&=&
\sum_{i,i'}
(\bar v M^{i'}(\vec p_3,\vec p_2;\vec p_1,\vec p_4)u)^*
(\bar v M^i_{\rho,\sigma}(\vec p_3,\vec p_2;\vec p_1,\vec p_4) u)
{\cal Q}_{ij}{\cal Q}^\dagger_{ji'},
\label{eq: Sommerfeld S-wave}
\en
where the additional term (${\cal Q}_{ij}{\cal Q}^\dagger_{ji'}$) is the Sommerfeld enhancement factors, which cannot be factorized in this work.
In the $V_{\cal I}=0$ limit, we have $\psi_{\cal I}(\vec r=0)=1$ giving ${\cal Q}_{ij}{\cal Q}^\dagger_{ji'}=\delta_{ij}\delta_{ji'}$ and the above equation is just a trivial identity.

\subsection{Hulth\'{e}n approximation}

Following~\cite{Cassel}, we use the following Hulth\'{e}n potential,
\be
V(\vec r)=-\alpha_{\cal I} \frac{(\pi^2 m_W/6) e^{-\pi^2 m_W r/6}}{1-e^{-\pi^2 m_W r/6}},
\en
to approximate the Yukawa potential. 
The parameters are chosen to match the wave functions in the Yukawa potential and the above potential in the Born approximation for $p, r\to 0$ ~\cite{Cassel}. 
The solution of the Schr\"odinger equation
with $\psi_{lm}=R_l Y_{lm}$ is found to be ~\cite{Cassel} 
\be
R_l=i^{l+1}\frac{t^{l+1}}{2pr} e^{-i pr}\frac{\Gamma(\lambda-a^-)\Gamma(\lambda-a^+)}{\Gamma(\lambda)\Gamma(\lambda-a^+-a^-)}
\,
{}_2F_1(a^-,a^+;\lambda;t),
\label{eq: Rl}
\en
with $p=\mu_\chi v=m_\chi v/2$,
\be
t=1-e^{-pr/\omega},
\quad
\omega=\frac{3v}{\pi^2}\frac{m_\chi}{m_W},
\quad
\lambda=2+2l,
\quad
a^{\pm}=1+l+i\omega(1\pm \sqrt{1-2\alpha_{\cal I}/(v\omega)}).
\en
Note that the phase of $R_l$ is different from \cite{Cassel}. 
It is determined from the asymptotic behavior of $R_l$ as shown below.

For a large $r$, the ${}_2F_1(a^-,a^+;\lambda;t)$ behaves asymptotically as \cite{Cassel, Lebedev}
\be
\frac{\Gamma(\lambda)\Gamma(\lambda-a^+-a^-)}{\Gamma(\lambda-a^-)\Gamma(\lambda-a^+)}
+e^{i2pr} \frac{\Gamma(\lambda)\Gamma(a^++a^--\lambda)}{\Gamma(a^+)\Gamma(a^-)},
\en
which is, in fact, equal to
\be
\frac{\Gamma(\lambda)\Gamma(\lambda-a^+-a^-)}{\Gamma(\lambda-a^-)\Gamma(\lambda-a^+)}
 +e^{i2pr} \left(\frac{\Gamma(\lambda)\Gamma(\lambda-a^+-a^-)}{\Gamma(\lambda-a^-)\Gamma(\lambda-a^+)}\right)^*.
\en
As usual we consider an incident wave along the positive $z$-direction, the radial wave function is related to $\psi(\vec r)$ as (see, for example~\cite{Sakurai})
\be
\psi(\vec r)=\sum_l i^l (2l+1) R_l(r) P_l(\cos\theta). 
\label{eq: psiRl}
\en
In the large $r$ limit, the radial wave function $R_l(r)$ for an incident plane wave with a out-going spherical wave can be expressed as,~\cite{Sakurai}  
\be
R_l(r)\to (-i)^l\,\frac{{e^{i(pr+2\delta_l)}}-{e^{-i (pr-l\pi)}}}{2ipr},
\label{eq: Rllimit}
\en
which reduces to the asymptotic form of $j_0(pr)\simeq\sin(pr-l\pi/2)/pr$ in the plane wave case when the phase shift $\delta_l$ is switched off. 
The normalization of $R_l$ as shown in Eq. (\ref{eq: Rl}) is determined accordingly with the phase shift given by
\be
\delta_l=\arg\left(\frac{\Gamma(\lambda-a^-)\Gamma(\lambda-a^+)}{\Gamma(\lambda)\Gamma(\lambda-a^+-a^-)}\right)+(l+1)\frac{\pi}{2}.
\en
Note that from Eq. (\ref{eq: psiRl}), the wave function at $\vec r=0$, i.e. $\psi(\vec r =0)$, can be obtained from
\be
\psi(\vec r =0)=\lim_{r\to 0} \int (d\Omega/4\pi)\psi_{\cal I}(\vec r)=R_{l=0}(r=0).
\label{eq: psi0Rl0}
\en
Using the above equation and the explicit form of $R_0(r)$ in Eq.~(\ref{eq: Rl}), we finally obtain Eq.~(\ref{eq: psi0Hulthen}).
Note that the phase shift $\delta_0$ is just the phase of $\psi(\vec r=0)$. This is a general result (see below).

\subsection{Solving $\psi(\vec r=0)$ numerically}

We follow \cite{Iengo:2009ni} to solve for $\psi(\vec r=0)$ numerically. 
We will pay attention to the phase of $\psi_(\vec r=0)$ as well as its size, as the interference between $\psi_{{\cal I}=0}(0)$ and $\psi_{{\cal I}=2}(0)$ is important in this work [see Eqs. (\ref{eq: SWW}) and (\ref{eq: SV0V0})].
Defining $\Phi_l(\rho)=N \rho^l R_l(r)$, where $\rho=p r$, where the normalization $N$ is to be determined later. 
The function $\Phi_l(\rho)$ satisfies the Schr\"{o}dinger equation as:
\be
\Phi^{\prime\prime}_l+\frac{2(l+1)}{\rho}\Phi'_l+(-2\frac{1}{p v}V(r)+1)\Phi_l=0,
\en
where the initial conditions are taken to be~\cite{Iengo:2009ni}
\be
\Phi_l(0)=1, 
\quad
\Phi'_l(0)=\frac{\rho V(r)}{pv (l+1)}\bigg|_{\rho\to {0}}\Phi_l(0),
\en
for a regular solution.
We now concentrate on the $l=0$ case. 
As one can see by taking $\rho\gg 1$, 
in the case that $|\rho V(r)|\ll1$,
the differential equation and its solution become
\be
\Phi^{\prime\prime}_0+\frac{2}{\rho}\Phi'_0+\Phi_0\bigg|_{\rho\gg 1}=0,
\qquad
\Phi_0(\rho)\to C\frac{\sin(\rho+\delta_0)}{\rho},
\en
with $C$ a real number. 
The above $\Phi_0$ is to be compared to $R_0(r)\to e^{i\delta_0}\sin(\rho+\delta_0)/\rho$
[see Eq. (\ref{eq: Rllimit})], as $\rho\gg 1$.
To work out the normalization $N$ it is useful noting, in the $\rho\gg 1$ region,
\be
\Phi_0(\rho-\pi/2)\to -C\frac{\cos(\rho+\delta_0)}{\rho-\pi/2},
\en
which can be used with $\Phi_0(\rho)$ to construct 
\be
\kappa\equiv\lim_{\rho\to\infty}e^{i\rho}\left[{-i\rho\Phi_0(\rho)-(\rho-\pi/2)\Phi_0(\rho-\pi/2)}\right]
=C e^{-i\delta_0}.
\label{eq: kappa}
\en
Consequently, we see that $R_0(r)$ can be obtained as
\be
R_0(r)=\kappa^{-1}\Phi_0(\rho),
\en 
since it satisfies the Schr\"{o}dinger equation and has the correct asymptotic behavior.
Finally, we have
\be
\psi(\vec r=0)=\kappa^{-1}\Phi_0(0)=\kappa^{-1}=\lim_{\rho\to\infty}\frac{e^{-i\rho}}{-i\rho\Phi_0(\rho)-(\rho-\pi/2)\Phi_0(\rho-\pi/2)}.
\label{eq: psikappa}
\en
Note that the phase of $\psi(\vec r=0)$ is just $\delta_0$ [see Eq. (\ref{eq: kappa})].

The differential equation for the case of the Yukawa potential 
can be solved numerically and we find that it is enough to take $\rho\simeq 200$ to obtain the limit in Eq. (\ref{eq: psikappa}). 
The Sommerfeld factors, $|\psi_{{\cal I}}(\vec r=0)|^2$ and $\delta^{{\cal I}=0}_0-\delta_0^{{\cal I}=2}$ using the above $\psi(\vec r=0)$ are shown in Fig.~\ref{fig:SWWSV0V0} and Fig.~\ref{fig:Deltadelta0}. They are compared with those obtained in the Hulth\'en potential. Both results agree with each other reasonably well.


\begin{thebibliography}{99}

\bibitem{RF}
  V.~C.~Rubin and W.~K.~Ford, Jr.,
  ``Rotation of the Andromeda Nebula from a Spectroscopic Survey of Emission Regions,''
  Astrophys.\ J.\  {\bf 159}, 379 (1970).

\bibitem{BBSnBxxx}
 K.~G.~Begeman, A.~H.~Broeils and R.~H.~Sanders,
  ``Extended rotation curves of spiral galaxies: Dark haloes and modified dynamics,''
  Mon.\ Not.\ Roy.\ Astron.\ Soc.\  {\bf 249}, 523 (1991).
  
 \bibitem{PDG1}
 J.~Beringer {\it et al.} [Particle Data Group Collaboration],
  ``Review of Particle Physics (RPP),''
  Phys.\ Rev.\ D {\bf 86}, 010001 (2012).

 \bibitem{Carroll}
   S.~M.~Carroll,
  ``Dark matter is real,''
  Nature Phys.\  {\bf 21}, 653 (2006).

 \bibitem{Cxxx}
  D.~Clowe, M.~Bradac, A.~H.~Gonzalez, M.~Markevitch, S.~W.~Randall, C.~Jones and D.~Zaritsky,
  ``A direct empirical proof of the existence of dark matter,''
  Astrophys.\ J.\  {\bf 648}, L109 (2006)
  [astro-ph/0608407].
  
 \bibitem{WMAPa}
    D.~N.~Spergel {\it et al.} [WMAP Collaboration],
  ``First year Wilkinson Microwave Anisotropy Probe (WMAP) observations: Determination of cosmological parameters,''
  Astrophys.\ J.\ Suppl.\  {\bf 148}, 175 (2003)
  [astro-ph/0302209].

 \bibitem{SDSS}
   M.~Tegmark {\it et al.} [SDSS Collaboration],
  ``Cosmological parameters from SDSS and WMAP,''
  Phys.\ Rev.\ D {\bf 69}, 103501 (2004)
  [astro-ph/0310723].
  
\bibitem{JKG} 
   G.~Jungman, M.~Kamionkowski and K.~Griest,
  ``Supersymmetric dark matter,''
  Phys.\ Rept.\  {\bf 267}, 195 (1996)
  [hep-ph/9506380].
  
\bibitem{PDG} 
  C.~Patrignani {\it et al.} [Particle Data Group],
  Chin.\ Phys.\ C {\bf 40}, no. 10, 100001 (2016) and 2017 update; http://pdg.lbl.gov.
  
\bibitem{DG}
  M.~Drees and G.~Gerbier,
  ``Mini-Review of Dark Matter: 2012,''
  arXiv:1204.2373 [hep-ph];
  C.~Patrignani {\it et al.} [Particle Data Group],
  ``Review of Particle Physics,''
  Chin.\ Phys.\ C {\bf 40}, no. 10, 100001 (2016).

\bibitem{Arcadi:2017kky} 
  G.~Arcadi, M.~Dutra, P.~Ghosh, M.~Lindner, Y.~Mambrini, M.~Pierre, S.~Profumo and F.~S.~Queiroz,
  ``The Waning of the WIMP? A Review of Models, Searches, and Constraints,''
  arXiv:1703.07364 [hep-ph].

\bibitem{LHC-0} 
  F.~Kahlhoefer,
  ``Review of LHC Dark Matter Searches,''
  Int.\ J.\ Mod.\ Phys.\ A {\bf 32}, no. 13, 1730006 (2017)
  [arXiv:1702.02430 [hep-ph]].
  
\bibitem{LHC-1} 
  S.~P.~Liew, M.~Papucci, A.~Vichi and K.~M.~Zurek,
  ``Mono-X Versus Direct Searches: Simplified Models for Dark Matter at the LHC,''
  JHEP {\bf 1706}, 082 (2017)
  [arXiv:1612.00219 [hep-ph]].
  
  \bibitem{LHC-2} 
  A.~Alves and K.~Sinha,
  ``Searches for Dark Matter at the LHC: A Multivariate Analysis in the Mono-$Z$ Channel,''
  Phys.\ Rev.\ D {\bf 92}, no. 11, 115013 (2015)
  [arXiv:1507.08294 [hep-ph]].
  
  \bibitem{LHC-2.5} 
  P.~Athron {\it et al.} [GAMBIT Collaboration],
  ``Status of the scalar singlet dark matter model,''
  arXiv:1705.07931 [hep-ph].
  
  \bibitem{LHC-3} 
  C.~H.~Chen and T.~Nomura,
  ``Searching for vector dark matter via Higgs portal at the LHC,''
  Phys.\ Rev.\ D {\bf 93}, no. 7, 074019 (2016)
  doi:10.1103/PhysRevD.93.074019
  [arXiv:1507.00886 [hep-ph]].
  
  \bibitem{LHC-4} 
  J.~Abdallah {\it et al.},
  ``Simplified Models for Dark Matter Searches at the LHC,''
  Phys.\ Dark Univ.\  {\bf 9-10}, 8 (2015)
  [arXiv:1506.03116 [hep-ph]].
  

\bibitem{Mitsou}
  V.~A.~Mitsou,
  ``Shedding Light on Dark Matter at Colliders,''
  Int.\ J.\ Mod.\ Phys.\ A {\bf 28}, 1330052 (2013)
  [arXiv:1310.1072 [hep-ex]].

\bibitem{GIST}
  J.~Goodman, M.~Ibe, A.~Rajaraman, W.~Shepherd, T.~M.~P.~Tait and H.~B.~Yu,
  ``Constraints on Dark Matter from Colliders,''
  Phys.\ Rev.\ D {\bf 82}, 116010 (2010)
  [arXiv:1008.1783 [hep-ph]].

\bibitem{FHKT}
  P.~J.~Fox, R.~Harnik, J.~Kopp and Y.~Tsai,
  ``Missing Energy Signatures of Dark Matter at the LHC,''
  Phys.\ Rev.\ D {\bf 85}, 056011 (2012)
  [arXiv:1109.4398 [hep-ph]].
  
 \bibitem{PandaX-II} 
  X.~Cui {\it et al.} [PandaX-II Collaboration],
  ``Dark Matter Results From 54-Ton-Day Exposure of PandaX-II Experiment,''
  arXiv:1708.06917 [astro-ph.CO].
 

 
 \bibitem{Xenon1T2017-SI} 
  E.~Aprile {\it et al.} [XENON Collaboration],
  ``First Dark Matter Search Results from the XENON1T Experiment,''
  arXiv:1705.06655 [astro-ph.CO].
  
  \bibitem{LUX2016-SI} 
  D.~S.~Akerib {\it et al.} [LUX Collaboration],
  ``Results from a search for dark matter in the complete LUX exposure,''
  Phys.\ Rev.\ Lett.\  {\bf 118}, no. 2, 021303 (2017)
  [arXiv:1608.07648 [astro-ph.CO]].
  
  \bibitem{LUX2016-SD} 
  D.~S.~Akerib {\it et al.} [LUX Collaboration],
  ``Results on the Spin-Dependent Scattering of Weakly Interacting Massive Particles on Nucleons from the Run 3 Data of the LUX Experiment,''
  Phys.\ Rev.\ Lett.\  {\bf 116}, no. 16, 161302 (2016)
  [arXiv:1602.03489 [hep-ex]].
  
\bibitem{HESS2016} 
  H.~Abdallah {\it et al.} [H.E.S.S. Collaboration],
  ``Search for dark matter annihilations towards the inner Galactic halo from 10 years of observations with H.E.S.S,''
  Phys.\ Rev.\ Lett.\  {\bf 117}, no. 11, 111301 (2016)
  [arXiv:1607.08142 [astro-ph.HE]].
  
\bibitem{HESS2017rr} 
  L.~Rinchiuso {\it et al.} [HESS Collaboration],
  PoS ICRC {\bf 2017}, 893 (2017)
  [arXiv:1708.08358 [astro-ph.HE]].

\bibitem{Fermi-LAT2015} 
  M.~Ackermann {\it et al.} [Fermi-LAT Collaboration],
  ``Searching for Dark Matter Annihilation from Milky Way Dwarf Spheroidal Galaxies with Six Years of Fermi Large Area Telescope Data,''
  Phys.\ Rev.\ Lett.\  {\bf 115}, no. 23, 231301 (2015)
  [arXiv:1503.02641 [astro-ph.HE]].

  
 \bibitem{WMAPb}
   G.~Hinshaw {\it et al.} [WMAP Collaboration],
  ``Nine-Year Wilkinson Microwave Anisotropy Probe (WMAP) Observations: Cosmological Parameter Results,''
  Astrophys.\ J.\ Suppl.\  {\bf 208}, 19 (2013)
  [arXiv:1212.5226 [astro-ph.CO]].

 \bibitem{Plank}
   P.~A.~R.~Ade {\it et al.} [Planck Collaboration],
  ``Planck 2013 results. XVI. Cosmological parameters,''
  Astron.\ Astrophys.\  {\bf 571}, A16 (2014)
  [arXiv:1303.5076 [astro-ph.CO]].
  
\bibitem{BPV} 
  C.~P.~Burgess, M.~Pospelov and T.~ter Veldhuis,
  ``The Minimal model of nonbaryonic dark matter: A Singlet scalar,''
  Nucl.\ Phys.\ B {\bf 619}, 709 (2001)
  [hep-ph/0011335].
  
  \bibitem{PW}
  B.~Patt and F.~Wilczek,
  ``Higgs-field portal into hidden sectors,''
  hep-ph/0605188.

  \bibitem{He1}
  X.~G.~He, T.~Li, X.~Q.~Li, J.~Tandean and H.~C.~Tsai,
  ``The Simplest Dark-Matter Model, CDMS II Results, and Higgs Detection at LHC,''
  Phys.\ Lett.\ B {\bf 688}, 332 (2010)
  [arXiv:0912.4722 [hep-ph]].
  
 \bibitem{BBEG}
  Y.~Bai, V.~Barger, L.~L.~Everett and G.~Shaughnessy,
  ``Two-Higgs-doublet-portal dark-matter model: LHC data and Fermi-LAT 135 GeV line,''
  Phys.\ Rev.\ D {\bf 88}, 015008 (2013)
  [arXiv:1212.5604 [hep-ph]].
  
\bibitem{He2}
  X.~G.~He and J.~Tandean,
  ``Low-Mass Dark-Matter Hint from CDMS II, Higgs Boson at the LHC, and Darkon Models,''
  Phys.\ Rev.\ D {\bf 88}, 013020 (2013)
  [arXiv:1304.6058 [hep-ph]].
  
\bibitem{DS}
  E.~M.~Dolle and S.~Su,
  ``Dark Matter in the Left Right Twin Higgs Model,''
  Phys.\ Rev.\ D {\bf 77}, 075013 (2008)
  [arXiv:0712.1234 [hep-ph]].

\bibitem{GWZ}
  W.~L.~Guo, Y.~L.~Wu and Y.~F.~Zhou,
  ``Exploration of decaying dark matter in a left-right symmetric model,''
  Phys.\ Rev.\ D {\bf 81}, 075014 (2010)
  [arXiv:1001.0307 [hep-ph]].

   
\bibitem{Haber}
   H.~E.~Haber and G.~L.~Kane,
  ``The Search for Supersymmetry: Probing Physics Beyond the Standard Model,''
  Phys.\ Rept.\  {\bf 117}, 75 (1985).
   
\bibitem{RSW}
  L.~Roszkowski, E.~M.~Sessolo and A.~J.~Williams,
  ``Prospects for dark matter searches in the pMSSM,''
  JHEP {\bf 1502}, 014 (2015)
  [arXiv:1411.5214 [hep-ph]].

\bibitem{Fowlie}
  A.~Fowlie, K.~Kowalska, L.~Roszkowski, E.~M.~Sessolo and Y.~L.~S.~Tsai,
  ``Dark matter and collider signatures of the MSSM,''
  Phys.\ Rev.\ D {\bf 88}, 055012 (2013)
  [arXiv:1306.1567 [hep-ph]].

 
  \bibitem{CS} 
  M.~Cirelli and A.~Strumia,
  ``Minimal Dark Matter: Model and results,''
  New J.\ Phys.\  {\bf 11}, 105005 (2009)
  [arXiv:0903.3381 [hep-ph]].

\bibitem{Cirelli2006} 
  M.~Cirelli, N.~Fornengo and A.~Strumia,
  ``Minimal dark matter,''
  Nucl.\ Phys.\ B {\bf 753}, 178 (2006)
  [hep-ph/0512090].

\bibitem{BB}
  Y.~Bai and J.~Berger,
  ``Fermion Portal Dark Matter,''
  JHEP {\bf 1311}, 171 (2013)
  [arXiv:1308.0612 [hep-ph]].
 
   
  
  \bibitem{ChuaWong} 
  C.~K.~Chua and G.~G.~Wong,
  ``Study of Majorana Fermionic Dark Matter,''
  Phys.\ Rev.\ D {\bf 94}, no. 3, 035002 (2016)
  [arXiv:1512.01991 [hep-ph]].
  
  \bibitem{chua}
   C.~K.~Chua and R.~C.~Hsieh,
  ``Study of Dirac fermionic dark matter,''
  Phys.\ Rev.\ D {\bf 88}, no. 3, 036011 (2013)
  [arXiv:1305.7008 [hep-ph]].
  
  \bibitem{APQ}
  A.~Alves, S.~Profumo and F.~S.~Queiroz,
  ``The dark $Z^{'}$ portal: direct, indirect and collider searches,''
  JHEP {\bf 1404}, 063 (2014)
  [arXiv:1312.5281 [hep-ph]].

 \bibitem{Dutra} 
  M.~Dutra, C.~A.~de S. Pires and P.~S.~Rodrigues da Silva,
  ``Majorana dark matter through a narrow Higgs portal,''
  JHEP {\bf 1509}, 147 (2015)
  [arXiv:1504.07222 [hep-ph]].
  
\bibitem{Yaguna} 
  C.~E.~Yaguna,
  ``Singlet-Doublet Dirac Dark Matter,''
  Phys.\ Rev.\ D {\bf 92}, no. 11, 115002 (2015)
  [arXiv:1510.06151 [hep-ph]].
  
\bibitem{Dedes} 
  A.~Dedes, D.~Karamitros and V.~C.~Spanos,
  ``Effective Theory for Electroweak Doublet Dark Matter,''
  Phys.\ Rev.\ D {\bf 94}, no. 9, 095008 (2016)
  [arXiv:1607.05040 [hep-ph]].
  
  \bibitem{Arcadi} 
  G.~Arcadi, M.~D.~Campos, M.~Lindner, A.~Masiero and F.~S.~Queiroz,
  ``The Dark Sequential Z' Portal: Collider and Direct Detection Experiments,''
  arXiv:1708.00890 [hep-ph].


\bibitem{Agrawal}
  P.~Agrawal, Z.~Chacko, C.~Kilic and R.~K.~Mishra,
  ``A Classification of Dark Matter Candidates with Primarily Spin-Dependent Interactions with Matter,''
  arXiv:1003.1912 [hep-ph].
  
  \bibitem{DFT} 
  A.~DiFranzo, P.~J.~Fox and T.~M.~P.~Tait,
  ``Vector Dark Matter through a Radiative Higgs Portal,''
  JHEP {\bf 1604}, 135 (2016)
  [arXiv:1512.06853 [hep-ph]].
  
  
  \bibitem{KMY} 
  J.~Kumar, D.~Marfatia and D.~Yaylali,
  ``Vector dark matter at the LHC,''
  Phys.\ Rev.\ D {\bf 92}, no. 9, 095027 (2015)
  [arXiv:1508.04466 [hep-ph]].

\bibitem{BKMNT} 
  G.~Bambhaniya, J.~Kumar, D.~Marfatia, A.~C.~Nayak and G.~Tomar,
  ``Vector dark matter annihilation with internal bremsstrahlung,''
  Phys.\ Lett.\ B {\bf 766}, 177 (2017)
  [arXiv:1609.05369 [hep-ph]].
  
  \bibitem{BBP} 
  B.~Barman, S.~Bhattacharya, S.~K.~Patra and J.~Chakrabortty,
  ``Non-Abelian Vector Boson Dark Matter, its Unified Route and signatures at the LHC,''
  arXiv:1704.04945 [hep-ph].

\bibitem{DGM} 
  M.~Duch, B.~Grzadkowski and M.~McGarrie,
  ``A stable Higgs portal with vector dark matter,''
  JHEP {\bf 1509}, 162 (2015)
  [arXiv:1506.08805 [hep-ph]].
  
  \bibitem{CT} 
  S.~Di Chiara and K.~Tuominen,
  ``A minimal model for SU(N ) vector dark matter,''
  JHEP {\bf 1511}, 188 (2015)
  [arXiv:1506.03285 [hep-ph]].
  
  \bibitem{KMNO} 
  S.~Kanemura, S.~Matsumoto, T.~Nabeshima and N.~Okada,
  ``Can WIMP Dark Matter overcome the Nightmare Scenario?,''
  Phys.\ Rev.\ D {\bf 82}, 055026 (2010)
  [arXiv:1005.5651 [hep-ph]].
  
  \bibitem{LLM} 
  O.~Lebedev, H.~M.~Lee and Y.~Mambrini,
  ``Vector Higgs-portal dark matter and the invisible Higgs,''
  Phys.\ Lett.\ B {\bf 707}, 570 (2012)
  [arXiv:1111.4482 [hep-ph]].

  \bibitem{KT1} 
  A.~Karam and K.~Tamvakis,
  ``Dark Matter from a Classically Scale-Invariant $SU(3)_X$,''
  Phys.\ Rev.\ D {\bf 94}, no. 5, 055004 (2016)
  [arXiv:1607.01001 [hep-ph]].
  
  \bibitem{KT2} 
  A.~Karam and K.~Tamvakis,
  ``Dark matter and neutrino masses from a scale-invariant multi-Higgs portal,''
  Phys.\ Rev.\ D {\bf 92}, no. 7, 075010 (2015)
  [arXiv:1508.03031 [hep-ph]].
  %
  

  
   %

  
  %
  \bibitem{Cirelli2009} 
  M.~Cirelli and A.~Strumia,
  ``Minimal Dark Matter: Model and results,''
  New J.\ Phys.\  {\bf 11}, 105005 (2009)
  [arXiv:0903.3381 [hep-ph]].
  
  %
  \bibitem{HINT} 
  J.~Hisano, K.~Ishiwata, N.~Nagata and T.~Takesako,
  ``Direct Detection of Electroweak-Interacting Dark Matter,''
  JHEP {\bf 1107}, 005 (2011)
  [arXiv:1104.0228 [hep-ph]].
  
  %
  \bibitem{Essig} 
  R.~Essig,
  ``Direct Detection of Non-Chiral Dark Matter,''
  Phys.\ Rev.\ D {\bf 78}, 015004 (2008)
  [arXiv:0710.1668 [hep-ph]].
  
\bibitem{Menendez}
  J.~Menendez, D.~Gazit and A.~Schwenk,
  ``Spin-dependent WIMP scattering off nuclei,''
  Phys.\ Rev.\ D {\bf 86}, 103511 (2012)
  [arXiv:1208.1094 [astro-ph.CO]].
 
 \bibitem{Cheng1}
 T.~P.~Cheng,
  ``Chiral Symmetry and the Higgs Nucleon Coupling,''
  Phys.\ Rev.\ D {\bf 38}, 2869 (1988).
  
 \bibitem{Cheng2}
  H.~Y.~Cheng,
  ``Low-energy Interactions of Scalar and Pseudoscalar Higgs Bosons With Baryons,''
 Phys.\ Lett.\ B {\bf 219}, 347 (1989).
  
63
  \bibitem{GLS}
  J.~Gasser, H.~Leutwyler and M.~E.~Sainio,
  ``Sigma term update,''
  Phys.\ Lett.\ B {\bf 253}, 252 (1991).
  
  \bibitem{Alarcon2011} 
  J.~M.~Alarcon, J.~Martin Camalich and J.~A.~Oller,
  Phys.\ Rev.\ D {\bf 85}, 051503 (2012)
  [arXiv:1110.3797 [hep-ph]].
  
\bibitem{Alarcon2012} 
  J.~M.~Alarcon, L.~S.~Geng, J.~Martin Camalich and J.~A.~Oller,
  Phys.\ Lett.\ B {\bf 730}, 342 (2014)
  [arXiv:1209.2870 [hep-ph]].
  
  
\bibitem{Cheng3} 
  H.~Y.~Cheng and C.~W.~Chiang,
  ``Revisiting Scalar and Pseudoscalar Couplings with Nucleons,''
  JHEP {\bf 1207}, 009 (2012)
  [arXiv:1202.1292 [hep-ph]].
  
\bibitem{EFO}
  J.~R.~Ellis, A.~Ferstl and K.~A.~Olive,
  ``Reevaluation of the elastic scattering of supersymmetric dark matter,''
  Phys.\ Lett.\ B {\bf 481}, 304 (2000)
  [hep-ph/0001005].

\bibitem{Mallot}
  G.~K.~Mallot,
  ``The Spin structure of the nucleon,''
  Int.\ J.\ Mod.\ Phys.\ A {\bf 15S1}, 521 (2000)
  [eConf C {\bf 990809}, 521 (2000)]
  [hep-ex/9912040].



 \bibitem{SVZ}
  M.~A.~Shifman, A.~I.~Vainshtein and V.~I.~Zakharov,
  ``Remarks on Higgs Boson Interactions with Nucleons,''
  Phys.\ Lett.\ B {\bf 78}, 443 (1978).
  
\bibitem{Hisano2012} 
  J.~Hisano, K.~Ishiwata and N.~Nagata,
  Phys.\ Rev.\ D {\bf 87}, 035020 (2013)
  [arXiv:1210.5985 [hep-ph]].
  
  
 \bibitem{Ressell}
  M.~T.~Ressell, M.~B.~Aufderheide, S.~D.~Bloom, K.~Griest, G.~J.~Mathews and D.~A.~Resler,
  ``Nuclear shell model calculations of neutralino - nucleus cross-sections for Si-29 and Ge-73,''
  Phys.\ Rev.\ D {\bf 48}, 5519 (1993).

%
\bibitem{LS} 
  J.~D.~Lewin and P.~F.~Smith,
  ``Review of mathematics, numerical factors, and corrections for dark matter experiments based on elastic  nuclear recoil,''
  Astropart.\ Phys.\  {\bf 6}, 87 (1996).

%
\bibitem{VKMHS} 
  L.~Vietze, P.~Klos, J.~Menendez, W.~C.~Haxton and A.~Schwenk,
  ``Nuclear structure aspects of spin-independent WIMP scattering off xenon,''
  Phys.\ Rev.\ D {\bf 91}, no. 4, 043520 (2015)
  [arXiv:1412.6091 [nucl-th]].


  
\bibitem{Hisano2015} 
  J.~Hisano, K.~Ishiwata and N.~Nagata,
  ``QCD Effects on Direct Detection of Wino Dark Matter,''
  JHEP {\bf 1506}, 097 (2015)
  [arXiv:1504.00915 [hep-ph]].
  
\bibitem{Owens2012} 
  J.~F.~Owens, A.~Accardi and W.~Melnitchouk,
  ``Global parton distributions with nuclear and finite-$Q^2$ corrections,''
  Phys.\ Rev.\ D {\bf 87}, no. 9, 094012 (2013)
  [arXiv:1212.1702 [hep-ph]].
  
\bibitem{XENON100SD}
  E.~Aprile {\it et al.} [XENON100 Collaboration],
  ``Limits on spin-dependent WIMP-nucleon cross sections from 225 live days of XENON100 data,''
  Phys.\ Rev.\ Lett.\  {\bf 111}, no. 2, 021301 (2013)
  [arXiv:1301.6620 [astro-ph.CO]].
  
  %
\bibitem{Engel:1992bf} 
  J.~Engel, S.~Pittel and P.~Vogel,
  ``Nuclear physics of dark matter detection,''
  Int.\ J.\ Mod.\ Phys.\ E {\bf 1}, 1 (1992).
  
%
\bibitem{CKW} 
  B.~Cabrera, L.~M.~Krauss and F.~Wilczek,
  ``Bolometric Detection of Neutrinos,''
  Phys.\ Rev.\ Lett.\  {\bf 55}, 25 (1985).

%
\bibitem{MF} 
  J.~Monroe and P.~Fisher,
  ``Neutrino Backgrounds to Dark Matter Searches,''
  Phys.\ Rev.\ D {\bf 76}, 033007 (2007)
  [arXiv:0706.3019 [astro-ph]].
  
  %
\bibitem{BSF} 
  J.~Billard, L.~Strigari and E.~Figueroa-Feliciano,
  ``Implication of neutrino backgrounds on the reach of next generation dark matter direct detection experiments,''
  Phys.\ Rev.\ D {\bf 89}, no. 2, 023524 (2014)
  [arXiv:1307.5458 [hep-ph]].
  
\bibitem{Kolb}
   E.~W.~Kolb and M.~S.~Turner,
  ``The Early Universe,''
  Front.\ Phys.\  {\bf 69}, 1 (1990).
  
\bibitem{CR:03}
  T.~S.~Coleman and M.~Roos,
  ``Effective degrees of freedom during the radiation era,''
  Phys.\ Rev.\ D {\bf 68}, 027702 (2003)
  [astro-ph/0304281].
  
\bibitem{GG}
  P.~Gondolo and G.~Gelmini,
  ``Cosmic abundances of stable particles: Improved analysis,''
  Nucl.\ Phys.\ B {\bf 360}, 145 (1991).
  
\bibitem{Hisano:2004} 
  J.~Hisano, S.~Matsumoto and M.~M.~Nojiri,
  ``Explosive dark matter annihilation,''
  Phys.\ Rev.\ Lett.\  {\bf 92}, 031303 (2004)
  [hep-ph/0307216].
  
\bibitem{Hisano:2005} 
  J.~Hisano, S.~Matsumoto, M.~M.~Nojiri and O.~Saito,
  ``Non-perturbative effect on dark matter annihilation and gamma ray signature from galactic center,''
  Phys.\ Rev.\ D {\bf 71}, 063528 (2005)
  [hep-ph/0412403].

\bibitem{Hisano:2006nn} 
  J.~Hisano, S.~Matsumoto, M.~Nagai, O.~Saito and M.~Senami,
  ``Non-perturbative effect on thermal relic abundance of dark matter,''
  Phys.\ Lett.\ B {\bf 646}, 34 (2007)
  [hep-ph/0610249].

%
\bibitem{AFSW} 
  N.~Arkani-Hamed, D.~P.~Finkbeiner, T.~R.~Slatyer and N.~Weiner,
  ``A Theory of Dark Matter,''
  Phys.\ Rev.\ D {\bf 79}, 015014 (2009)
  [arXiv:0810.0713 [hep-ph]].
  
 \bibitem{Chun1} 
  E.~J.~Chun, J.~C.~Park and S.~Scopel,
  JCAP {\bf 1212}, 022 (2012)
  [arXiv:1210.6104 [astro-ph.CO]].
  
  \bibitem{Chun2} 
  E.~J.~Chun and J.~C.~Park,
  ``Electro-Weak Dark Matter: non-perturbative effect confronting indirect detections,''
  Phys.\ Lett.\ B {\bf 750}, 372 (2015)
  [arXiv:1506.07522 [hep-ph]].
  
 \bibitem{Chun3} 
  E.~J.~Chun, S.~Jung and J.~C.~Park,
  ``Very Degenerate Higgsino Dark Matter,''
  JHEP {\bf 1701}, 009 (2017)
  [arXiv:1607.04288 [hep-ph]].
  
 \bibitem{Kochanek:1995xv} 
  C.~S.~Kochanek,
  ``The Mass of the Milky Way galaxy,''
  Astrophys.\ J.\  {\bf 457}, 228 (1996)
  [astro-ph/9505068].

\bibitem{Strumia} 
  A.~Strumia,
  ``Sommerfeld corrections to type-II and III leptogenesis,''
  Nucl.\ Phys.\ B {\bf 809}, 308 (2009)
  [arXiv:0806.1630 [hep-ph]].

\bibitem{Cassel} 
  S.~Cassel,
  ``Sommerfeld factor for arbitrary partial wave processes,''
  J.\ Phys.\ G {\bf 37}, 105009 (2010)
  [arXiv:0903.5307 [hep-ph]].

  \bibitem{Feng:2010zp}
  J.~L.~Feng, M.~Kaplinghat and H.~-B.~Yu,
  ``Sommerfeld Enhancements for Thermal Relic Dark Matter,''
  Phys.\ Rev.\ D {\bf 82}, 083525 (2010)
  [arXiv:1005.4678 [hep-ph]];
  T.~R.~Slatyer,
  ``The Sommerfeld enhancement for dark matter with an excited state,''
  JCAP {\bf 1002}, 028 (2010)
  [arXiv:0910.5713 [hep-ph]].
  
 \bibitem{Coulomb}
  L.~D.~Landau and E.~M.~Lifshitz,
  ``Quantum Mechanics : Non-Relativistic Theory,''
  (Course of theoretical physics III, 3rd ed.), Pergamon Press (1977).
  
   \bibitem{Lebedev0}
N. N. Lebedev, translated by R. Silverman, ``Special Functions \& Their Applications," (Dover Books on Mathematics),
Dover Publications, Revised ed. edition (1972), pp. 15, 261. 
  
   \bibitem{Landau}
 E.~M.~Lifshitz and L.~P.~Pitaevskii, ``Relativistic Quantum Theory,'' 
 ( Volume 4 part 1 of A Course of Theoretical Physics ), 
  Pergamon Press (1973). 

  
  \bibitem{Lebedev}
N. N. Lebedev, translated by R. Silverman, ``Special Functions \& Their Applications," (Dover Books on Mathematics),
Dover Publications, Revised ed. edition (1972), p. 249.

\bibitem{Sakurai}
 J.~J.~Sakurai and J.~Napolitano,
  ``Modern quantum physics,''
  Boston, USA: Addison-Wesley (2011). 

\bibitem{Iengo:2009ni} 
  R.~Iengo,
  ``Sommerfeld enhancement: General results from field theory diagrams,''
  JHEP {\bf 0905}, 024 (2009)
  [arXiv:0902.0688 [hep-ph]].
  

  

\end{thebibliography}
\end{document}